\documentclass[aps,twocolumn,floats,epsfig,pdflatex,longbibliography]{revtex4-2}
\usepackage{color,ulem,verbatim,float}
\usepackage{amssymb,amsbsy,amsmath,mathrsfs}
\usepackage{amssymb}
\usepackage{amsbsy}
\usepackage{amsmath}
\usepackage{float}
\usepackage{graphicx}
\usepackage{xcolor}
\usepackage{bm}
\usepackage{natbib}
\usepackage[plainpages=false,pdfpagelabels,colorlinks=true,linkcolor=red,urlcolor=blue,citecolor=blue,pdftitle={},pdfauthor={},pdfdisplaydoctitle={},pdfduplex=DuplexFlipLongEdge]{hyperref}

\newcommand\bea{\begin{eqnarray}}
\newcommand\eea{\end{eqnarray}}
\newcommand\beq{\begin{equation}}
\newcommand\eeq{\end{equation}}

\newcommand{\non}{\nonumber}
\newcommand{\al}{\alpha}

\newcommand{\ka}{\kappa}

\newcommand{\si}{\sigma}

\newcommand{\om}{\omega}

\newcommand{\St}{\mathcal{S}}
\newcommand{\hd}{h_{D}^{x}}

\newcommand{\Sgn}[0]{\mathrm{Sgn}}

\begin{document}


\title{Dynamical freezing in the thermodynamic limit: the strongly driven ensemble}



\author{Asmi Haldar$^{1,4}$, Anirban Das$^2$, Sagnik Chaudhuri$^2$, 
Luke Staszewski$^1$, Alexander Wietek$^1$, Frank Pollmann$^3$,
Roderich Moessner$^1$, and Arnab Das$^2$}

\affiliation{$^1$Max Planck Institute for the Physics of Complex Systems, N\"{o}thnitzer Stra\ss e 38, Dresden 01187, Germany \\
$^2$Indian Association for the Cultivation of Science (School of Physical Sciences), P.O. Jadavpur University\\ 2A \& 2B Raja S. C. Mullick Road, Kolkata 700032, West Bengal, India \\
$^3$Technical University of Munich
85748 Garching b. M\"{u}nchen, James-Franck-Str. 1/I, Germany\\
$^4$ Laboratoire de Physique
Théorique - IRSAMC 
Paul Sabatier University,
118, route de Narbonne building 3R1B4
31400 Toulouse, France
}

\date\today

\begin{abstract}
The ergodicity postulate, a foundational pillar of Gibbsian statistical mechanics predicts that a periodically driven (Floquet) system in the absence of any conservation law heats to a featureless  `infinite temperature' state. Here, we find--for a clean and interacting generic spin chain subject to a {\it strong} driving field--that this can be prevented by the emergence of {\it approximate but stable} conservation-laws not present in the undriven system. We identify their origin: they do not necessarily owe their stability to familiar protections by symmetry, topology, disorder, or even high energy costs. We show numerically, {\it in the thermodynamic limit,} that when required by these emergent conservation-laws, the entanglement-entropy density of an infinite subsystem remains zero over our entire simulation time of several decades in natural units. We further provide a recipe for designing such conservation laws with high accuracy. Finally, we present an ensemble description, which we call the strongly driven ensemble incorporating these constraints. This provides a way to control many-body chaos through stable Floquet-engineering. Strong signatures of these conservation-laws should be experimentally accessible since they manifest in all length and time scales. Variants of the spin model we have used, have already been realized using Rydberg-dressed atoms.
\end{abstract}

\maketitle

Thermodynamics is based on maximizing entropy subject to the constraints imposed by conservation laws~\cite{Jaynes}. The `ergodicity postulate' of equal a priori probability (see, e.g.,~\cite{Ma_Stat_Mech}),
on which the entire structure of statistical mechanics rests, connects this macroscopic description to the microscopic world.    In the context of quantum many-body systems, a counterpart of the ergodicity postulate is  
the eigenstate thermalization hypothesis (see review:~\cite{dalessio_kafri_16}). Its cousin in periodically-driven settings,
Floquet-thermalization~\cite{LDM_PRE, Rigol_Infinite_T}, is very simple: it states that a driven system without conservation laws will heat up until its entropy is maximized, and its state is {\it entirely} featureless.

Prominent exceptions to thermalization
are systems with strong disorder resulting in non-ergodic phases like quantum spin glasses~\cite{Binder_Young_RMP, Parisi_SpinGlass_Book}, quantum many-body localized (MBL) states (see reviews:~\cite{MBL_AdP, Alet_MBL_Review, Abanin_Bloch_MBL_RMP}) and its Floquet version (the Floquet-MBL)~\cite{FlqMBL1,FlqMBL2,Flq_MBL_Sierant}. These systems are believed not to thermalize, though the extent of the MBL phase and its stability in infinite systems are still subject to current research~\cite{Prosen_Suntaj_NoMBL_1,Anatoli_Sels_NoMBL,Morningstar_NoMBL,deroeck2024}. Recently, Hilbert-space fragmentation observed in finite systems also bears the promise of an independent route to ergodicity breaking~\cite{Frank_Sala_Hilbert_Space_Fragmentation_PRX, Vedika_Nandkishore_Hilbert_Space_Fragmentation}. \\

\noindent
Here we define the `strongly driven ensemble' to describe a different route to breaking ergodicity which captures Dynamical Freezing (DF) in an infinite, closed interacting quantum system that is subject to strong periodic driving. In a nutshell, this phenomenon encompasses
the generation by the strong driving field of new global conservation-laws that are not present in the undriven system. We refer to these as {\bf Emergent Conserved Operators (ECO)}.

Unlike usual conservation laws, the conservation of ECOs are
{\it approximate}, i.e., they display small fluctuations
and a steady average slightly different from their initial values, yet they are {\it stable}, i.e., the fluctuations do not grow with time.
Both of these deviations (fluctuations and the difference between the average and the initial values) can be reduced at will by increasing the drive strength. ECOs thus break ergodicity, and, as we show here, dominate the steady-state ensemble for local sub-systems observed {stroboscopically} (i.e., at a fixed time within each cycle). 
The ECOs appear when the drive-amplitude crosses a threshold defined as the point 
beyond which the accuracy of the conservation of the ECO grows monotonically with system size and
saturates to a small value as $L\to\infty$.

\noindent
{\it Existing Scenario:} Dynamical Freezing~\cite{AD-DMF} 
has been studied in integrable systems~\cite{AD-DMF, AD-SDG, Mahesh_Freezing, 
 Russomanno_Dynamical_Freezing, Kris-Periodic, Analabha_Dynamical_Freezing, Naveen_Dynamical_Freezing}, and in interacting systems~\cite{Onset, Asmi_DF_PRX_2021, Bhaskar_DF,Diptiman_DF, Analabha_Mori_Rehman_DF, Debanjan_DF_QDot, Debanjan_DF_Transmon,Krishanu_DF}. However, studies for the latter have remained restricted to small system-sizes. Also, only one ECO has been identified so far -- the strong drive term itself~\cite{AD-DMF,Onset,Asmi_DF_PRX_2021}, whose conservation was
 shown to be perpetual in those finite systems.\\

\begin{figure*}[t!]
\begin{center}
\includegraphics[width=0.32\linewidth]{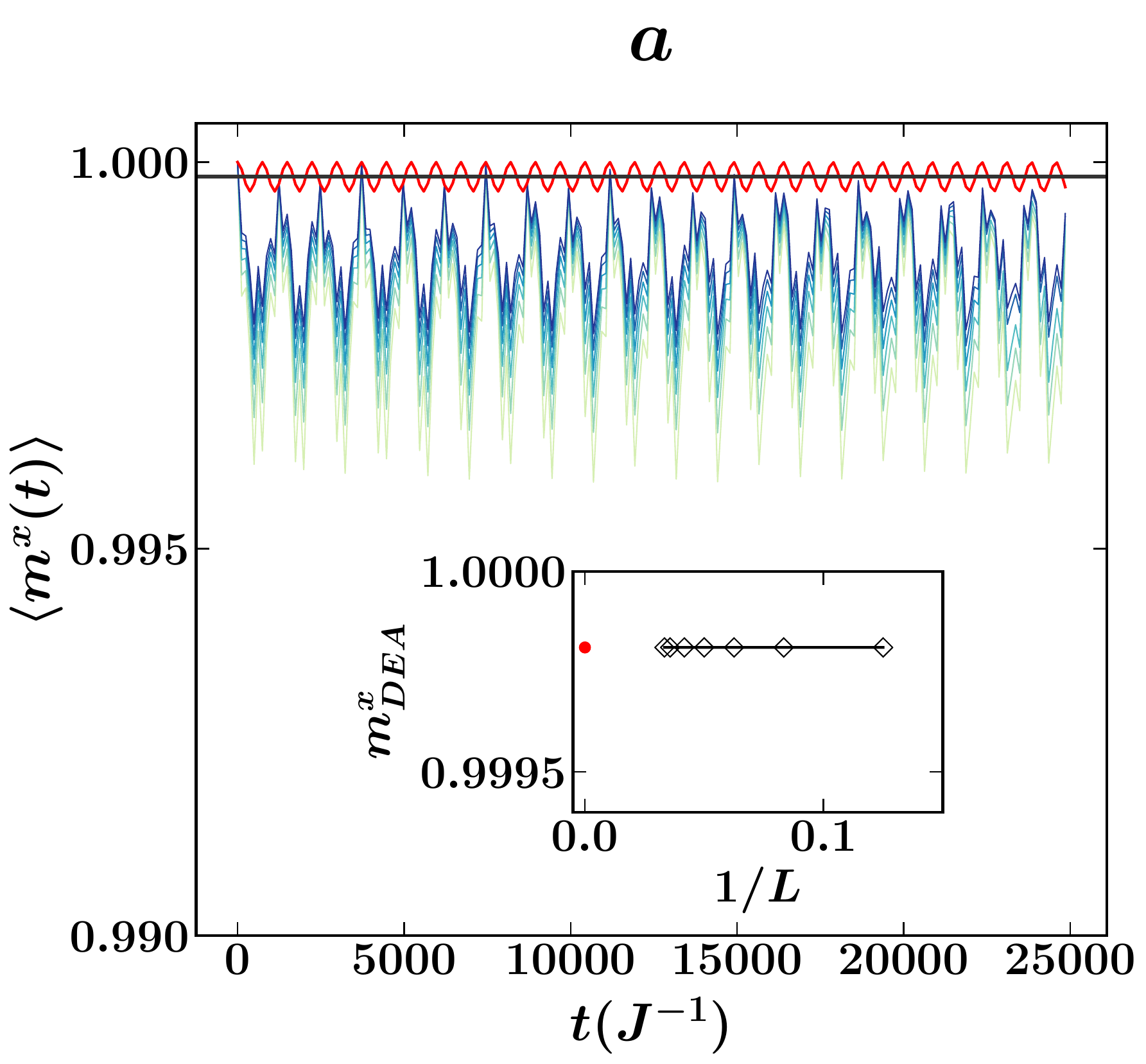}
\includegraphics[width=0.32\linewidth]{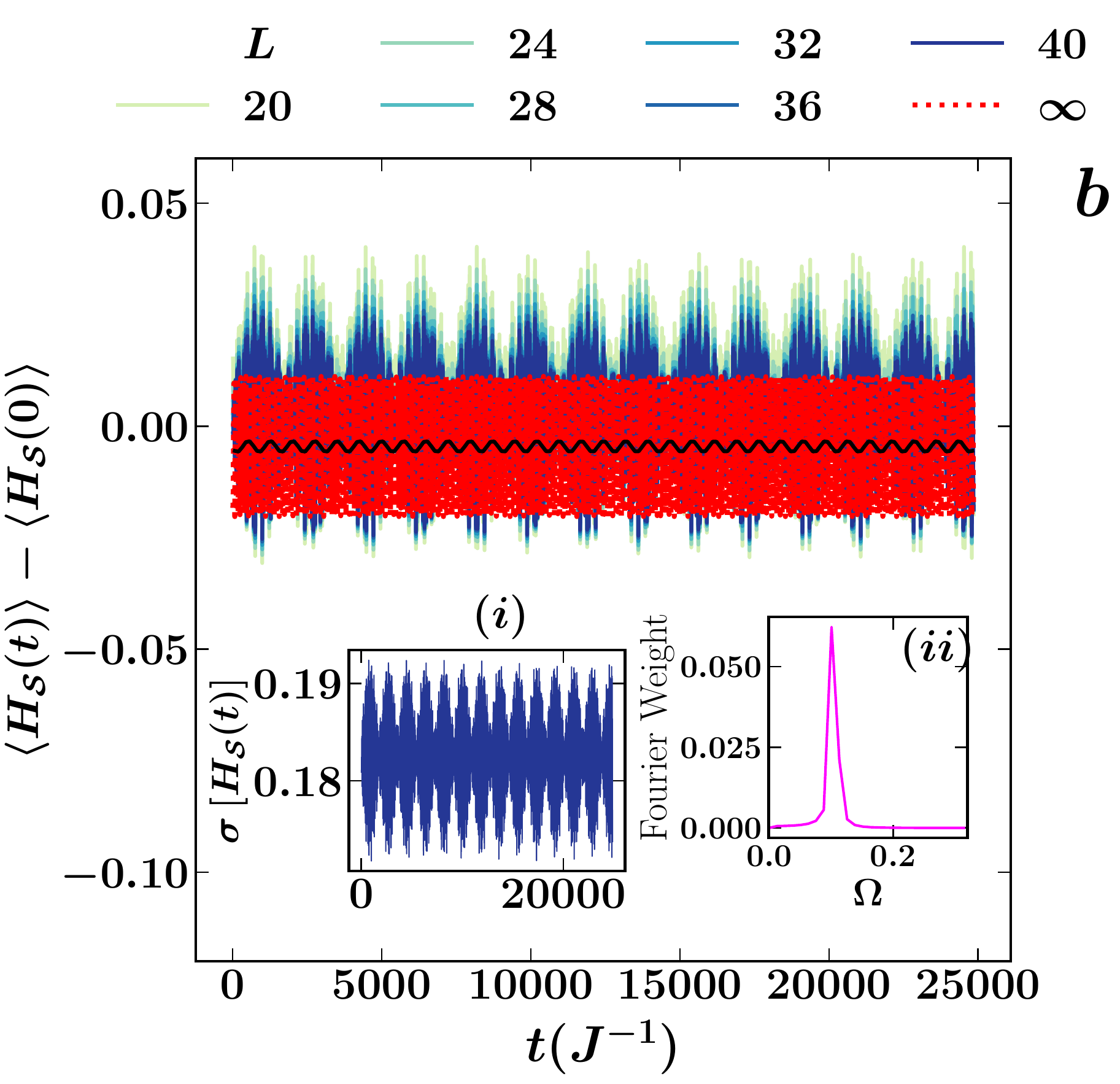}
\includegraphics[width=0.32\linewidth]{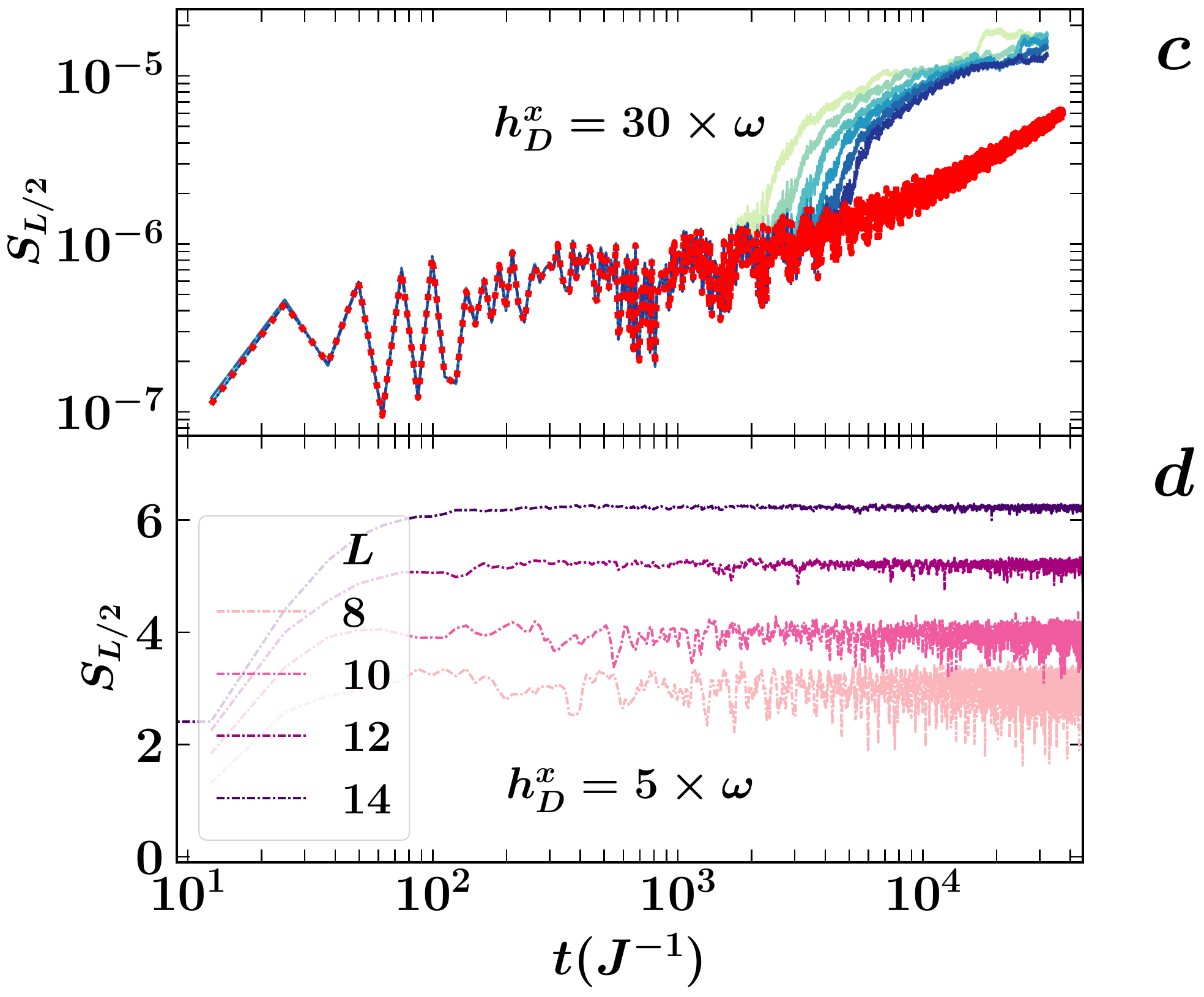}
\end{center}
\caption{{\bf Emergent Conservation
of $m^x$ in an infinite-system:} The legend common to all frames is given above. {\bf (a)} 
Real-time dynamics starting from a fully polarized state in $x-$direction: $|\uparrow\uparrow\uparrow ...\rangle_{x}$ for parameter values $J=2.0, \kappa = 0.5, h^{x}_{0} = 0.15, h^{z}=\sqrt{3}/1.5,  \omega = \phi/1.6, 
h^{x}_{D} = 30\times\omega,$ where $\phi = (\sqrt{5}+1)/2 =$ the Golden Mean. Inset: DEA ($t\to\infty$ limit) of $m^x(t)$ vs $1/L$ showing no 
$L-$dependence - the $L\to\infty$ extrapolation is shown with the red dot. In the main panel, this DEA value (plotted in a solid black horizontal line) is compared with the dynamics of the infinite system (red continuous line). 
We see that {\it these results obtained in the two different limits ($L\to\infty,$ finite $t$) and ($t\to\infty,$ finite $L$) agree with astonishing accuracy.} {\bf (b)} The energy absorption $H_{\St}(t) - H_{\St}(0)$ (main). The thick black line shows the running average over 10 cycles for the
infinite system. Inset~{\bf (i):} Variance ${\Large{\sigma}}\left[H_{\St}(t)\right]= \sqrt{\langle H_{\St}^{2}(t)\rangle - \langle H_{\St} (t)\rangle^{2}}/L$ for $L = 40$ exhibiting no net growth with time. 
Inset~{\bf (ii)} displays Fourier transform of $H_{\St}(t) - H_{\St}(0)$
for $L=\infty$, showing that the dominant time-scales of energy exchange are very short compared to the evolution time, with no perceptible weight around zero frequency.
{\bf (c)} $S_{L/2}$ as a function of time in the frozen regime ($L=20-40,\infty; \hd = 30\times\omega$). This is contrasted with the thermalized dynamics
($L=8,10,12,14; \hd = 5\times\omega$), where a rapid $L-$dependent growth is visible {\bf (d)}. Crucially, the trend of $L-$dependence of $S_{L/2}$ in {\bf (c)} is opposite to that in {\bf (d)}.
}
\label{Fig:1:Real_Time_Dyn_mx_I}
\end{figure*}

\noindent
{\it A Summary of Our Three Main Findings:}
{\bf Firstly} (Fig.~\ref{Fig:1:Real_Time_Dyn_mx_I}), we show that the driven magnetization continues to serve as an ECO without any sign of decay even in an infinite system over several decades of evolution time. The density of the half-chain entanglement entropy remains zero throughout the entire evolution. We show that the threshold field strength above which this freezing is observed is finite in the infinite system; and further that it is consistent with the finite-size threshold estimated from exact-diagonalization ({\bf ED}) results for the $t\to\infty$ limit 
 (Fig.~\ref{Fig:1:Real_Time_Dyn_mx_II}{\bf a}).
{\bf Secondly}, we show that there are further local ECOs that do not require the usual protections of symmetry, topology, disorder, or a high energy cost (Fig.~\ref{Fig:4:Other_Conservations}). We show that their existence sheds light on the intricate pattern of ergodicity breaking as reflected in the long-time entanglement-growth in finite systems (Fig.~\ref{Fig:3:EE_Spectrum}), and show how new ECOs can be designed (Fig.~\ref{Fig:5:Cr_ECO}). {\bf Thirdly,} we show that the conservation-laws of the ECOs are 
respected across the entire Hilbert space, 
and hence in the dynamics with any generic initial state (Fig.~\ref{Fig:1:Real_Time_Dyn_mx_II}{\bf b}, example in~\ref{Fig:1:Real_Time_Dyn_mx_II}{\bf c}). Consequently, instead of the Gibbsian expectation of Floquet thermalisation~\cite{LDM_PRE,Rigol_Infinite_T}, the long-time-average of local operators is given by a Gibbs-like description which we term {\it strongly driven ensemble}, with the ECOs as the effective constraints (Fig.~\ref{Fig:6:DF_Ensemble}).
Furthermore, with exact numerical results for finite systems, we show that
the occurrence of ECOs is not a fine-tuned property of special points in parameter space, but that it is generic for a strong drive (Fig.~\ref{Fig:1:Real_Time_Dyn_mx_II}{\bf d}). 
\\

\section{Emergent Conservation of $m^x$ in the Thermodynamic Limit}
\label{Sec_II:Real_Time_Dynamics}
 We focus on the dynamics of the periodically driven,
 non-integrable Ising spin chain of the following form.
 \bea H(t) &=& H_{0}(t) ~+~ V, ~~ {\rm where} \non \\
H_{0} (t) &=& H_{0}^{x} ~+~ \Sgn (\sin (\om t)) ~H_{D}, ~~ {\rm with} 
\non \\
H_{0}^{x} = &-& ~\sum_{n=1}^L ~J \si_n^x \si_{n+1}^x + \sum_{n=1}^L ~\ka 
\si_n^x \si_{n+2}^x - h_{0}^{x}~\sum_{n=1}^L\si_n^x, \non \\
H_{D} = &-& ~ h_D^x ~\sum_{n=1}^L \si_n^x, ~~ {\rm and} \non \\
V = &-& ~ h^z \sum_{n=1}^L \si_n^z, 
\label{Eq:Hamiltonian_1} 
\eea
\noindent where, $\si_{n}^{x/y/z}$ are the Pauli matrices, and $\Sgn{( ~ )}$ denotes the sign of its
argument.\\

\noindent
{\it Recapitulation of Finite-$L$ Results:}
For finite-size systems amenable to exact numerics, it was
shown that the drive term itself is an ECO for strong drive~\cite{AD-DMF, Onset,Asmi_DF_PRX_2021}. According to those, in this case
the longitudinal magnetization 
\begin{equation}
      m^{x} = \frac{1}{L}\sum_{i}^{L}\sigma_{i}^{x}
    \label{Eq:m^x}
\end{equation}
\noindent 
is an ECO for a finite system for $\hd \gg h^{z}$. The
conservation would be maximally accurate at the {\it freezing peaks}~\cite{Asmi_DF_PRX_2021}, in this case, given by 
\beq
\hd = n\omega,
\label{Freezing_Cond}
\eeq
\noindent
where $n$ is an integer. 
The strong drive term ($m^x$) is the only ECO identified so far, and it was found to be
stably conserved as $t\to\infty.$ Those results are based on the Diagonal-Ensemble-Average discussed below.
Here we explore this phenomenon in an infinite system, uncover the associated phenomenology
, and compare it with finite-size results.

\noindent {\it Diagonal-Ensemble-Average}~({DEA}):
  DEA is the infinite-time limit of the dynamics of a many-body system when all local
  observables reach a steady state. 
  For a periodically driven system observed stroboscopically, if
  $|\mu_{\alpha}\rangle$ denotes the $\alpha$-th-eigenstates of the evolution operator $U(T,0)$ ({\it Floquet Eigenstates}) then the late-time expectation value of any local observable at times $t = nT$ in the $n\to\infty$ limit is given by~\cite{Rigol_Nature,Reimann,Asmi_Flq_Rev} 
  \begin{eqnarray}
  \langle {\cal O} \rangle_{\infty} = \lim_{t\to\infty} \langle \psi(t) |{\cal O}|\psi(t)\rangle &=& \sum_{\alpha}|C_{\alpha}|^2\langle \mu_{\alpha} |{\cal O}| \mu_{\alpha}\rangle, \nonumber \\
  &=& \sum_{\alpha}|C_{\alpha}|^2 {\cal O}_{\alpha},
      \label{Eq:DEA}
   \end{eqnarray}
  \noindent
  where ${\cal O}_{\alpha}$s are the Floquet expectation 
  values, and
  $|\psi(0)\rangle = \sum_{\alpha}C_{\alpha}|\mu_{\alpha}\rangle.$ Thus, the 
  ${\cal O}_{\alpha}$s contain
  all the information about the long-time fate of $\langle{\cal O}(t)\rangle.$ However, for systems with local conservation-laws or finite-size, the limit might not exist in a strict sense for certain initial-states due to the presence of small oscillations about the DEA~\cite{Abanin_Bloch_MBL_RMP,Asmi_DF_PRX_2021}. In those cases (like the present one with ECOs), the DEA accurately gives the limit of the time-averaged dynamics of local observable as $t \to \infty$~\cite{dalessio_kafri_16}. \\

\begin{figure*}[t!]
\begin{center}
\includegraphics[width=0.4\linewidth]{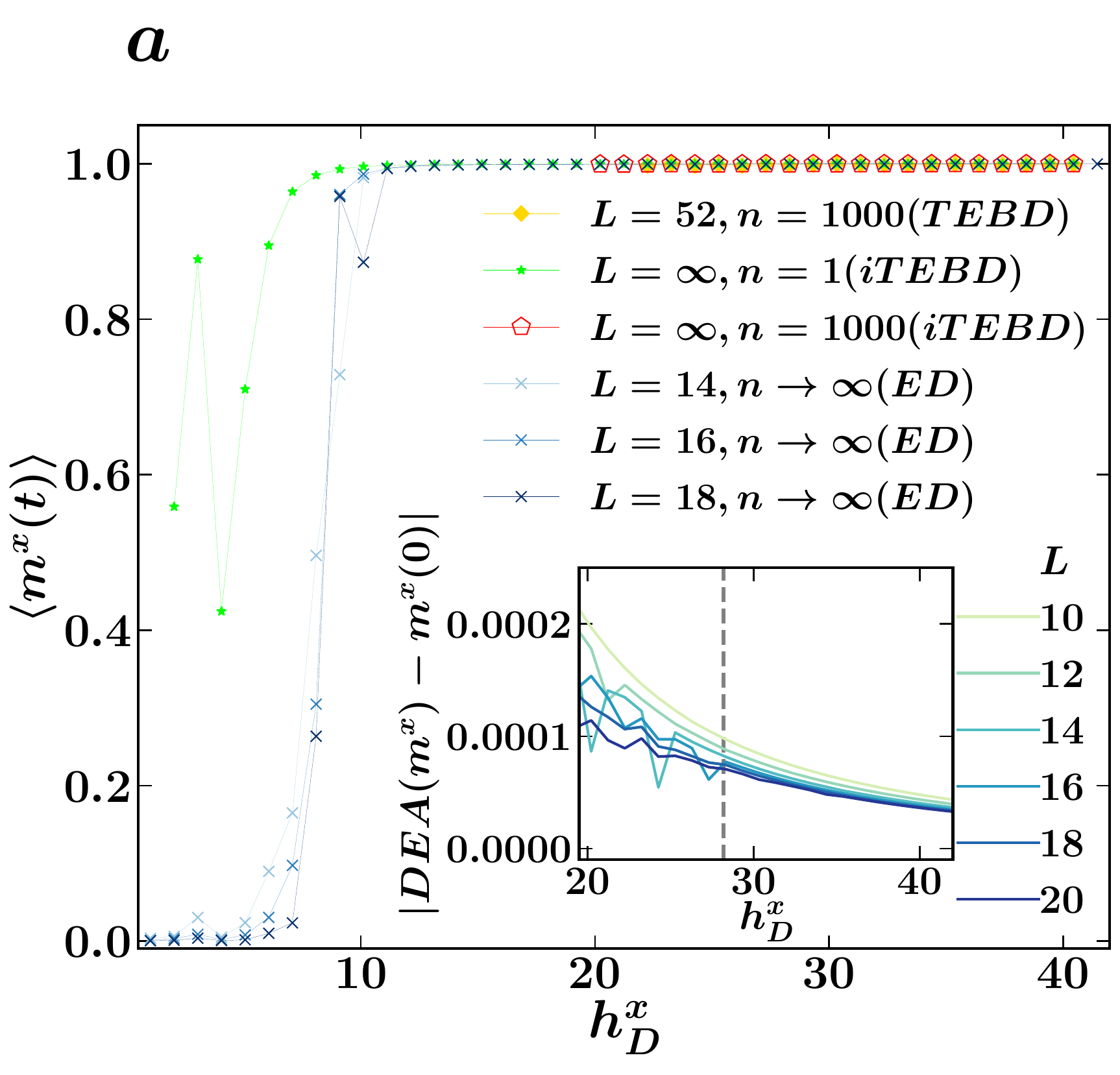}
\includegraphics[width=0.4\linewidth]{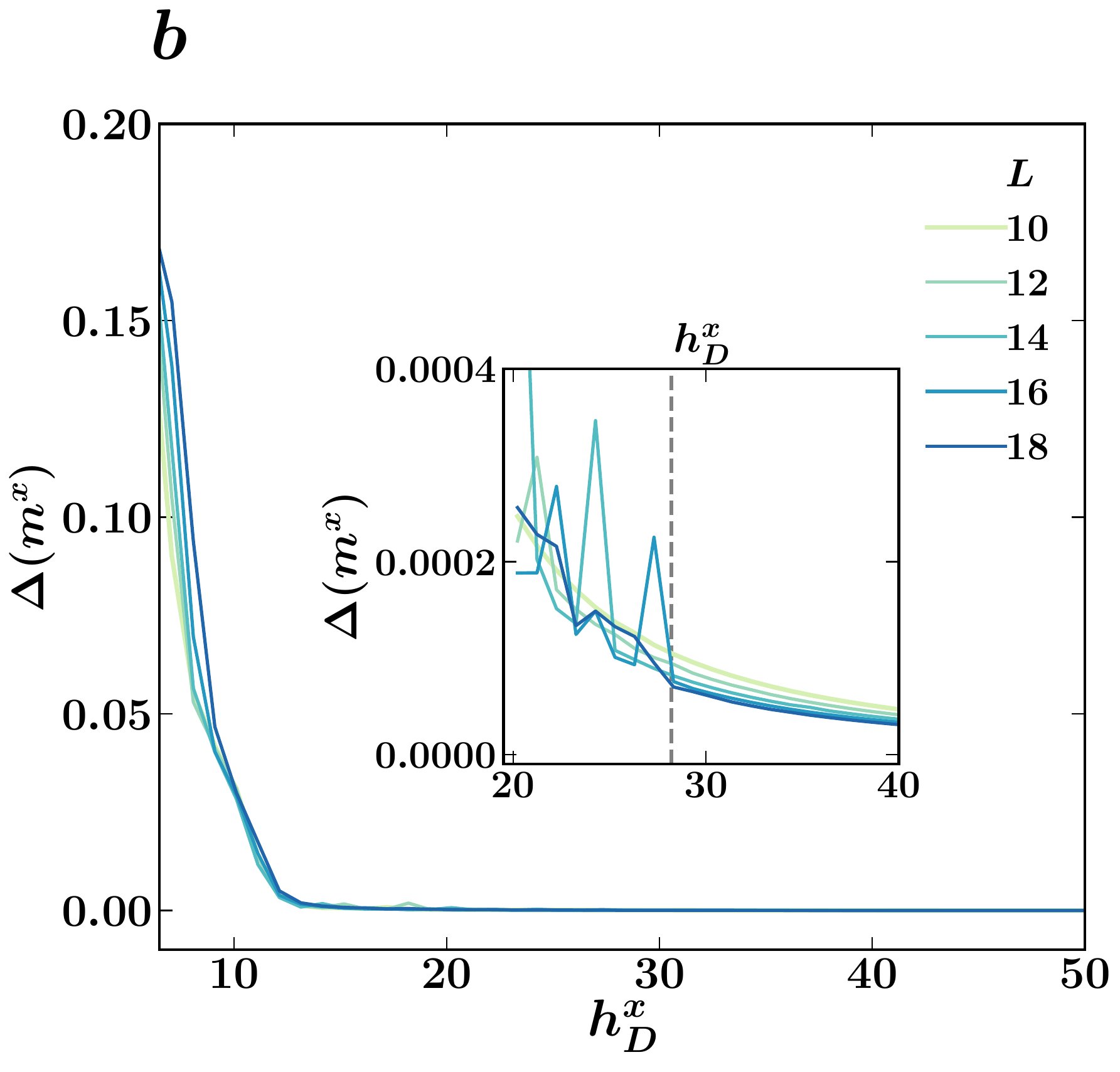}
\includegraphics[width=0.4\linewidth]{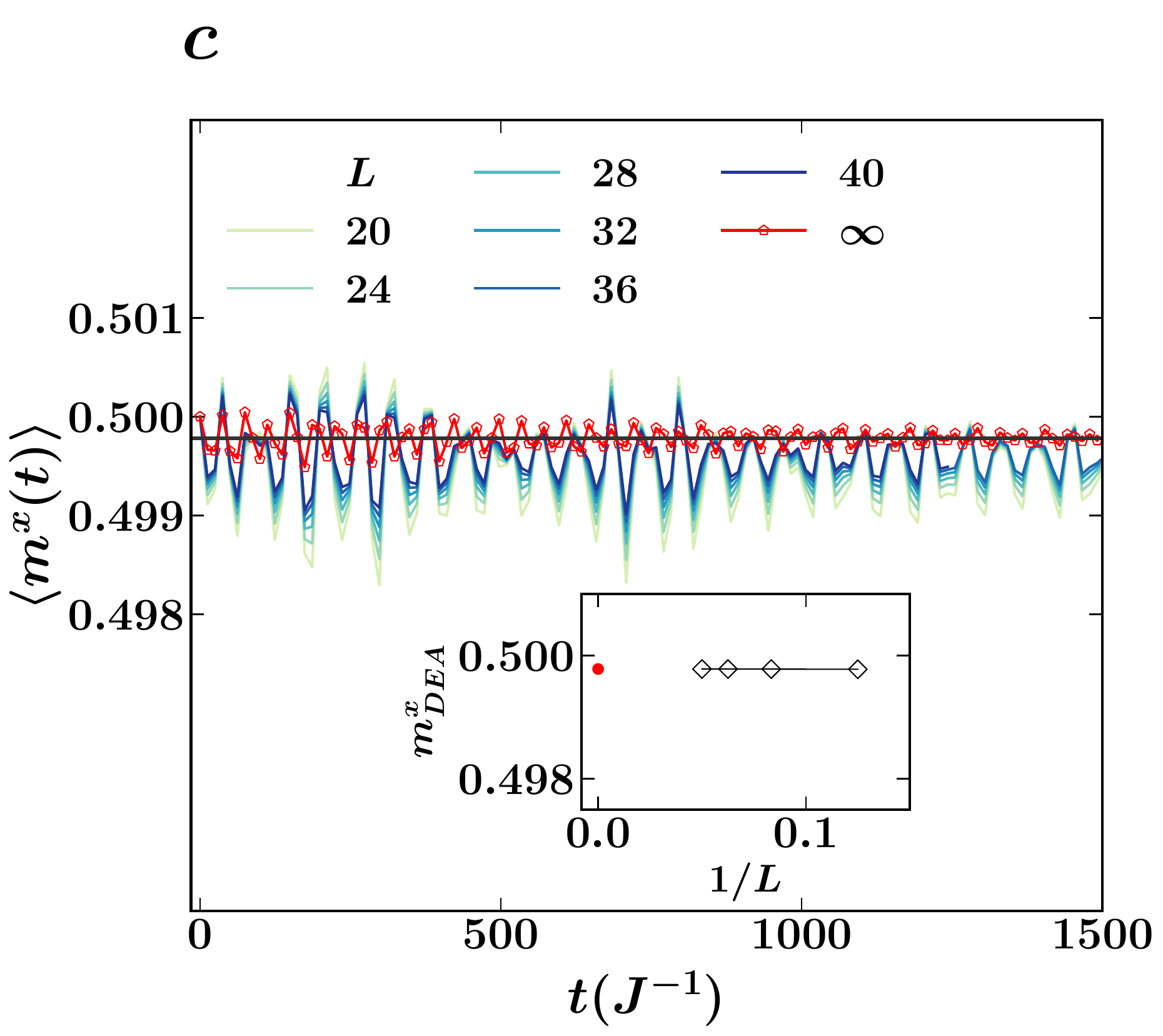}
\includegraphics[width=0.4\linewidth]{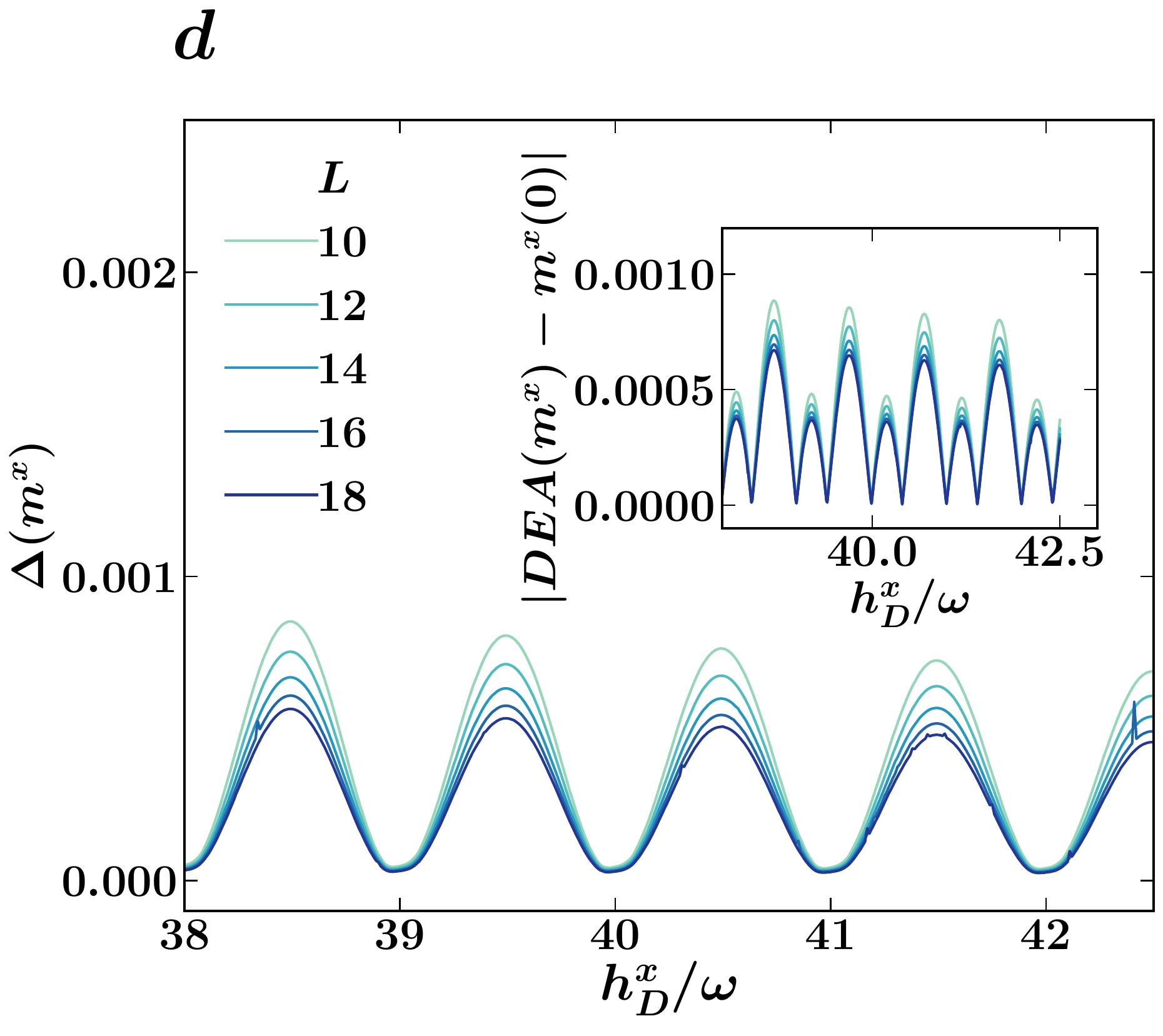}
\end{center}
\caption{
{\bf Emergent Conservation
of $m^x$: threshold estimation for $L=\infty$, stability across the spectrum, and freezing away from the peak:}
{\bf (a)} Comparison of the freezing thresholds (an overestimation) for finite and infinite systems for the real-time dynamics starting from a fully polarized state in $x-$direction: $|\uparrow\uparrow\uparrow ...\rangle_{x}$ and DEA (for $L \le 18$). 
{\bf Inset}: estimated threshold for a high-temperature initial-state with
inverse-temperature $\beta=10^{-2}$ for a
Hamiltonian $H_{\cal X},$ which is $H(0)$ with $h^{z}=1.2, h_{D}^{x} + h_{0}^{x}=5.1, J=1.0, \kappa=0.7$,  
The absolute difference between the DEA and the
initial value of $m^{x}$ is used to
mark the threshold where the $L$-dependence
of the quantity changes trend (marked in the inset with a vertical line).
{\bf{(b)}} Main: $\Delta(m^{x})$ (Eq.~\ref{Eq:Delta_O})
vs $h_{D}^{x}.$ Inset: a zoom-in, with the 
estimated threshold ($h_{D}^{x} \approx 28\times\omega$) marked by 
a vertical line. 
{\bf (c)} Stability of $m^x$ starting from the {\it mid-spectrum state of $H(0):$} $|...\uparrow\uparrow\downarrow\uparrow ... \uparrow\uparrow\downarrow\uparrow ...\rangle_{x}$, with the value of $m^{x} = 0.5$ for $L=\infty$ (iTEBD).
{\bf Inset:} DEA for the same for various values of $L.$
{\bf(d)} Stability of $m^x$ away from the freezing peaks, Eq.~\ref{Freezing_Cond}, (dips in $\Delta(m^x)$), 
continuously through the freezing-valleys (peaks in $\Delta(m^x)$). Stronger freezing for larger $L$ is shown over the entire regime. {\bf Inset:} the same using the absolute difference between the DEA and $m^{x}(0)$; the DEA is for the same thermal initial-state as in {\bf (b)}.  Parameter values: 
 $J=2.0, \kappa = 0.5, h^x_0 = 0.15, h^{z}=\sqrt{3}/1.5,  \omega = \phi/1.6,$ where $\phi =$ Golden-mean.}
\label{Fig:1:Real_Time_Dyn_mx_II}
\end{figure*}    
\noindent
\begin{center}
    {\bf Real-time Dynamics of $m^x$ in an infinite system:}\\
\end{center}

\noindent
In an infinite one-dimensional spin system, the magnetic polarization ($m^x$) is expected to be fragile against a global periodic drive, with the drive steadily increasing the energy density with time, thereby steadily reducing $m^x.$ Yet, here we see a stable freezing/conservation of magnetization in states under periodic drives in an infinite system.
Fig.~\ref{Fig:1:Real_Time_Dyn_mx_I}{\bf a} shows 
$m^{x}(t)$ starting from the initial-state fully polarized in $x-$direction. $m^x$ exhibits  freezing over several decades of time-evolution for several system-sizes $L$ (using time-evolving block decimation - TEBD~\cite{Vidal_2}), and also for the $L\to\infty$ limit (using iTEBD)~\cite{Vidal_3} for $\hd = k\omega$. 
$h_{D}^{x}$ is chosen to be large as suggested from finite-size studies~\cite{Onset}. The inset shows 
no $L-$dependence of the DEA ($t\to\infty$ limit) of $m^x(t)$. The consistency of stability in the $L\to\infty$ limit and $t\to\infty$ limit is visible from the accurate coincidence of the iTEBD dynamics (red line) and the DEA (black horizontal line). $m^x$ exhibits only small fluctuations around its DEA. We have used bond-dimension up to $\chi=1000$ and ensured that the results do not change with increasing $\chi$ (see methods). Note that TEBD can only be done with open boundary conditions, hence the finite size results exhibit a weak $L-$dependence,  which decreases with increasing $L$ and disappears in the $L\to\infty$ limit, while the DEA is shown for periodic boundary conditions.
We followed the finite-size prescription of choosing the parameters
to avoid accidental, isolated resonances~\cite{Asmi_DF_PRX_2021}. 
\\

\begin{center}
   {\bf Coherent Counter-balance in Energy Exchanges:}\\ 
\end{center}
\noindent
The above stability of $m^x$ is explained by
the surprising absence of any net 
energy absorption by the system, 
as measured by the undriven part 
$$
H_{\St} = H_{0}^{x} + V
$$ 
\noindent
of the Hamiltonian (Fig.~\ref{Fig:1:Real_Time_Dyn_mx_I}b, main). 
It shows $\langle H_{\St} \rangle$ for various $L$ including $L=\infty$. 
The Fourier spectrum of 
$\langle H_{\St} \rangle (t)$ for $L=\infty$
(inset {\bf (ii)}) exhibits a single sharp peak at a finite frequency, with no perceptible weight around zero-frequency. This rules out any slow growth of $\langle H_{\St} \rangle$. Here, unlike in prethermalization, the heating is averted because the energy absorbed by the system is counterbalanced {\it accurately and coherently} by the energy lost by it. This coherence (as opposed to Markov-like randomness) is also manifested in the absence of growth of the variance ${\Large{\sigma}}\left[H_{\St}(t)\right]= \sqrt{\langle H_{\St}^{2}(t)\rangle - \langle H_{\St} (t)\rangle^{2}}/L$ for $L = 40$  (inset {\bf (i)}). This sharply contrasts the widespread intuition of the inevitability of the occurrence of a net finite positive heating rate leading to steady energy-growth, based on a Fermi's Golden Rule type expectation for periodically driven systems~\cite{Sakurai_Book} and its several variants including those derived for strongly driven ones~\cite{Tatsuhiko_FGR, Takashi_FGR}.\\
\begin{figure}[htb]
\begin{center}
\includegraphics[width=0.7\linewidth]{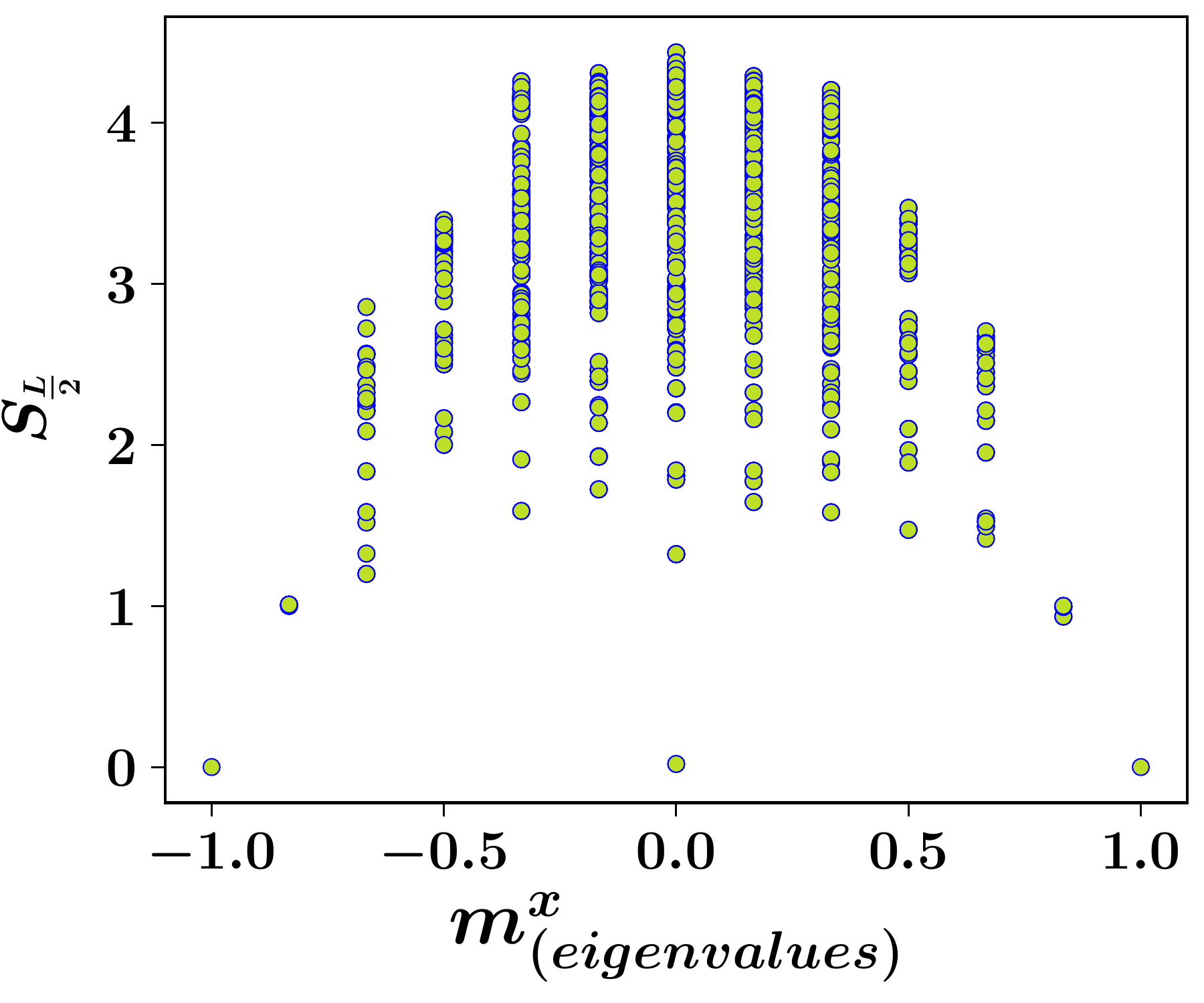}
\end{center}
\caption{
Half chain entanglement entropy $S_{L/2}$ starting from all eigenstates
of $\{\sigma_{i}^{x}\}$ as the initial state, after evolution for $10^8$ cycles, arranged according to the value of $m^x$ of the initial states.
The spread of $S_{L/2}$ within each eigen-subspace of $m^x$ indicates additional dynamical constraints over and above the emergent conservation of $m^x.$ Parameter values: 
 $J=2.0, \kappa = 0.5, h^x_0 = 0.15, h^{z}=\sqrt{3}/1.5, h_{D}^{x} =30\times\omega, \omega = \phi/1.6,$ where $\phi =$ Golden-mean, $L=12$.
}
\label{Fig:3:EE_Spectrum}
\end{figure}    
\begin{figure*}[ht]
\begin{center}
\includegraphics[width=0.4\linewidth]{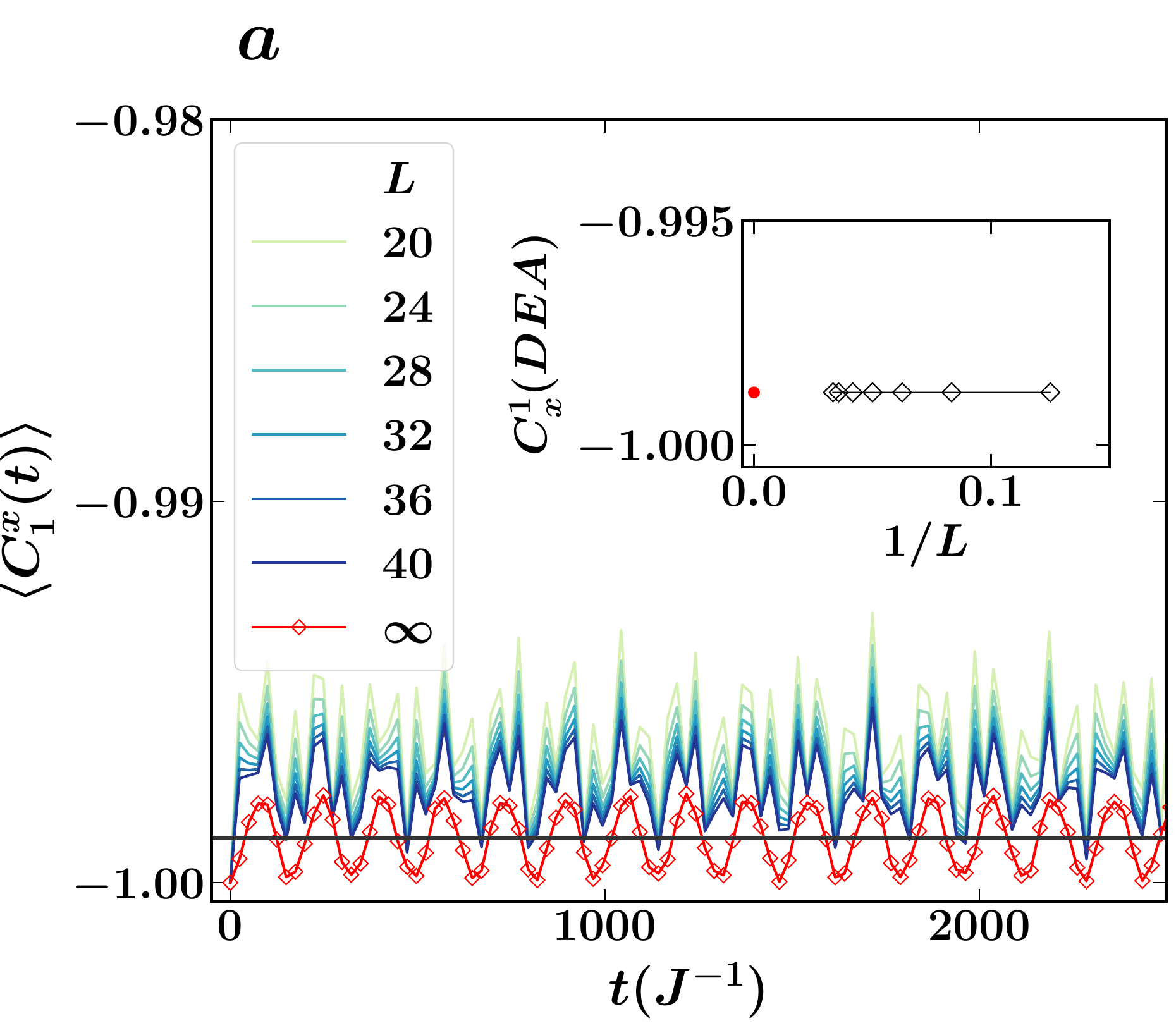}
\includegraphics[width=0.4\linewidth]{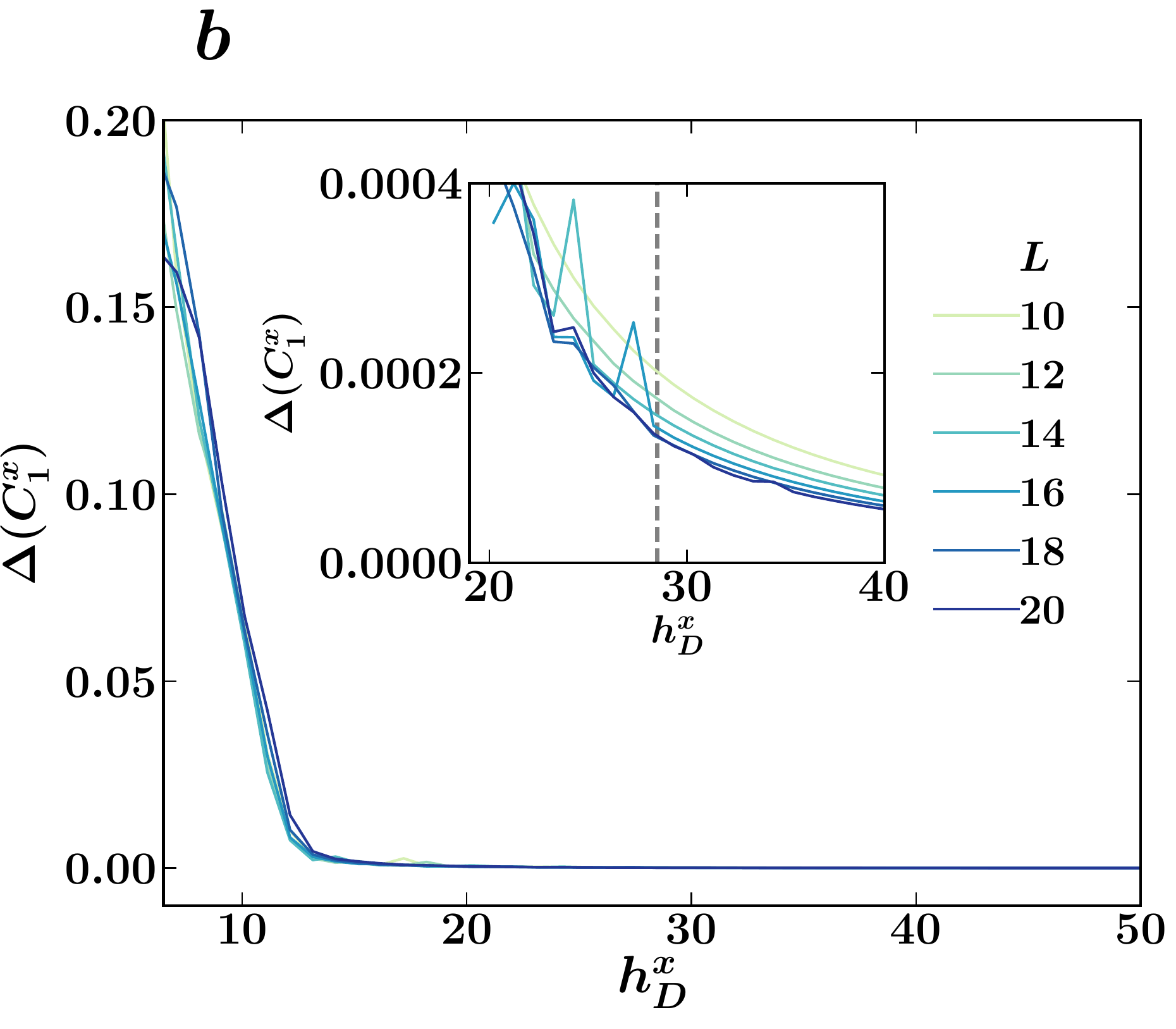}
\includegraphics[width=0.4\linewidth]{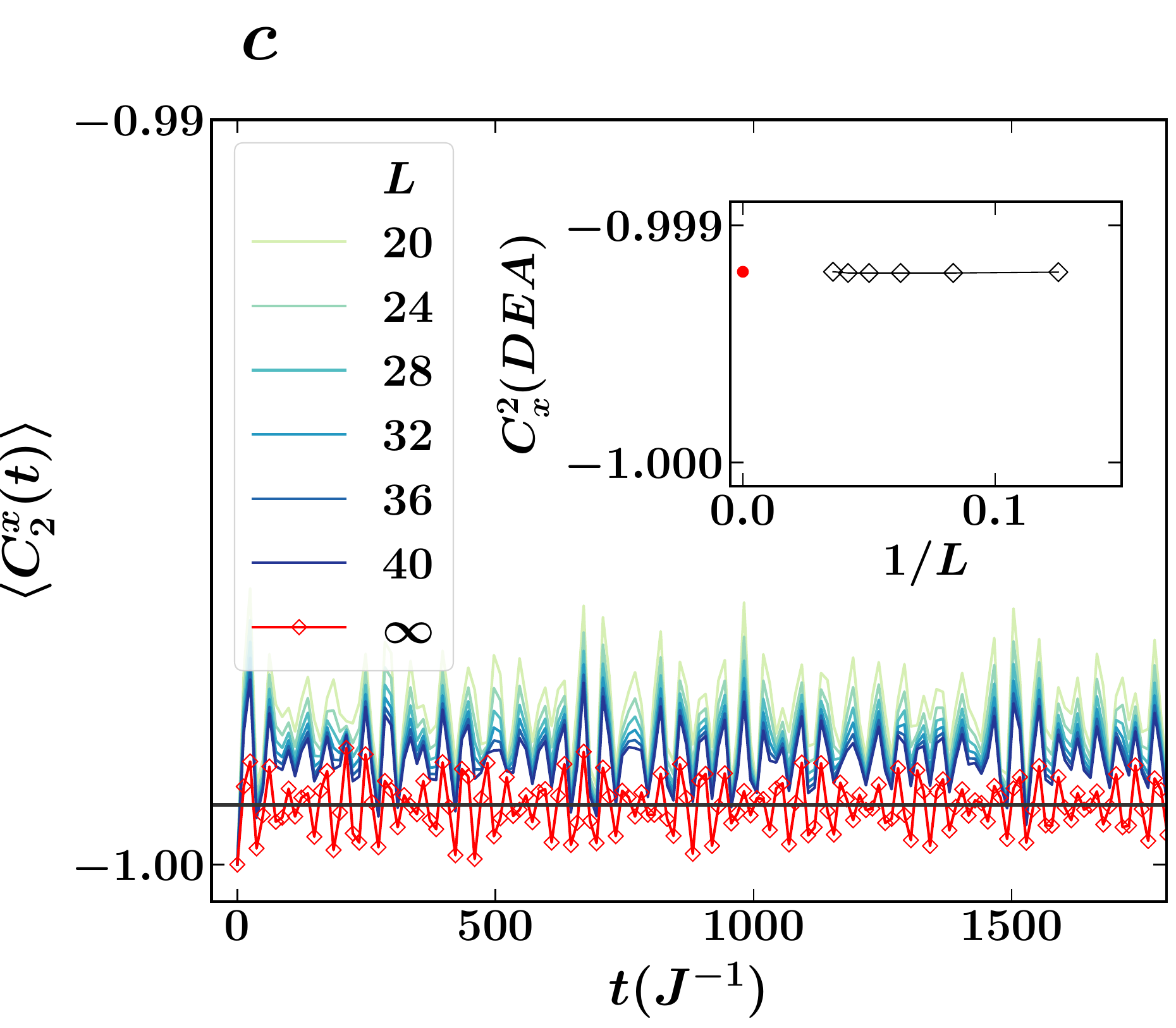}
\includegraphics[width=0.4\linewidth]{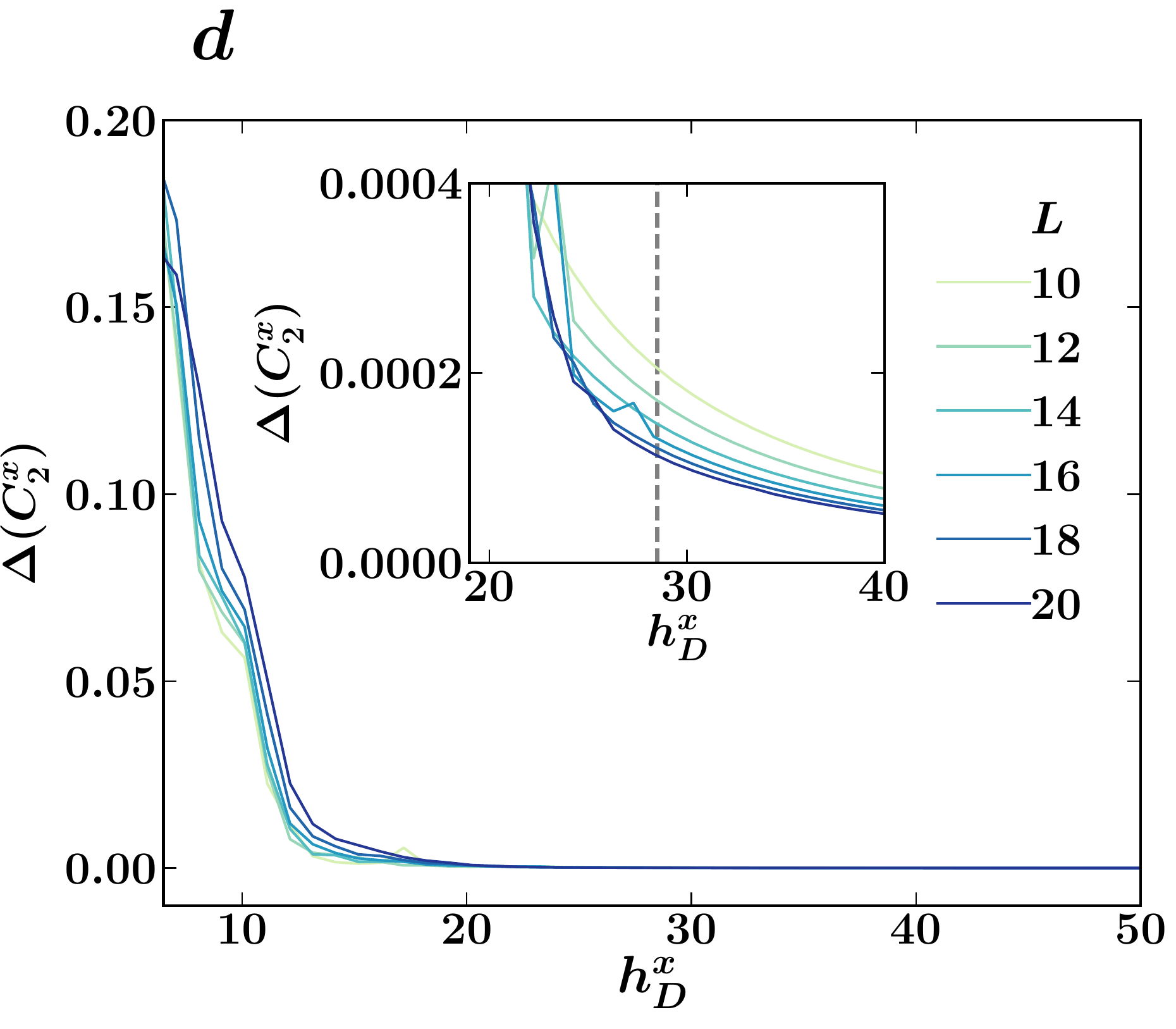}
\end{center}
\caption{
{\bf Energy unprotected conservation laws:} {\bf{(a)}}
The real-time dynamics of $C_{1}^{x}$ for various $L$, including $L=\infty$, starting from the N\'{e}el state
$|\uparrow\downarrow\uparrow... \rangle_{x}$. Inset: DEA of $C_{1}^{x}$ shows no perceptible $L-$dependence - the $L\to\infty$ extrapolation is shown with the red dot. This DEA value is compared with the dynamics in the main panel  (solid black line). {\bf(b)} shows the degree of conservation across the entire Hilbert-space via $\Delta (C_{1}^{x})$.
The inset zooms in.
{\bf (c)} and {\bf(d)}: same as {\bf{(a)}} and {\bf (b)} respectively, but for
$C_{2}^{x}$. For {\bf (c)}, the initial-state is
$|\uparrow\uparrow\downarrow\downarrow ... \uparrow\uparrow\downarrow\downarrow... \rangle_{x}.$
Parameter values: $J=2.0, \kappa = 0.5, h^x_0 = 0.15, h^{z}=\sqrt{3}/1.5,  \omega = \phi/1.6,$ where $\phi =$ Golden-mean, ($\hd = 30\times\omega$ for {\bf (a), (c)}).
}
\label{Fig:4:Other_Conservations}
\end{figure*}    
\begin{center}
    {\bf Subsystem Entanglement Entropy:} \\
\end{center}

\noindent
Figure~\ref{Fig:1:Real_Time_Dyn_mx_I}{\bf c} compares the half-chain entanglement entropy $S_{L/2}$ in the thermalizing regime for $h_{D}^{x}= 5\times\omega$ (small) for $L=8,10,12,14$  with that in the frozen regime for $h_D^{x} = 30\times\omega$ (large) for $L=20-40, \infty.$ While in the thermalizing regime $S_{L/2}$  shows a rapid growth to a saturation value proportional to $L$, in the frozen regime $S_{L/2}$ exhibits no perceptible growth, and remains $L$-independent (area-law) up to infinite size. The small growth in $S_{L/2}$ (still maintaining an area law) is consistent with the approximate nature of the conservation of the ECOs. The absence of entropy generation in an infinite quantum chaotic many-body system under an external drive is contrary to the notions of many-body chaos and thermalization.
The thermalizing and frozen regimes are separated by a threshold as discussed next.
\\
\noindent
\begin{center}
    {\bf A Measure of Stability of the ECOs Across the Hilbert-Space}\\
\end{center}

\noindent
To show that an operator ${\cal O}$ is an ECO, in light of Eq.~(\ref{Eq:DEA}) it is
sufficient to show that each $|\mu_{\alpha}\rangle$ is its approximate eigenstate, with 
${\cal O}_{\mu_{\alpha}} = \langle\mu_{\alpha}|{\cal O}|\mu_{\alpha}\rangle$ close to the $\alpha-$th
eigenvalue of ${\cal O}$ (both arranged, say, in descending order). The difference between the two
measures the inaccuracy of ${\cal O}$ as an ECO, and we hence define the following to measure
the inaccuracy over the entire Hilbert space:
\beq
\Delta ({\cal O}) = \frac{1}{D_{_{H}}}
\sum_{\al=1}^{D_{_{H}}}|\langle{\cal O}\rangle_{\mu_{\al}} -  \lambda_{\al}|,
\label{Eq:Delta_O}
\eeq
\noindent where both 
$\langle{\cal O}\rangle_{\mu_{\al}}$  and  $\lambda_{\al}$ are arranged in decreasing order, and $D_{H}$ is the Hilbert space dimension. A $\Delta {\cal O}$ decreasing systematically to zero with increasing system size signals stability of the conservation of ${\cal O}$ in the
thermodynamic limit, while the opposite trend indicates an instability, e.g.\ for  Floquet-thermalization. \\
\noindent
\begin{center}
    {\bf The Freezing Threshold} \\
\end{center}
\noindent
{\it The $L\to\infty$ and $t\to\infty$ Limits:}
We estimate the threshold (field-strength beyond which 
 stable freezing is observed) from the numerics as follows. From the TEBD/iTEBD dynamics for 
$|\psi(0)\rangle = |\uparrow\uparrow...\uparrow\rangle_{x}$, we see (Fig.~\ref{Fig:1:Real_Time_Dyn_mx_II}{\bf a}), that 
there is a field strength ($h_{D}^{x} \approx 20$), above which the TEBD/iTEBD results coincide for various system sizes and
saturate with respect to $h_{D}^{x}.$ Further,
for $h_{D}^{x} > 20$: 
(A) the results for various system sizes coincide irrespective of the number of cycles, and (B) that value also coincides with the exact $t\to\infty$ value (DEA) for $L=18$ (blue line). We hence take $h_{D}^{x} \approx 20$  as the threshold for $|\psi(0)\rangle = |\uparrow\uparrow...\uparrow\rangle_{x}$ in the $L\to\infty$ and $t\to\infty$ limit.\\

\noindent
{\it High-Temperature behavior:}
This is estimated for an 
initial-state with $\beta=10^{-2}$ (inset),
from the trend in $L-$dependence of the absolute difference between DEA and the initial value of $m^{x}$ (Fig.~\ref{Fig:1:Real_Time_Dyn_mx_II}a). The value of $\hd$ at which this quantity starts exhibiting monotonically decreasing behavior with increasing $L$ is a safe estimate (overestimation) of the threshold around $h_{D}^{x} \approx 28\times \omega$ (marked with a vertical line in the inset). The trend is the opposite on the thermalizing side.\\

\noindent
{\it For an arbitrary initial state:}
Fig.~\ref{Fig:1:Real_Time_Dyn_mx_II}{\bf b} shows $\Delta (m^x)$
vs $\hd$ for various $L$. The plots show a change in 
the trend of the $L-$dependence of $\Delta(m^x)$ as
a function of $\hd.$ The transition region
contains  large fluctuations in
$\Delta (m^x)$ with $L$, whence the precise location
of the transition is hard to determine. For our parameters,
$\Delta (m^x)$ shows a clear and systematic decline to zero with increasing $L$ from 
$\hd  \lessapprox 28\times\omega$ (marked with vertical lines in Insets (a), (b)  Fig.~\ref{Fig:1:Real_Time_Dyn_mx_II}). 
This marks the freezing threshold (vertical line in the inset showing a zoom-in). Below it, $\Delta (m^x)$ increases with $L,$ indicating instability in the conservation of $m^x$ over the
entire Hilbert space, while above it, this trend is reversed,  
indicating stability. 
As an example of this stability, $m^x(t)$
starting from a {\it mid-spectrum state} of $H(0),$ namely, 
$|...\uparrow\uparrow\downarrow\uparrow ... \uparrow\uparrow\downarrow\uparrow ...\rangle_{x}$ 
is shown in Fig.~\ref{Fig:1:Real_Time_Dyn_mx_II}{\bf c} for $L\to\infty$. The corresponding sector of 
the ECOs are large in this case, and with time
the mixing within the sector outweighs the mixing
with the neighboring sector, resulting in decreased
fluctuations at longer times.
The inset shows the $L-$dependence of the DEA.
\\

\noindent
\begin{center}
    {\bf Stability Away from the Freezing Peaks}\\
\end{center}

\noindent
The robustness of the freezing of $m^x$
away from the peaks, which occur for integer $h_{D}^{x}/\omega$ (Eq.~\ref{Freezing_Cond}), 
is shown in
Fig.~\ref{Fig:1:Real_Time_Dyn_mx_II}{\bf d}.  A zoomed-in view shows the stability
in terms of the $L-$dependence
of the deviation $\Delta(m^x)$ from their exact conservation (main): 
the larger system shows a smaller deviation from the exact conservation. The stability persists continuously as a function of $h_{D}^{x}$ through several freezing peaks ($h_{D}^{x}/\omega=$integers) and valleys between them. Inset shows the absolute difference between the initial value of $m^{x}$ and its DEA (same initial-state as in Fig.~\ref{Fig:1:Real_Time_Dyn_mx_II}a, Inset), with the same trend as $\Delta(m^x).$


\begin{figure*}[htb]
\begin{center}
\includegraphics[width=0.4253\linewidth]{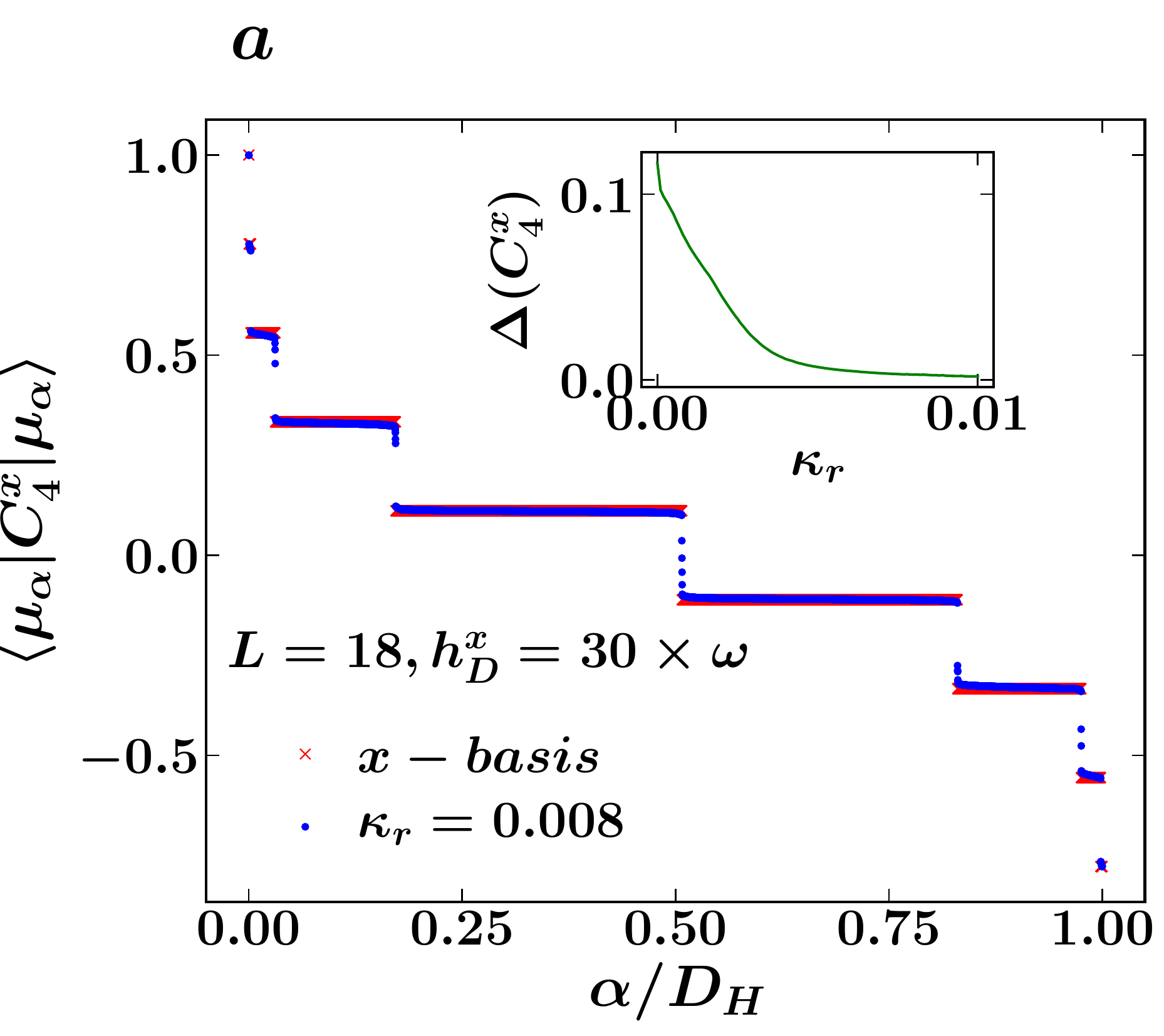}
\includegraphics[width=0.4253\linewidth]{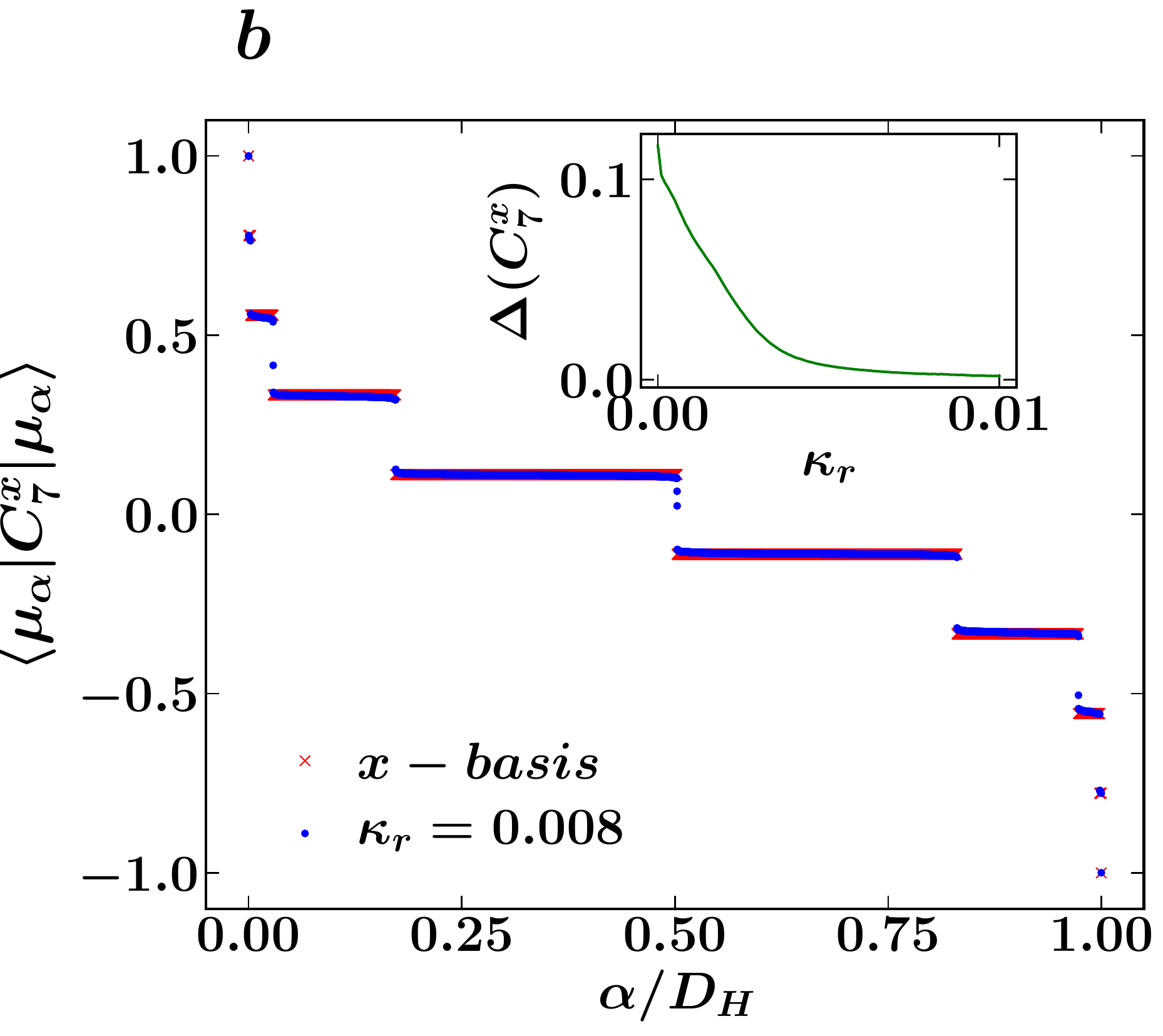}
\end{center}
\caption{{\bf Designing short and long-ranged ECOs $C_{r}^{x}$:} 
Replacing the static $C_{2}^{x}$ term in $H(t)$ by $C_{r}^{x}$
with a small coupling elevates $C_{r}^{x}$ to the status of an ECO (see Sec.\ref{Sec:Tailoring_ECO}). In each frame,
the main plot shows the step-like structure of the 
Floquet expectation-values of $C_{r}^{x},$ 
compared with the eigenvalues of $C_{r}^{x}.$ Insets show the rapid decline of 
$\Delta(C_{r}^{x})$ as a the function of strength $\kappa_{r}$ of the coupling of $C_{r}^{x}$
in the Hamiltonian (see~\cite{suppl} for plots for more values of $r$).
Parameter values: $J=2.0, h^x_0 = 0.15, h^{z}=\sqrt{3}/1.5, \omega = \phi/1.6,$ where $\phi =$ Golden-mean, $L=18$. }
\label{Fig:5:Cr_ECO}
\end{figure*}    
\begin{figure*}[htb]
\begin{center}
\includegraphics[width=0.325\linewidth]{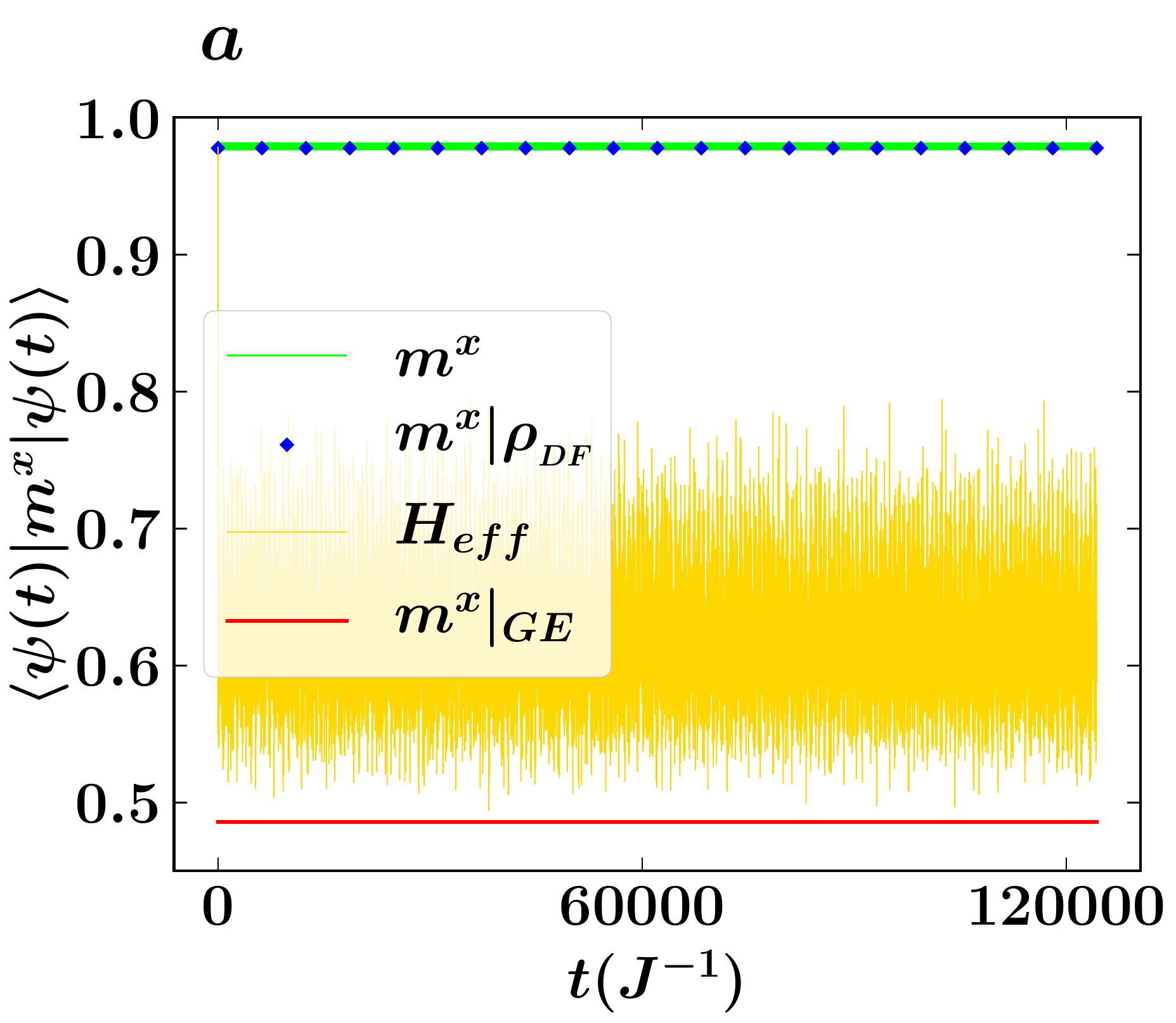}
\includegraphics[width=0.325\linewidth]{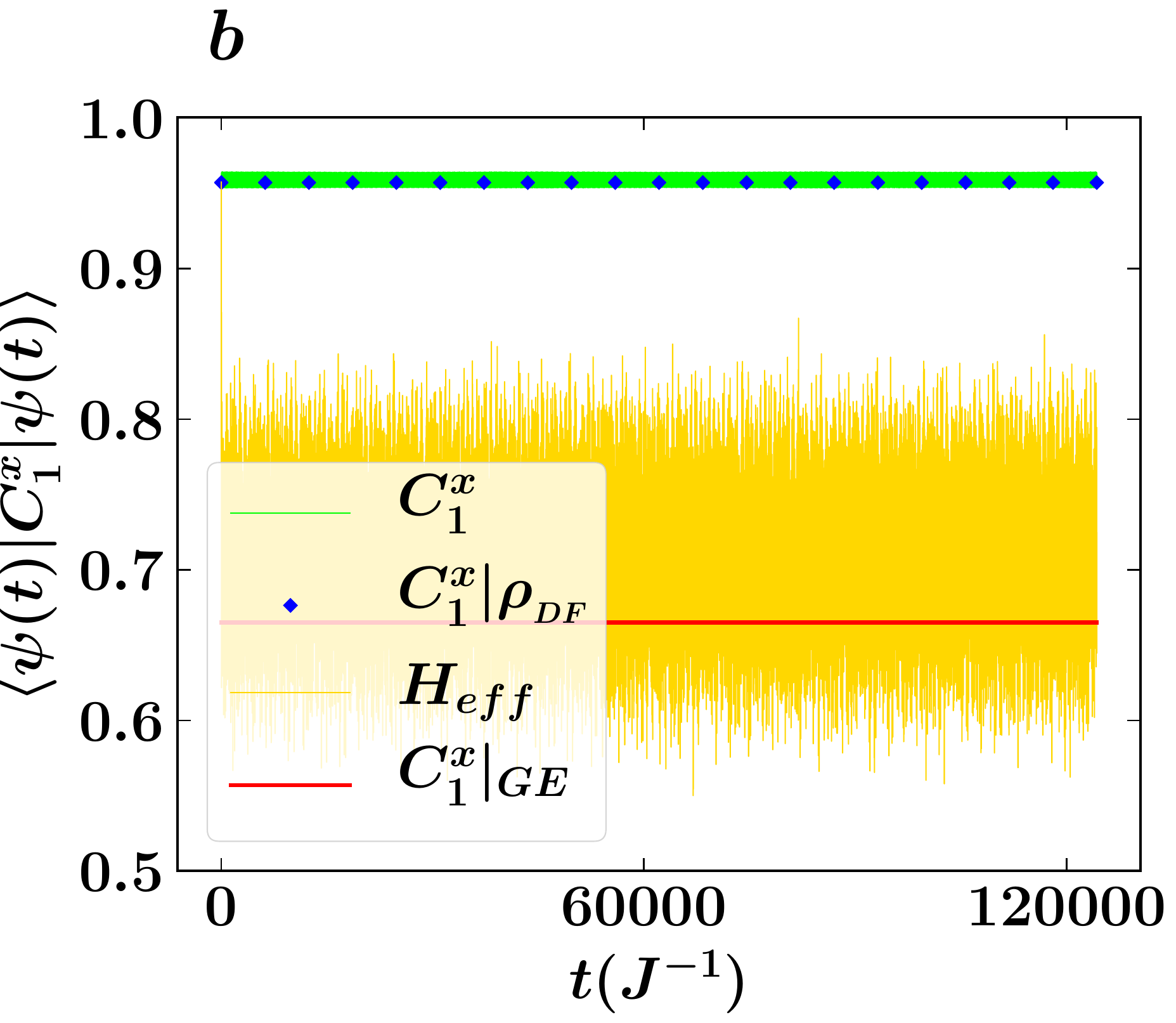}
\includegraphics[width=0.325\linewidth]{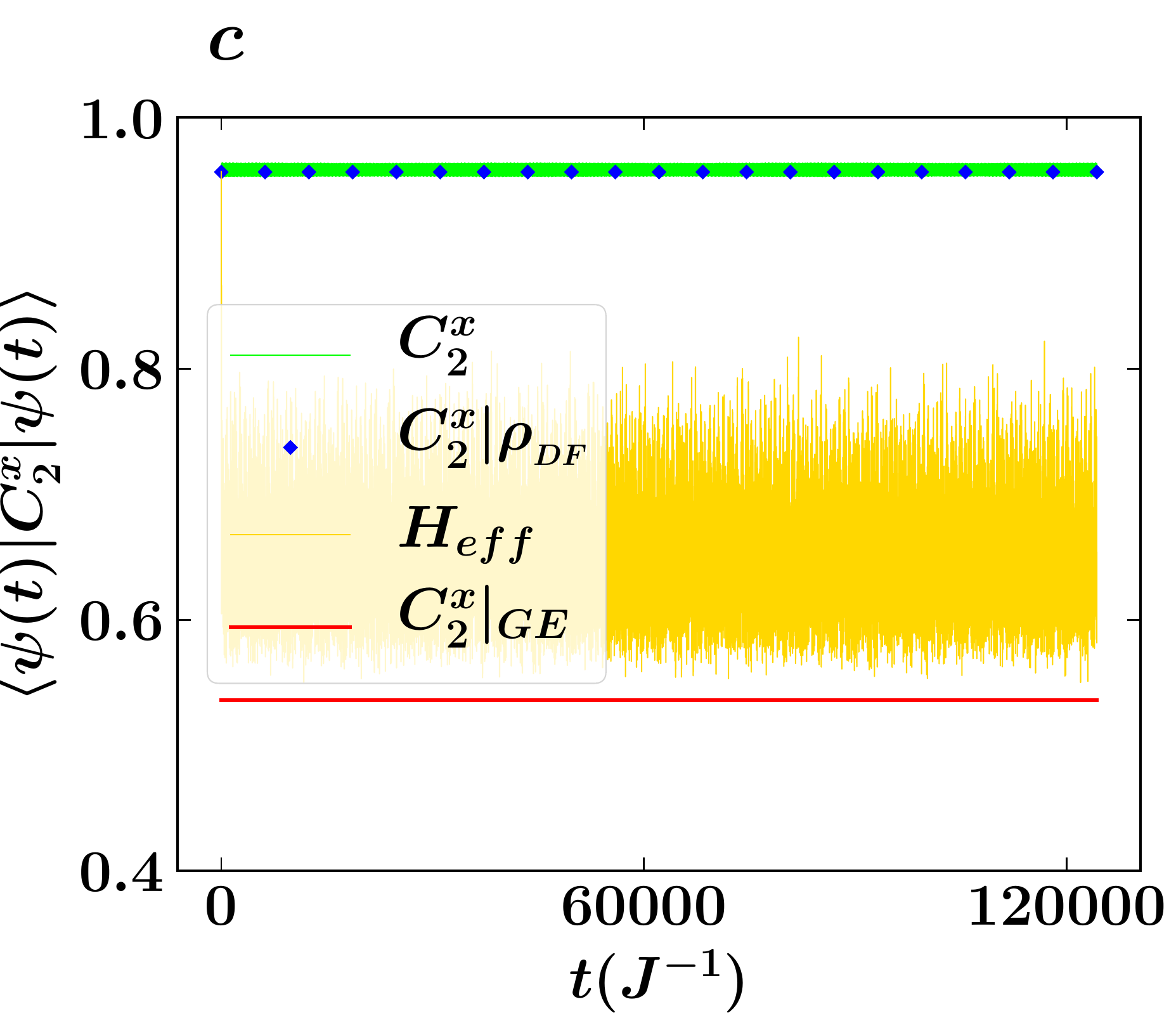}
\includegraphics[width=0.325\linewidth]{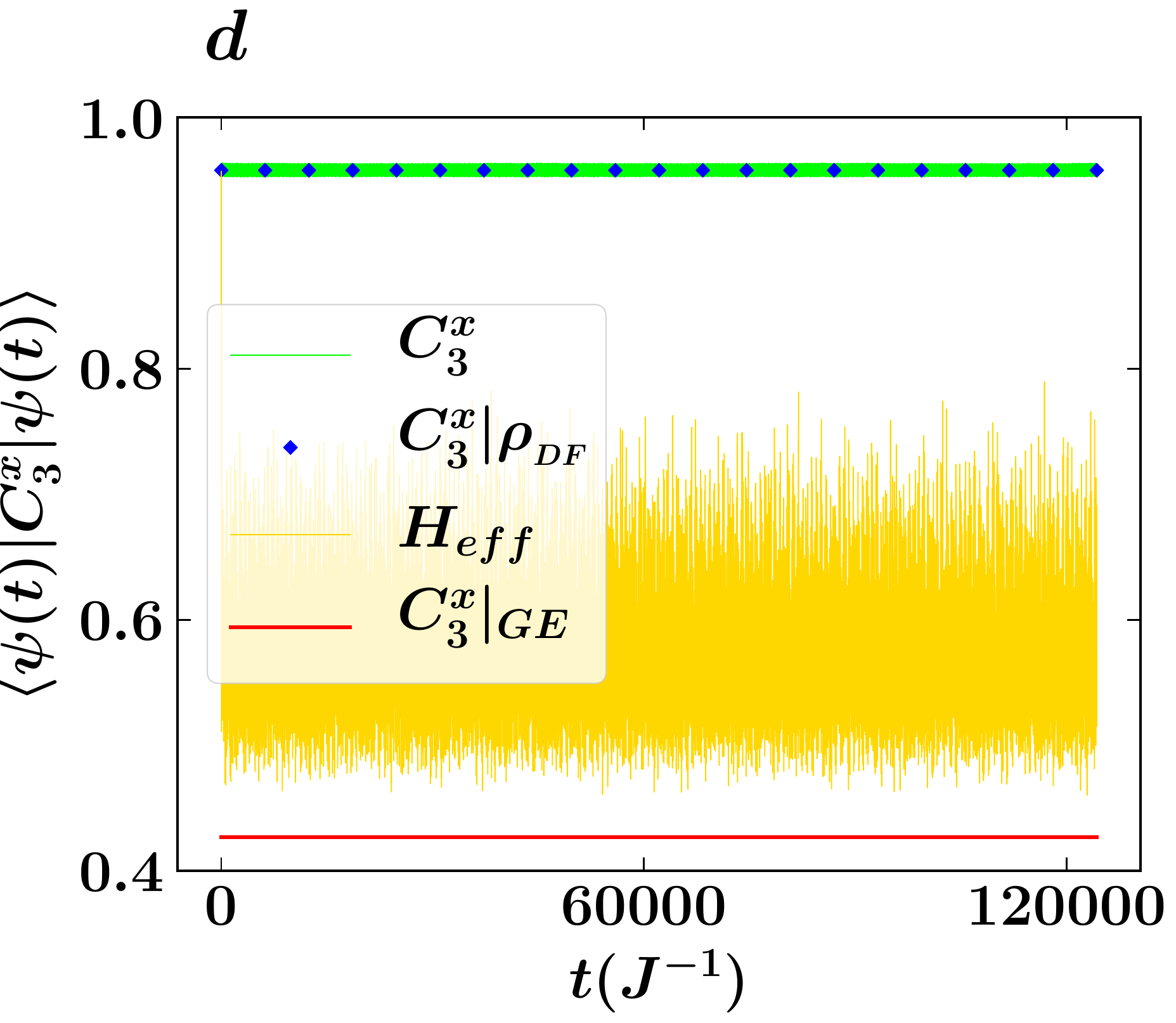}
\includegraphics[width=0.325\linewidth]{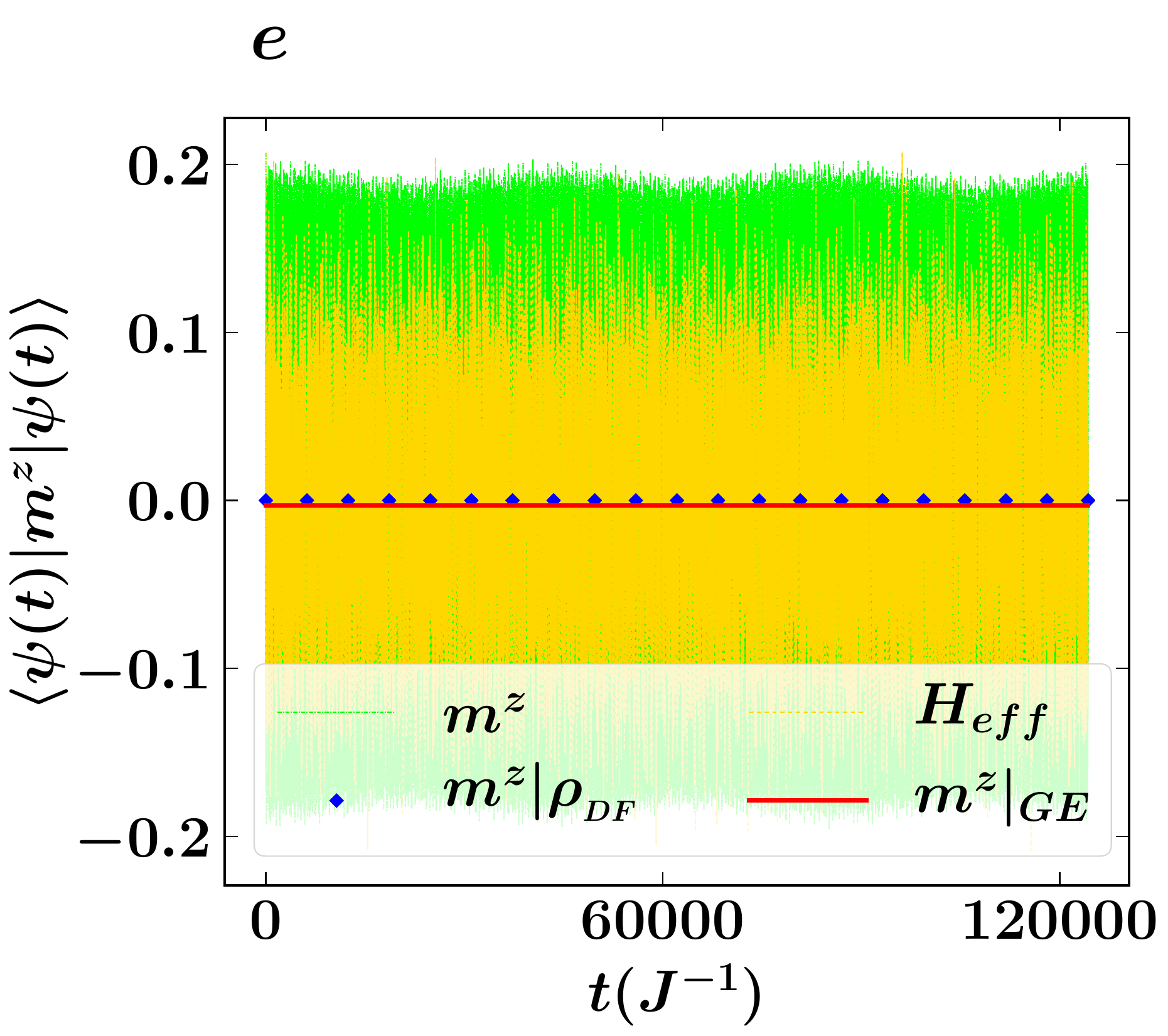}
\includegraphics[width=0.325\linewidth]{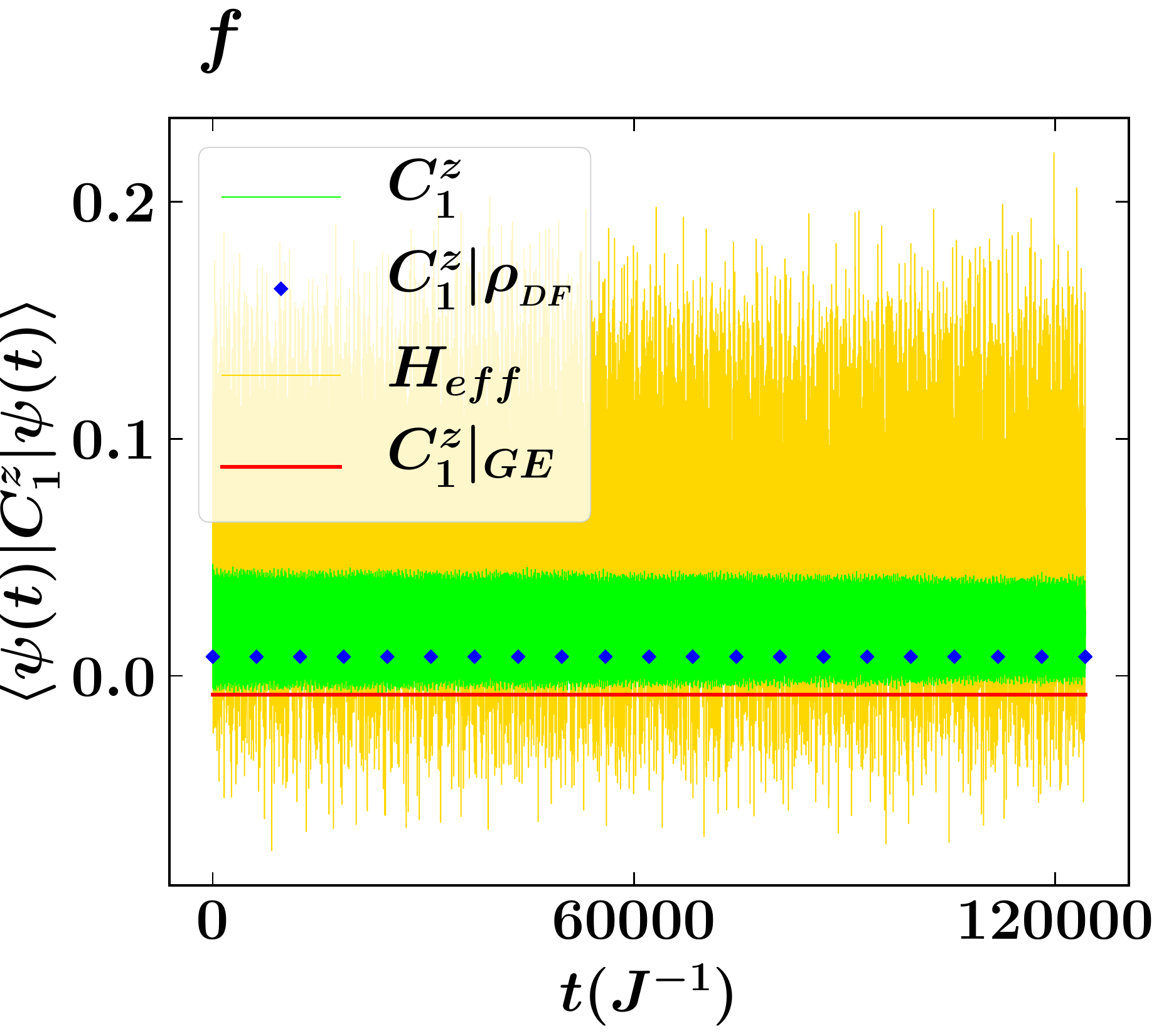}
\end{center}
\caption{{\bf The Strongly Driven Ensemble}:
Comparison of the exact dynamics of three quantities which are ECOs 
($m^{x}(nT), C_{1}^{x}(nT), C_{2}^{x}(nT);$ the {\bf top panel}),
and three quantities which are not ($C_{3}^{x}, m^{z},$ and $C_{1}^{z};$ {\bf bottom panel})  
with (i)  the prediction of the 
DF ensemble (Eq.~\ref{Eq:rho_DF}), (ii) the dynamics
by the effective Hamiltonian $H_{eff}$ up to the 3rd-order of a rotating frame Magnus expansion (see~\cite{suppl}), 
and (iii) the thermal (Gibbs') ensemble with $H_{eff}$ 
(denoted by $|_{GE}$). The initial-state is the ground state of $H_{\cal X} = H(0)$ with
with $h^{z}=1.2, h_{D}^{x} + h_{0}^{x}=5.1, J=1.0, \kappa=0.7.$
The results clearly show the leading role
of the ECO in determining the Statistical Mechanical ensemble describing the time-averaged behavior of the observables regardless of their commutation
with the drive: the average of the real-time exact dynamics (green line) is
best approximated by prediction of the DF-ensemble (blue dots). 
Parameter values: $J=2.0, \kappa = 0.5, h^x_0 = 0.15, h^{z}=\sqrt{3}/1.5, 
h_{D}^{x} = 30 \times \omega, 
\omega = \phi/1.6,$ where $\phi =$ Golden-mean ($L=18$).}
\label{Fig:6:DF_Ensemble}
\end{figure*}    

\section{Conservation laws unprotected by large energy costs}
\noindent
The immediate question that springs to mind is whether $m^x$ is
the only emergent conservation law. 
Fig.~\ref{Fig:3:EE_Spectrum} shows
$S_{L/2}$ of {\it all} final states after $10^{8}$ cycles, starting from  
all the $2^L$ eigenstates of $\{ \sigma_{i}^{x} \}$ as initial states and plotted them against
the eigenvalues of $m^x$ for the respective initial states.
If the emergent conservation of $m^x$ was the only constraint, then we would have got a unique value of $S_{L/2}$ corresponding to the size of the eigen-subspace of a given eigenvalue of $m^x$. But instead, the final $S_{L/2}$ shows a large variation depending on the details of the initial-states within a given eigen-subspace, indicating the presence of further constraints.\\

At its extreme, states like $|\uparrow\downarrow\uparrow ...\rangle_{x}$ (and its spatially translated partner) and $|\uparrow\uparrow\downarrow\downarrow ... \rangle$ (and its four translated partners), which lie in the $m^{x}=0$ subspace that grows $\sim$ exponentially with $L$, show no significant growth of $S_{L/2}.$  Analyzing $S_{L/2}$ carefully in each eigen-subspace, we uncover
at least two new strongly conserved ECOs of the form:
\begin{equation}
C_{r}(x) = 
 \frac{1}{L}\sum_{i}\sigma_{i}^{x}\sigma_{i+r}^{x}, 
\label{Eq:Cr_Def}
\end{equation}
\noindent
with $r=1$ and $r=2$.
We emphasize the following regarding these two quantities: unlike $m^x,$
the conservation of $C_{(1,2)}^{x}$ are not directly protected by the largeness of $\hd$ and associated energy cost. 

For example, the process
$
|\uparrow\downarrow\rangle_{x} ~ \leftrightarrow
 ~ |\downarrow\uparrow\rangle_{x}
$
can violate the 
conservation of $C_{(1,2)}^{x}$ but it 
does not
change $m^x$, thence incurring a possible
energy cost of order of the smaller terms $J/h^{z},\kappa/h^{z}, h_{0}^{x}/h^{z}$, but not of the large field $h_{D}^{x}/h^{z}.$ 
However, surprisingly, from Fig.~\ref{Fig:4:Other_Conservations} {\bf a-d}, these processes do not destabilize the conservation of $C_{1}$ and $C_{2}$ respectively, even in an infinite system. \\

In detail, Fig.~\ref{Fig:4:Other_Conservations}{\bf a} shows  dynamics of 
$C_{1}^{x}(t)$  starting from the N\'{e}el state 
$
|\psi(0)\rangle = |\uparrow\downarrow ... \uparrow\downarrow ... \rangle_{x}, 
$
which is an eigenstate of $C_{1}^{x}$ with eigenvalue $-1.$
For the entire time, $C_{1}^{x}$ remains close to its initial value. The DEA of $C_{1}^{x}$ for finite systems is shown in the inset.
Similarly, Fig.~\ref{Fig:4:Other_Conservations}{\bf c} shows the stability 
of the initial-state 
$|\uparrow\uparrow\downarrow\downarrow ...\rangle,$ due to the
appearance of $C_{2}^{x}$ as an ECO.
Figs.~\ref{Fig:4:Other_Conservations}{\bf b} and \ref{Fig:4:Other_Conservations}{\bf d} shows the stability of $C_{1}^{x}$ 
and $C_{2}^{x}$  via their respective spectral deviation $\Delta$
across the Hilbert space.
\\
\section{Designing conservation laws}
\label{Sec:Tailoring_ECO}
What is it that makes an operator ${\cal O}_{x}$ commuting with the drive an accurate ECO? It turns out one can stabilize $C_{r}^{x}$ of any range -- long or short -- as an ECO to great accuracy just by including it 
in the Hamiltonian with a tiny (much smaller than the strong drive) prefactor. Then,  $\Delta({\cal O}_{x})$ decreases rapidly
with the increase in the magnitude of the coupling.
To show this, we use the same Hamiltonian of 
Eq.~(\ref{Eq:Hamiltonian_1}), except, 
we replace the next-neighbour interaction term $\kappa\sum_{i=1}^{L}\sigma^{x}_{i}\sigma_{i+2}^{x}$ by a further-neighbour interaction term
$-L\kappa_{r} C_{r}^{x}$ of Eq.~(\ref{Eq:Cr_Def}). We denote the resulting Hamiltonian by $H_{r}(t)$ 
($r=2$ with a change of sign of $\kappa_{r}$ gives $H(t)$ of Eq.~(\ref{Eq:Hamiltonian_1})).
Fig.~\ref{Fig:5:Cr_ECO} shows, for various $r,$ $C_{r}$ emerges rapidly as 
a stronger ECO with increasing $|\kappa_{r}|.$  
The agreement of the eigenvalues of $C_{r}$ and its Floquet expectation values shown as the main plots in Fig.~\ref{Fig:5:Cr_ECO}, is just for $\kappa_{r} = 0.008$ (for all $r$), three orders of magnitude smaller than $\hd.$ The insets show the rapid fall of $\Delta(C^{x}_{r})$ with increasing $\kappa_{r}$.

\section{The Strongly Driven Ensemble}
Our final central result is the identification of the strongly driven ensemble which captures the above results quantitatively.  
This is obtained in the spirit of generalized periodic Gibbs ensembles~\cite{Jaynes, PGE} which describe the stroboscopically observed late-time synchronized state of the system.
This then takes the form of a Gibbsian ``equilibrium" ensemble which crucially includes  the 
three independent ECOs, namely, $m^x, C_{1,2}^{x}$ as the relevant
conservation laws. The local properties can be described by 
\beq
\rho_{_{DF}} = \frac{1}{\cal Z}\sum_{\alpha}e^{-(\beta_{0}m^{x} 
+ \beta_{1}C_{1}^{x} + \beta_{2}C_{2}^{x})}|x_{\alpha}\rangle\langle x_{\alpha}|,
\label{Eq:rho_DF}
\eeq
\noindent
where $\{|x_{\alpha}\rangle\}$ are the eigenstates of 
$\{\sigma_{i}^{x}\},$ and $\beta_{0,1,2}$ are suitable Lagrange
multipliers and the partition function ${\cal Z}$ is the normalization factor. 
Note that this differs fundamentally
from Gibbsian premise, where only exact conservation laws are considered, as we include  the {\it emergent} ECOs, which  are perpetual but approximate.\\

Fig.~\ref{Fig:6:DF_Ensemble} compares prediction of $\rho_{_{DF}}$ (blue dots) with: (i) the exact real-time dynamics (green line); (ii) the dynamics with the 3rd-order approximation 
of $H_{eff}$ in the appropriate frame~(\cite{Asmi_DF_PRX_2021,suppl}, yellow line); and (iii) the Gibbs' ensemble with an appropriate  effective Floquet-Hamiltonian
$H_{eff}$ (red line), obtained from a truncated Magnus-like expansion in a rotating frame for strong drive~\cite{Anatoli_Rev, Asmi_DF_PRX_2021, suppl} which is only constrained by one temporarily conserved quantity~\cite{Kuwahara_Mori_Saito_Prethermal, Dima_Prethermal, Anatoli_Rev}.
The initial-state is the ground state of $H_{\cal X}=H(0)$ with $h^{z}=1.2, h^{x}=5.1, J=1.0, \kappa=0.7, \hd = 30\times\omega.$ 

This shows that $\rho_{_{DF}}$ accounts well for the time-averaged dynamics in the long run, while the 3rd-order description fails in general. This conclusively demonstrates the role of ECOs in the statistical Mechanics of DF. \\

\noindent
\begin{center}
    {\bf Dynamical Freezing vs Prethermal Stability} \\
\end{center}
We begin by noting that in the $\omega\to\infty$ limit,
our system does not support any freezing of ECOs, since
the average Hamiltonian does not commute with them and is non-integrable with no large term. By contrast, this is the limit where prethermal stability is absolute~\cite{Ho_Mori_Abanin_Pretherm_Rev}. \\

First, we estimate the timescale $\tau_{pre}$ of the canonical prethermal stability~\cite{Ho_Mori_Abanin_Pretherm_Rev}. When 
applied in appropriate frame to our case (see Fig.~\ref{Fig:1:Real_Time_Dyn_mx_I}), it gives an estimate of
$\tau_{pre} \approx 20~ J^{-1}$ (see Methods). The simulation reported here reaches $t=25000 ~ J^{-1}$ -- almost three orders of magnitude longer than the estimated prethermalization time, 
and there is no sign of any degradation of the ECO for an infinite system, while $S_{L/2}$ tends to saturate to a finite area-law value with $L.$ 
Also, contrasting the  
exponential suppression of heating with 
the largest energy-scale in prethermalization, here the heating is a non-monotonic function of
both $h_{D}^{x}$ and $\omega$ in the DF regime.

Secondly, the stability of $C_{1,2}^{x}$ is not underwritten by the smallness of an energy scale in the Hamiltonian compared to the driving frequency $\omega$, and thus falls outside the purview of Floquet prethermalization.

Finally, we note that the first two orders of the expansion of $H_{eff}$ (see~\cite{Asmi_DF_PRX_2021, suppl}), predicting complete freezing of $m^x, C_{1,2}^{x},$ would be much closer to the time-average of the actual dynamics than that also including the 3rd-order (Fig.~\ref{Fig:6:DF_Ensemble}). This would imply the putative prethermal dynamics of those operators, in this case, should only be driven by the first two terms in $H_{eff},$ and hence be completely frozen.\\

\noindent
{\bf Strong Experimental Signatures} of DF and ECOs should be readily realizable in various quantum simulator platforms because
DF {\it is not merely a low-energy phenomenon, and hence its signatures are manifest also in experimentally-accessible length and time scales}. Variants of the spin model we have used have already been realized using quantum simulators based on Rydberg-dressed atoms~\cite{Bloch_Rydberg}. The dynamics can also be simulated easily in the Google sycamore processor as done in ~\cite{Google_Sycamore_DTC}.\\

\centerline{\bf Acknowledgments}

\vspace{.3cm} A.H., R.M., and A.D. are grateful to Diptiman Sen for previous collaboration \cite{Asmi_DF_PRX_2021}. 
A.H. was supported by the Marie Sk\l{}odowska-Curie grant agreement No. 101110987 (from 01.11.2023). A.D. thanks the MPI-PKS visitor's program for hosting collaborative visits during this project and extending the
computational facility for this work.
This research was financially supported by the European Research Council (ERC) under the European Union’s Horizon 2020 research and innovation program under grant agreement No. 771537. F.P. acknowledges the support of the Deutsche Forschungsgemeinschaft (DFG, German Research Foundation) under Germany’s Excellence Strategy EXC-2111-390814868 and DFG Research Unit FOR 5522 (project-id 499180199). F.P.’s research is part of the Munich Quantum Valley, which is supported by the Bavarian state government with funds from the Hightech Agenda Bayern Plus. This work was in part supported by the Deutsche Forschungsgemeinschaft under grants SFB 1143 (project-id 247310070) and the cluster of excellence ct.qmat (EXC 2147, project-id 390858490). A.W.\ acknowledges support by the DFG through the Emmy Noether program (Grant No.\ 509755282).
All the Numerical Work was performed on the Computing Cluster at MPI-PKS.

\bibliography{main_V0} 
\widetext
\section*{Methods}
\setcounter{section}{0}
\setcounter{equation}{0}
\setcounter{figure}{0}
\setcounter{table}{0}
\makeatletter
\renewcommand{\theequation}{M\arabic{equation}}
\renewcommand{\bibnumfmt}[1]{[S#1]}
\renewcommand{\citenumfont}[1]{S#1}

\section{Estimate of corresponding prethermal Timescale}

\subsection{Formula for the Estimate}
We follow the estimate of the prethermal time following the approach described in Annals of Physics {\bf 454}, 169297 (2023).
The prethermal time scale $\tau_{pre}$ is given as follows.
\begin{equation}
    \tau_{pre} = \left(\frac{A}{\Lambda}\right)e^{C(\Omega / \Lambda)},
\label{EqMeth:TakDef_TauPre}     
\end{equation}
\noindent
where $\Omega$ is the driving frequency, and $\Lambda$ is a measure of
the local bandwidth. Following the prescription in Annals of Physics {\bf 454}, 169297 (2023), here this is estimated from the norm of the Hamiltonian defined below,
and $C$ and $A$ are positive numbers that do not depend on $\Omega,$ but can depend on other parameters of the Hamiltonian.\\

In detail, $\Lambda$ is defined in the following manner. One considers a quantum spin (or fermion) system on a $d$-dimensional regular lattice. Each lattice site is labeled by $i = 1, 2, . . . , N$, $N$ being the total number of lattice sites. For a  Hamiltonian of the form such as ours
\begin{equation}
    H(t) = \sum_{X:|X| \leq k}h_{X}(t),
\label{EqMeth:HamGenForm} 
\end{equation}
\noindent
$X$ denotes a subset of the sites of the lattice and $h_{X}(t)$ is an operator acting non-trivially only on region $X$. The condition $|X| \leq k$ means that $X$ contains at most $k$ different sites, i.e., the Hamiltonian is such that it has at most $k$-site interactions. The local bandwidth $\Lambda$ of a time-periodic Hamiltonian $H(t)$ is then defined as
\begin{equation}
    \Lambda = \underset{t \in [0,T]}{\max} \Lambda (t),
\label{EqMeth:Lambda_Def} 
\end{equation}
\noindent
where the instantaneous bound $\Lambda (t)$ is given by
\begin{equation}
    \Lambda (t) = \underset{i \in [1,2,...,N]}{\max} \sum_{X:|X| \leq k, i \in X} ||h_{X}(t)||,
\label{EqMeth:Lambda_t_Def} 
\end{equation}
\noindent
where $|| . ||$ denotes the operator norm and the sum runs over all subsets of sites that include the site $i$. 
Here we take the square root of the largest eigenvalue of $A^{\dagger}A$ as $||A||$ for an operator $A.$ {We evaluate $\Lambda$ by maximizing $\Lambda(t)$ over the time interval $[0,T]$ numerically.}

\subsection{System Hamiltonian}
For our  system Hamiltonian 
\begin{eqnarray}
 H(t) &=& H_{0}(t) ~+~ V, ~~ {\rm where} \nonumber \\
H_{0} (t) &=& H_{0}^{x} ~+~ \mathrm{Sgn}(\sin{(\omega t)}) ~H_{D}, ~~ {\rm with} 
\nonumber \\
H_{0}^{x} = &-& ~\sum_{n=1}^L ~J \sigma_n^x \sigma_{n+1}^x + \sum_{n=1}^L ~\kappa 
\sigma_n^x \sigma_{n+2}^x - h_{0}^{x}~\sum_{n=1}^L\sigma_n^x, \nonumber \\
H_{D} = &-& h_D^{x}\sum_{n=1}^L\sigma_n^x, ~~ {\rm and} ~~ 
V = - ~ h^z \sum_{n=1}^L \sigma_n^z, \nonumber
\label{EqMeth:Hamiltonian_1} 
\end{eqnarray}
which can be written as
\begin{eqnarray}
       H(t) &=& H_{0} + r(t)H_{D}, ~ {\rm where} \label{EqMeth:Ham_Stat_Drive} \\ 
H_{0} &=& H_{0}^{x} + V, ~ {\rm and}
\label{EqMeth:H0_Sys_Def} \\
r(t) &=& \mathrm{Sgn}(\sin (\omega t))
\label{EqMeth:SquarePulse}
\end{eqnarray}
\subsection{Hamiltonian in the moving frame}
Now, if we naively use the bare drive frequency $\omega$ 
for estimating $\tau_{pre},$
this will yield an underestimate, since $\omega$ is not the largest scale here. Here we want to get the strictest estimate for $\tau_{pre},$
and hence we switch to the frame where the largest scale in the problem appears 
as the drive frequency. {We work in this frame and
call the drive frequency in this frame the {\it effective frequency}, as in this frame the inverse of the
largest scale serves as the small parameter in a
Magnus expansion, and we get the largest estimate of $\tau_{pre}$.
\begin{equation}
    H^{mov}(t) = W\textsuperscript{\textdagger}(t)H_{0}W(t)
\label{EqMeth:HamRot} 
\end{equation}
where the rotation operator is 
\begin{equation}
    W(t) = \exp\left[ -i \int_{t_{0}}^{t} r(t')H_{D}dt' \right]
\label{EqMeth:RotOp} 
\end{equation}
Using equations~(\ref{EqMeth:Hamiltonian_1}), (\ref{EqMeth:SquarePulse}) and (\ref{EqMeth:RotOp}) we get
\begin{equation}
    W(t) = \exp\left[ ih^{x}_{D}\sum_{j}\sigma^{x}_{j}\int_{t_{0}}^{t}\mathrm{Sgn}(\sin (\omega t'))dt' \right] = \prod_{j}\exp\left[ ih^{x}_{D}\sigma^{x}_{j}\int_{t_{0}}^{t}\mathrm{Sgn}(\sin (\omega t'))dt' \right]
\label{EqMeth:RotOpSys_1} 
\end{equation}
Now we define
\begin{equation}
    \theta (t) = h^{x}_{D}\int_{t_{0}}^{t}\mathrm{Sgn}(\sin (\omega t'))dt'
\label{EqMeth:Theta_Def} 
\end{equation}
Putting these all together, we get
\begin{equation}
    H^{mov}(t) = \prod_{i}\exp[-i\sigma^{x}_{i}\theta (t)]H_{0} \exp[i\sigma^{x}_{i}\theta (t)] = H^{x}_{0} - h^{z}\sum_{i}\exp[-i\sigma^{x}_{i}\theta (t)]H_{0} \exp[i\sigma^{x}_{i}\theta (t)].
\label{EqMeth:HamSysRot_1} 
\end{equation}
On further simplification, we get
\begin{equation}
    H^{mov}(t) = H^{x}_{0} - h^{z}\cos(2\theta )\sum_{i}\sigma^{x}_{i} + h^{z}\sin(2\theta )\sum_{i}\sigma^{y}_{i}.
\label{EqMeth:HamSysRot_Fin} 
\end{equation}
\subsection{Estimate of prethermal time for our case}
Now, we want to find the the effective frequency $\omega_{eff}$ for this moving frame Hamiltonian. Let us denote by $T_{eff} = 2\pi/\omega_{eff}$ the corresponding effective time period. Let us choose two instants of time $t_1$ and $t_2$ which are a period apart, i.e., $t_{2} - t_{1}$ = $T_{eff}$. Now, from the form of $H^{mov}(t)$ [Eq. (\ref{EqMeth:HamSysRot_Fin})], it is clear that for  $t_{2}$ - $t_{1}$ = $T_{eff}$ to be true,  $t_1$ and $t_2$ must be such that they satisfy
\begin{equation}
    \theta (t_{2}) - \theta (t_{1}) = \pi
\label{EqMeth:Theta_Condition} 
\end{equation}
\noindent
From Eq. (\ref{EqMeth:Theta_Def}), it is clear that (for simplicity, considering $t_{0}$ = 0 in Eq.~(\ref{EqMeth:Theta_Def})) the above condition (Eq. (\ref{EqMeth:Theta_Condition})) can be satisfied only if
\begin{equation}
    h^{x}_{D}\frac{T}{2} \geq \pi\quad \implies\quad h^{x}_{D} \geq \omega
\label{EqMeth:hxD_Om_Condition} 
\end{equation}
Now, considering $t_{1}$ $<$ $t_{2}$ $<$ T/2, Eq. (\ref{EqMeth:Theta_Condition}) takes the form
\begin{equation}
    h^{x}_{D}(t_{2} - t_{1}) = \pi\quad \implies\quad T_{eff} = \frac{\pi}{h^{x}_{D}}
\label{EqMeth:TmPrd_Eff} 
\end{equation}
Therefore, the effective frequency for the moving frame Hamiltonian is
\begin{equation}
    \omega_{eff} = 2\pi/T_{eff} = 2h^{x}_{D}
\label{EqMeth:Om_Eff} 
\end{equation}
\noindent In this frame, the drive frequency is thus proportional to
$h_{D}^{x}$ (the drive amplitude in the lab frame) which is the
largest coupling/scale for us (see, e.g., Phys. Rev. X 11, 021008 (2021).). Since the
dynamics of the operators {that commute with the drive}  (e.g., $m^{x}$) are identical in
both the lab frame and the rotating frame, we extract $\tau_{pre}$
from direct numerics in the lab frame, and track their stability
on time scales compared to it.\\

\noindent
We substitute $\omega_{eff}$ in Eq. (\ref{EqMeth:TakDef_TauPre}) to obtain our prethermal time scale
\begin{equation}
    \tau_{pre} 
= \frac{A}{\Lambda}
\exp{\left(\frac{2C~h_{D}^{x}}{\Lambda}\right)}.
\label{EqMeth:TauPre_Sys_Params1} 
\end{equation}
We calculate
 $\Lambda$ using Eq.~(\ref{EqMeth:Lambda_t_Def}),
 while $C$ and $A$ are evaluated
 from exact numerics as follows. For a given $h_{D}^{x}$
 (keeping all other parameters fixed), we fit
 $m^{x}(t)$ (obtained by exact solution of the time-dependent Schr\"{o}dinger equation) by a decaying exponential  $e^{-t/\tau}.$ In the thermalizing regime, the growth of this time-scale $\tau$ is expected to be lower-bounded by $\tau_{pre}$ (see, e.g. Annals of Physics
{\bf 454} 169297 (2023)}). 
 
 We focus on the parameters used in 
 Fig.1(a) of the main text: $J=2.0$, $\kappa=0.5$, $h_0^x = 0.15$, $h^z=\sqrt{3}/1.5$, $\omega = \phi/1.6$ where $\phi$ is the Golden Mean and $h^{x}_{D} = 30\times\omega$.
 In general, $\tau$ vs $h_{D}^{x}$ exhibits non-monotonic behavior, and since we are interested in a prethermal bound, we concentrate on the lower envelope of the $\tau$ vs $h_{D}^{x}$ curve (the locus of the local minima of the curve) as a strict estimate.
 We consider the stretch between $h^x_D$ = $3.5\omega$ and $h^x_D$ = $5.0\omega,$ as for smaller  $h_{D}^{x}$ the thermalization is too weak (and even possibly absent altogether -- see Phys. Rev. Lett. {\bf 121}, 264101 (2018)). This makes extracting numerically reliable results within our simulation time scale difficult, if not impossible. On the other hand, when $h_{D}^{x}$ is too large, DF appears, and the nice exponential dependence of $\tau_{pre}$ on $h_{D}^{x}$ is lost. In the chosen regime, the local minima of $\tau_{pre}$  vs $h_{D}^{x}$ (i.e., its lower envelope) is well approximated by an exponentially growing function of $h_{D}^{x}.$
 A linear fit of these local minima (for $L = 12$) yields a straight line with a slope 
$S \approx 0.08$, intercept 
$I \approx 0.01$ and the fitting-error
$\chi^{2} \approx 0.00152.$
Using the values of $S$ and $I$  and Eq.~(\ref{EqMeth:TauPre_Sys_Params1}), we get: 
$C \approx 0.24$  and 
$A \approx 6.23.$
Substituting $C$ = 0.24, 
$A$ = 6.23, 
$\Lambda$ = 6.16
for $h^{x}_{D} = 30\times\omega$ in Eq.~(\ref{EqMeth:TakDef_TauPre}), we get the value of the prethermal time to be
\begin{equation*}
   \tau_{pre} \approx 21.5 ~ J^{-1}.
\end{equation*}

\section{On Accuracy of numerical results using TEBD}
\noindent
{\bf Trotter Approximation:}
The TEBD algorithm uses Suzuki-Trotter (ST) decomposition 
(J. Math. Phys. {\bf 32}, 400–407 (1991))
to approximate the time-evolution operator. We have used the first order ST decomposition with a time-step size $\delta t=0.01$ - for which our results
(expectation values of local operators) converged up to 
$\sim 10^{-7}$ when compared to the corresponding results with $\delta t=0.001$.

Since at any time our Hamiltonian is of the form $H = H_{1} + H_{2}$  with each of $H_{(1,2)}$ containing only mutually commuting terms but $[H_{1},H_{2}] \ne 0.$ Hence this error does not increase with evolution time since we have kept our $\delta t$ fixed throughout the simulation (see, Science Advances {\bf 5}: eaau8342  (2019)). Hence, our Trotter error
is well below the resolution throughout the entire simulation time.\\

\noindent
{\bf Truncation Error:}
For the update scheme of matrix product states (MPS), we have truncated the MPS discarding Schmidt values ($s$) which are smaller than $10^{-12}$, keeping the maximum number of Schmidt values (bond-dimension) $\chi =1000$. The total truncation error - $\sum_{i(discarded)} s_i^2$ - accumulated at the end of the long time simulation for the fully polarized state is $\sim 10^{-20}$, while the maximal bond-dimension is never saturated.

\newpage

\widetext
\section*{Supplemental Material for ``Dynamical freezing in the thermodynamic limit: the strongly driven ensemble''}
\setcounter{section}{0}
\setcounter{equation}{0}
\setcounter{figure}{0}
\setcounter{table}{0}
\makeatletter
\renewcommand{\theequation}{S\arabic{equation}}
\renewcommand{\thefigure}{S\arabic{figure}}
\renewcommand{\bibnumfmt}[1]{[S#1]}
\renewcommand{\citenumfont}[1]{S#1}

\section{Quantum Mechanics of a Periodically-driven Quantum Matter and the $t\to\infty$ limit: the Diagonal Ensemble Average (DEA)
}

We consider stroboscopic observations at equally spaced times, $t = nT,$ where
$T$ is the drive-period and $n$ an integer. To that end, one
concentrates on the evolution operator $U(T,0)$ over one period, 
$U(T,0)|\psi(0)\rangle = |\psi(T)\rangle.$ Thus, the wave-function after $n$ drive-cycles will be given by 
$|\psi(nT)\rangle = \left[U(T;0)\right]^{n}|\psi(0)\rangle.$ Now if
$|\mu_{\alpha}\rangle; \alpha = 1 \ldots N$ ($N$ is the Hilbert space dimension) are a complete ortho-normalized set of eigenstates of $U(T,0)$ (with respective eigenvalues 
$e^{-i\mu_{\alpha}}$), then they form a complete basis, and we can express $|\psi(0)\rangle = \sum_{\alpha}C_{\alpha}|\mu_{\alpha}\rangle,$ and
any observable ${\cal O}$ as 
${\cal O} = \sum_{\alpha,\beta}
O_{\alpha,\beta}|\mu_{\alpha}\rangle\langle\mu_{\beta}|.
$ Then after $n$ cycles, the expression for the expectation value of ${\cal O}$ is
\begin{eqnarray}
\langle{\cal O}(nT)\rangle &=&  \langle \psi(nT) |{\cal O}| \psi(nT)\rangle \nonumber\\
&=& \sum_{\alpha,\beta} C_{\alpha}^{\ast}C_{\beta} 
e^{-inT(\mu_{\beta} - \mu_{\alpha})} O_{\alpha\beta} |\mu_{\beta}\rangle\langle\mu_{\alpha}|.
\nonumber
    \label{EqSup:O_nT}
\end{eqnarray}
\noindent
Since $n$ is in the phase, as $n\to\infty,$ 
the terms in the above sum will be oscillating with infinite rapidity about zero as a function of $\alpha$ and $\beta,$ so the terms will cancel each other unless $\alpha = \beta,$ in which case the phase vanishes (see, e.g.~\cite{Reimann}).
Hence at late times, we have 
\begin{equation}
\lim_{n\to\infty}\langle{\cal O}(nT)\rangle
= \sum_{\alpha}|C_{\alpha}|^{2}O_{\alpha\alpha}.
    \label{EqSup:DEA}
\end{equation}
\noindent This is the limiting value to which, 
generally speaking, $\langle{\cal O}(nT)\rangle$ converges in the limit $n\to\infty$~\cite{Rigol_Nature,Reimann}. This limit
is called the {\bf Diagonal Ensemble Average} or {\bf DEA}. 

\section{The Dyson Series Expansion for $H_{eff}$}

This particular perturbation theory allows us to calculate the Floquet unitary time evolution operator perturbatively. Given a time-dependent Hamiltonian H(t) (which may not commute with itself at different times), we split the Hamiltonian into the following two parts 
\begin{equation}
H(t) = H_{0}(t) + V
\end{equation}
where $H_{0}(t)$ is time-dependent but exactly solvable, and V is a time-independent term which we want to treat perturbatively.\\

We denote the time evolution operator corresponding to $H_{0}(t)$ as $U_{0}(t,0)$ and it satisfies 
\begin{equation}
i\frac{\partial{U_{0}(t,0)}}{\partial{t}} = H_{0}(t)U_{0}(t,0)
\end{equation}
The states in the interaction picture are defined as 
\begin{equation}
\psi^{I}(t) = U_{0}^{\dagger}(t,0)\psi(t)
\end{equation}
and satisfy the Schr\"odinger equation
\begin{eqnarray}
    i\frac{\partial{\psi^{I}(t)}}{\partial{t}} &=&V^{I}(t)\psi^{I}(t)\\
V^{I}(t) &=&U_{0}^{\dagger}(t,0)VU_{0}(t,0)
\end{eqnarray}
The corresponding time evolution operator satisfies the equation
\begin{equation}
i\frac{\partial{U^{I}(t,0)}}{\partial{t}} = V^{I}(t)U^{I}(t,0)
\end{equation}
Assuming the initial condition $U^{I}(0,0)$ = $\mathbb{I}$, the solution of the above equation 
\begin{equation}
U^{I}(t,0) = \mathbb{I} - i\int_{0}^{t}dt'V^{I}(t')U^{I}(t',0)
\end{equation}
provides an iterative way of calculating $U^{I}(t,0)$ in powers of $V^{I}$ up to any given order:
\begin{equation}
U^{I}(t,0) = \mathbb{I} + \left(-i\right)\int_{0}^{t}dt_{1}V^{I}(t_{1}) + \left(-i\right)^{2}\int_{0}^{t}dt_{1}V^{I}(t_{1})\int_{0}^{t_{1}}dt_{2}V^{I}(t_{2}) + ...
\end{equation}
The first, second and third order perturbative corrections to the unitary time evolution operator are thus given by
\begin{eqnarray}
U^{I}_{1}(t,0) &=& \left(-i\right)\int_{0}^{t}dt_{1}V^{I}(t_{1})
\label{EqSup:UI_1stOrd_Def}\\
U^{I}_{2}(t,0) &=& \left(-i\right)^{2}\int_{0}^{t}dt_{1}V^{I}(t_{1})\int_{0}^{t_{1}}dt_{2}V^{I}(t_{2})
\label{EqSup:UI_2ndOrd_Def}\\
U^{I}_{3}(t,0) &=& \left(-i\right)^{3}\int_{0}^{t}dt_{1}V^{I}(t_{1})\int_{0}^{t_{1}}dt_{2}V^{I}(t_{2})\int_{0}^{t_{2}}dt_{3}V^{I}(t_{3})
\label{EqSup:UI_3rdOrd_Def}
\end{eqnarray}
Finally, the full time evolution operator is given by
\begin{equation}
U(t,0) = U_{0}(t,0)U^{I}(t,0)
\label{EqSup:Full_U_Def}
\end{equation}
Now, if $T$ is the time period of the periodic drive and we focus only on the stroboscopic dynamics (at $t = nT$, where n is an integer), then it is sufficient to calculate the Floquet unitary time evolution operator $U(T,0)$. The Floquet Hamiltonian $H_{F}$ is defined as 
\begin{equation}
U(T,0) = e^{-iH_{F}T} \implies H_{F} = \frac{i}{T}\ln[U(T,0)]
\label{EqSup:Heff_Def}
\end{equation}
In the cases where $U_{0}(T,0)$ = $\mathbb{I}$, the Floquet Hamiltonian $H_{F}$ can be obtained by setting $t = T$ in Eq. (\ref{EqSup:Full_U_Def}) and then substituting $U(T,0)$ in Eq. (\ref{EqSup:Heff_Def}). Using the expansion of $\ln(1 + x)$, one finds that the first, second and third order terms of the Floquet Hamiltonian $H_{F}$ are
\begin{eqnarray}
H_{F}^{\left(1\right)} &=& \frac{i}{T}U^{I}_{1}(T,0)
\label{EqSup:Heff_1stOrd_Def}
\\
H_{F}^{\left(2\right)} &=&  \frac{i}{T}\left[ U^{I}_{2}(T,0) - \frac{1}{2}\left(U^{I}_{1}(T,0)\right)^{2} \right]
\label{EqSup:Heff_2ndOrd_Def}
\\
H_{F}^{\left(3\right)} &=& \frac{i}{T}\left[ U^{I}_{3}(T,0) - U^{I}_{1}(T,0)U^{I}_{2}(T,0) + \frac{1}{3}\left(U^{I}_{1}(T,0)\right)^{3} \right]
\label{EqSup:Heff_3rdOrd_Def}
\end{eqnarray}
Now, we proceed to apply this Floquet perturbation theory to a case of our interest.

\subsection{The System Hamiltonian}

We consider  $L$ spins in a one-dimensional chain with time-dependent system Hamiltonian 
\begin{equation}
H(t) = H^{x}_{int} + H^{x}_{long} + H^{z}_{trans} + r(t)H^{x}_{drive}
\end{equation}
where 
\begin{eqnarray}
H^{x}_{int} &=&  - J\sum_{i}\sigma_{i}^{x}\sigma_{i+1}^{x} + K\sum_{i}\sigma_{i}^{x}\sigma_{i+2}^{x}
\\
H^{x}_{long} &=&  - h^{x}_{0}\sum_{i}\sigma_{i}^{x}
\\
H^{z}_{trans} &=&  - h^{z}\sum_{i}\sigma_{i}^{z}
\\
H^{x}_{drive} &=&  - h^{x}_{D}\sum_{i}\sigma_{i}^{x}
\label{EqSup:DriveTerm}
\\
r(t) &=&  Sgn(\sin(\omega t))
\label{EqSup:SquareDrive}
\end{eqnarray}
We also define
\begin{equation}
H^{x}_{0} = H^{x}_{int} + H^{x}_{long} = - J\sum_{i}\sigma_{i}^{x}\sigma_{i+1}^{x} + K\sum_{i}\sigma_{i}^{x}\sigma_{i+2}^{x} - h^{x}_{0}\sum_{i}\sigma_{i}^{x}
\end{equation}

\subsection{Calculation of the Floquet Unitary $U(T,0)$}
The one-dimensional spin chain is strongly driven in the longitudinal direction, and the driving field $h^{x}_{D}$ is the only large parameter in the Hamiltonian, compared to which all the other parameters are small. We thus split the Hamiltonian H(t) as follows
\begin{equation}
H(t) = H_{0}(t) + V
\end{equation}
where 
\begin{equation}
H_{0}(t) = - Sgn(sin(\omega t))h^{x}_{D}\sum_{i}\sigma_{i}^{x}
\end{equation}
and 
\begin{equation}
V = - J\sum_{i}\sigma_{i}^{x}\sigma_{i+1}^{x} + K\sum_{i}\sigma_{i}^{x}\sigma_{i+2}^{x} - h^{x}_{0}\sum_{i}\sigma_{i}^{x} - h^{z}\sum_{i}\sigma_{i}^{z}
\end{equation}
Now, as $H_{0}(t)$ commutes with itself at all times, we have
\begin{equation}
U_{0}(t,0) = exp\left(-i\int_{0}^{t}dt'H_{0}(t')\right)
\end{equation}
Now, we have 
\begin{eqnarray}
\int_{0}^{t}dt'H_{0}(t') &=& -h^{x}_{D}t\sum_{i}\sigma^{x}_{i},\quad 0 \leq t \leq T/2\nonumber \\ &=& -h^{x}_{D}\left(T - t\right)\sum_{i}\sigma^{x}_{i},\quad T/2 \leq t \leq T
\end{eqnarray}
So, $U_{0}(t,0)$ is given by
\begin{eqnarray}
U_{0}(t,0) &=& exp\left[ih^{x}_{D}t\sum_{i}\sigma^{x}_{i}\right],\quad 0 \leq t \leq T/2\nonumber \\ &=& exp\left[ih^{x}_{D}\left(T - t\right)\sum_{i}\sigma^{x}_{i}\right],\quad T/2 \leq t \leq T
\end{eqnarray}
So, we see that 
\begin{equation}
U_{0}(T,0) = \mathbb{I}
\end{equation}
and so an analytical form for the Floquet Hamiltonian can be written down for this case.
\newline
\newline
Now, we define
\begin{equation}
\theta(t) = h^{x}_{D}t
\end{equation}
and 
\begin{equation}
\phi(t) = h^{x}_{D}\left(T - t\right)
\end{equation}
So, we can write
\begin{eqnarray}
U_{0}(t,0) &=& exp\left[i\theta(t)\sum_{i}\sigma^{x}_{i}\right],\quad 0 \leq t \leq T/2\nonumber \\ &=& exp\left[i\phi(t)\sum_{i}\sigma^{x}_{i}\right],\quad T/2 \leq t \leq T
\label{EqSup:U0_Sys}
\end{eqnarray}
Now, let us calculate the perturbation Hamiltonian in the interaction picture. We know that
\begin{equation*}
V^{I}(t) = U_{0}^{\dagger}(t,0)VU_{0}(t,0)
\end{equation*}
Substituting $U_{0}(t,0)$ from Eq. (\ref{EqSup:U0_Sys}) in the above equation, we get
\begin{eqnarray}
V^{I}(t) &=& exp\left[-i\theta(t)\sum_{i}\sigma^{x}_{i}\right]Vexp\left[i\theta(t)\sum_{i}\sigma^{x}_{i}\right],\; 0 \leq t \leq T/2\nonumber \\ &=& exp\left[-i\phi(t)\sum_{i}\sigma^{x}_{i}\right]Vexp\left[i\phi(t)\sum_{i}\sigma^{x}_{i}\right],\; \frac{T}{2} \leq t \leq T
\end{eqnarray}
Using Eq. (\ref{EqSup:DriveTerm}) and Eq. (\ref{EqSup:SquareDrive}) and calculating $V^{I}(t)$ for 0 $\leq$ t $\leq$ T/2, we get 
\begin{eqnarray}
V^{I}(t) &=& exp\left[-i\theta(t)\sum_{i}\sigma^{x}_{i}\right]\left(H^{x}_{0} + H^{z}_{trans}\right)exp\left[i\theta(t)\sum_{j}\sigma^{x}_{j}\right]\nonumber \\ &=& H^{x}_{0} + \prod_{i}exp\left[-i\theta(t)\sigma^{x}_{i}\right]\left(-h^{z}\sum_{k}\sigma^{z}_{k}\right) \prod_{j}exp\left[i\theta(t)\sigma^{x}_{j}\right]\nonumber \\ &=& H^{x}_{0} - h^{z}\sum_{k}exp\left[-i\sigma^{x}_{k}\theta(t)\right]\sigma^{z}_{k}exp\left[i\sigma^{x}_{k}\theta(t)\right]
\end{eqnarray}
Similarly, for T/2 $\leq$ t $\leq$ T, we get 
\begin{equation}
V^{I}(t) = H^{x}_{0} - h^{z}\sum_{k}exp\left[-i\sigma^{x}_{k}\phi(t)\right]\sigma^{z}_{k}exp\left[i\sigma^{x}_{k}\phi(t)\right]
\end{equation}
Let us define
\begin{equation}
S^{y} = \sum_{i}\sigma^{y}_{i} 
\end{equation}
and
\begin{equation}
S^{z} = \sum_{i}\sigma^{z}_{i} 
\end{equation}
Now, using the identity
\begin{equation}
exp\left[\pm i\sigma^{x}_{k}\alpha\right] = cos\alpha \pm i\sigma^{x}_{k}sin\alpha
\end{equation}
and carrying out further simplification, we get
\begin{eqnarray}
V^{I}(t) &=& H^{x}_{0} - h^{z}cos\left(2\theta\right)S^{z} + h^{z}sin\left(2\theta\right)S^{y},\quad 0 \leq t \leq T/2\nonumber \\ &=& H^{x}_{0} - h^{z}cos\left(2\phi\right)S^{z} + h^{z}sin\left(2\phi\right)S^{y},\quad T/2 \leq t \leq T
\end{eqnarray}
Now, we proceed to calculate $U^{I}(T,0)$ order by order.

\subsubsection{First Order}
We know that
\begin{equation}
U^{I}_{1}(T,0) = -i\int_{0}^{T}dt_{1}V^{I}(t_{1})
\end{equation}
Now, evaluating the integrals, we get
\begin{equation}
    \int_{0}^{T}dt_{1}H^{x}_{0} = H^{x}_{0}T
\end{equation}
\begin{equation}
    -h^{z}S^{z}\left[ \int_{0}^{T/2}dt_{1}cos\left(2\theta\right) + \int_{T/2}^{T}dt_{1}cos\left(2\phi\right)  \right] = -\frac{h^{z}}{h^{x}_{D}}sin\left(h^{x}_{D}T\right)S^{z}
\end{equation}
\begin{equation}
    h^{z}S^{y}\left[ \int_{0}^{T/2}dt_{1}sin\left(2\theta\right) + \int_{T/2}^{T}dt_{1}sin\left(2\phi\right)  \right] = \frac{2h^{z}}{h^{x}_{D}}sin^{2}\left(\frac{h^{x}_{D}T}{2}\right)S^{y}
\end{equation}
So, the from the above, we get the first order term in the unitary time evolution operator to be
\begin{equation}
U^{I}_{1}(T,0) = -i\left[ \left(H^{x}_{0}\right)T - \frac{h^{z}}{h^{x}_{D}}S^{z}sin\left(h^{x}_{D}T\right) + \frac{2h^{z}}{h^{x}_{D}}S^{y}sin^{2}\left(\frac{h^{x}_{D}T}{2}\right) \right]
\label{EqSup:UI_Sys_1stOrd}
\end{equation}
Using Eq. (\ref{EqSup:Heff_1stOrd_Def}), we get the first order term of the Floquet Hamiltonian to be 
\begin{equation}
H_{F}^{(1)} = H^{x}_{0} - \frac{h^{z}}{h^{x}_{D}T}S^{z}sin\left(h^{x}_{D}T\right) + \frac{2h^{z}}{h^{x}_{D}T}S^{y}sin^{2}\left(\frac{h^{x}_{D}T}{2}\right)
\end{equation}
\noindent
The above Hamiltonian is exactly similar to the one obtained by Magnus expansion in the rotating frame (in zeroth order). We can see that if we put the freezing condition $h^{x}_{D}T$ = $2k\pi$ or $h^{x}_{D}$ = $k\omega$ (where k is an integer), the above Hamiltonian reduces to 
\begin{equation}
H_{F}^{(1)}|_{h^{x}_{D} = k\omega} = H^{x}_{0}
\end{equation}
\subsubsection{Second Order}
We know that
\begin{equation}
U^{I}_{2}(T,0) = \left(-i\right)^{2}\int_{0}^{T}dt_{1}V^{I}(t_{1})\int_{0}^{t_{1}}dt_{2}V^{I}(t_{2})
\end{equation}
We now denote
\begin{equation}
\theta(t_{1}) = \theta_{1}\quad and\quad \theta(t_{2}) = \theta_{2}
\end{equation}
Similarly,
\begin{equation}
\phi(t_{1}) = \phi_{1}\quad and\quad \phi(t_{2}) = \phi_{2}
\end{equation}
We note that
\begin{equation}
\theta_{1} = h^{x}_{D}t_{1}\quad and\quad \theta_{2} = h^{x}_{D}t_{2}
\end{equation}
and 
\begin{equation}
\phi_{1} = h^{x}_{D}\left(T - t_{1}\right)\quad and\quad \phi_{2} = h^{x}_{D}\left(T - t_{2}\right)
\end{equation}
We can write
\begin{equation}
U^{I}_{2}(T,0) = \left(-i\right)^{2}\left(I_{A} + I_{B} + I_{C}\right)
\end{equation}
where we need to calculate the following three integrals
\begin{equation}
I_{A} = \int_{0}^{T/2}\int_{0}^{t_{1}}dt_{1}dt_{2}U(t_{1})U(t_{2})
\end{equation}
\begin{equation}
I_{B} = \int_{T/2}^{T}\int_{0}^{T/2}dt_{1}dt_{2}W(t_{1})U(t_{2})
\end{equation}
\begin{equation}
I_{C} = \int_{T/2}^{T}\int_{T/2}^{t_{1}}dt_{1}dt_{2}W(t_{1})W(t_{2})
\end{equation}
where
\begin{equation}
U(t) = H^{x}_{0} - h^{z}cos\left(2\theta(t)\right)S^{z} + h^{z}sin\left(2\theta(t)\right)S^{y}
\end{equation}
\begin{equation}
W(t) = H^{x}_{0} - h^{z}cos\left(2\phi(t)\right)S^{z} + h^{z}sin\left( 2 \phi\left(t\right) \right) S^{y}
\end{equation}
Now, evaluating the integrals, we get
\begin{equation}
    \left(H^{x}_{0}\right)^{2}\left[ \int_{0}^{T/2}\int_{0}^{t_{1}}dt_{1}dt_{2} + \int_{T/2}^{T}\int_{0}^{T/2}dt_{1}dt_{2} + \int_{T/2}^{T}\int_{T/2}^{t_{1}}dt_{1}dt_{2} \right] = \frac{\left(H^{x}_{0}T\right)^{2}}{2}
\end{equation}
\begin{multline}
    -h^{z}\left(H^{x}_{0}S^{z}\right)\left[ \int_{0}^{T/2}\int_{0}^{t_{1}}dt_{1}dt_{2}cos\left(2\theta_{2}\right) + \int_{T/2}^{T}\int_{0}^{T/2}dt_{1}dt_{2}cos\left(2\theta_{2}\right)\right. \\ + \left. \int_{T/2}^{T}\int_{T/2}^{t_{1}}dt_{1}dt_{2}cos\left(2\phi_{2}\right)\right] = -\frac{h^{z}T}{2h^{x}_{D}}sin\left(h^{x}_{D}T\right)H^{x}_{0}S^{z}
\end{multline}
\begin{multline}
    h^{z}\left(H^{x}_{0}S^{y}\right)\left[ \int_{0}^{T/2}\int_{0}^{t_{1}}dt_{1}dt_{2}sin\left(2\theta_{2}\right) + \int_{T/2}^{T}\int_{0}^{T/2}dt_{1}dt_{2}sin\left(2\theta_{2}\right)\right. \\ + \left. \int_{T/2}^{T}\int_{T/2}^{t_{1}}dt_{1}dt_{2}sin\left(2\phi_{2}\right)\right] = \frac{h^{z}T}{h^{x}_{D}}sin^{2}\left(\frac{h^{x}_{D}T}{2}\right)H^{x}_{0}S^{y}
\end{multline}
\begin{multline}
    -h^{z}\left(S^{z}H^{x}_{0}\right)\left[ \int_{0}^{T/2}\int_{0}^{t_{1}}dt_{1}dt_{2}cos\left(2\theta_{1}\right) + \int_{T/2}^{T}\int_{0}^{T/2}dt_{1}dt_{2}cos\left(2\phi_{1}\right)\right. \\ + \left. \int_{T/2}^{T}\int_{T/2}^{t_{1}}dt_{1}dt_{2}cos\left(2\phi_{1}\right)\right] = -\frac{h^{z}T}{2h^{x}_{D}}sin\left(h^{x}_{D}T\right)S^{z}H^{x}_{0}
\end{multline}
\begin{multline}
    \left(h^{z}\right)^{2}\left(S^{z}\right)^{2}\left[ \int_{0}^{T/2}\int_{0}^{t_{1}}dt_{1}dt_{2}cos\left(2\theta_{1}\right)cos\left(2\theta_{2}\right) + \int_{T/2}^{T}\int_{0}^{T/2}dt_{1}dt_{2}cos\left(2\phi_{1}\right)cos\left(2\theta_{2}\right)\right. \\ + \left. \int_{T/2}^{T}\int_{T/2}^{t_{1}}dt_{1}dt_{2}cos\left(2\phi_{1}\right)cos\left(2\phi_{2}\right)\right] =  \frac{(h^{z})^{2}}{2(h^{x}_{D})^{2}}sin^{2}\left(h^{x}_{D}T\right)\left(S^{z}\right)^{2}
\end{multline}
\begin{multline}
    -\left(h^{z}\right)^{2}\left(S^{z}S^{y}\right)\left[ \int_{0}^{T/2}\int_{0}^{t_{1}}dt_{1}dt_{2}cos\left(2\theta_{1}\right)sin\left(2\theta_{2}\right) + \int_{T/2}^{T}\int_{0}^{T/2}dt_{1}dt_{2}cos\left(2\phi_{1}\right)sin\left(2\theta_{2}\right)\right. \\ + \left. \int_{T/2}^{T}\int_{T/2}^{t_{1}}dt_{1}dt_{2}cos\left(2\phi_{1}\right)sin\left(2\phi_{2}\right)\right] =  - \frac{(h^{z})^{2}}{(h^{x}_{D})^{2}}sin^{2}\left(\frac{h^{x}_{D}T}{2}\right)sin\left(h^{x}_{D}T\right) S^{z}S^{y}
\end{multline}
\begin{multline}
    h^{z}\left(S^{y}H^{x}_{0}\right)\left[ \int_{0}^{T/2}\int_{0}^{t_{1}}dt_{1}dt_{2}sin\left(2\theta_{1}\right) + \int_{T/2}^{T}\int_{0}^{T/2}dt_{1}dt_{2}sin\left(2\phi_{1}\right)\right. \\ + \left. \int_{T/2}^{T}\int_{T/2}^{t_{1}}dt_{1}dt_{2}sin\left(2\phi_{1}\right)\right] = \frac{h^{z}T}{h^{x}_{D}}sin^{2}\left(\frac{h^{x}_{D}T}{2}\right)S^{y}H^{x}_{0}
\end{multline}
\begin{multline}
    -\left(h^{z}\right)^{2}\left(S^{y}S^{z}\right)\left[ \int_{0}^{T/2}\int_{0}^{t_{1}}dt_{1}dt_{2}sin\left(2\theta_{1}\right)cos\left(2\theta_{2}\right) + \int_{T/2}^{T}\int_{0}^{T/2}dt_{1}dt_{2}sin\left(2\phi_{1}\right)cos\left(2\theta_{2}\right)\right. \\ + \left. \int_{T/2}^{T}\int_{T/2}^{t_{1}}dt_{1}dt_{2}sin\left(2\phi_{1}\right)cos\left(2\phi_{2}\right)\right] =  - \frac{(h^{z})^{2}}{(h^{x}_{D})^{2}}sin^{2}\left(\frac{h^{x}_{D}T}{2}\right)sin\left(h^{x}_{D}T\right) S^{y}S^{z}
\end{multline}
\begin{multline}
    \left(h^{z}\right)^{2}\left(S^{y}\right)^{2}\left[ \int_{0}^{T/2}\int_{0}^{t_{1}}dt_{1}dt_{2}sin\left(2\theta_{1}\right)sin\left(2\theta_{2}\right) + \int_{T/2}^{T}\int_{0}^{T/2}dt_{1}dt_{2}sin\left(2\phi_{1}\right)sin\left(2\theta_{2}\right)\right. \\ + \left. \int_{T/2}^{T}\int_{T/2}^{t_{1}}dt_{1}dt_{2}sin\left(2\phi_{1}\right)sin\left(2\phi_{2}\right)\right] =  2\frac{(h^{z})^{2}}{(h^{x}_{D})^{2}}sin^{4}\left(\frac{h^{x}_{D}T}{2}\right)\left(S^{y}\right)^{2}
\end{multline}

\paragraph{The Second Order Floquet Unitary and Effective Hamiltonian}
Collecting all the terms above and grouping them appropriately, we get the second order term in the unitary time evolution operator to be
\begin{equation}
U^{I}_{2}(T,0) =  \left(-i\right)^{2}\left[ S_{1} + S_{2} + S_{3} + S_{4} + S_{5} \right]
\label{EqSup:UI_Sys_2ndOrd}
\end{equation}
where 
\begin{equation}
S_{1} = \frac{\left(H^{x}_{0}T\right)^{2}}{2} 
\end{equation}
\begin{equation}
S_{2} = - \frac{h^{z}T}{2h^{x}_{D}}sin\left(h^{x}_{D}T\right)\left[H^{x}_{0}S^{z} + S^{z}H^{x}_{0}\right]
\end{equation}
\begin{equation}
S_{3} = \frac{h^{z}T}{h^{x}_{D}}sin^{2}\left(\frac{h^{x}_{D}T}{2}\right)\left[H^{x}_{0}S^{y} + S^{y}H^{x}_{0}\right]
\end{equation}
\begin{equation}
S_{4} = - \frac{(h^{z})^{2}}{(h^{x}_{D})^{2}}sin^{2}\left(\frac{h^{x}_{D}T}{2}\right)sin\left(h^{x}_{D}T\right)\left[ S^{y}S^{z} + S^{z}S^{y} \right] 
\end{equation}
\begin{equation}
S_{5} = \frac{(h^{z})^{2}}{2(h^{x}_{D})^{2}}\left(S^{z}\right)^{2}sin^{2}\left(h^{x}_{D}T\right) + 2\frac{(h^{z})^{2}}{(h^{x}_{D})^{2}}\left(S^{y}\right)^{2}sin^{4}\left(\frac{h^{x}_{D}T}{2}\right)
\end{equation}
It is readily seen that on imposing the freezing condition $h^{x}_{D}T$ = $2k\pi$ or $h^{x}_{D}$ = $k\omega$ (where k is an integer), all the terms in $U^{I}_{2}(T,0)$ (except $S_{1})$ become equal to zero. \\

Moreover, using Eq. (\ref{EqSup:UI_Sys_1stOrd}) and Eq. (\ref{EqSup:UI_Sys_2ndOrd}), one can easily verify that in this particular case we have
\begin{equation}
U^{I}_{2}(T,0) = \frac{1}{2}\left[U^{I}_{2}(T,0)\right]^{2}
\end{equation}
So, using Eq. (\ref{EqSup:Heff_2ndOrd_Def}), we get the second order term of the Floquet Hamiltonian to be 
\begin{equation}
H_{F}^{(2)} = 0
\end{equation}
Note that the second-order term of the Floquet Hamiltonian is always zero, even when the freezing condition is not satisfied. This result is also similar to the result obtained from first-order Magnus expansion in a rotating frame.

\subsubsection{Third Order}
We know that
\begin{equation}
U^{I}_{3}(t,0) = \left(-i\right)^{3}\int_{0}^{t}dt_{1}V^{I}(t_{1})\int_{0}^{t_{1}}dt_{2}V^{I}(t_{2})\int_{0}^{t_{2}}dt_{3}V^{I}(t_{3})
\end{equation}
We now denote
\begin{equation}
\theta(t_{1}) = \theta_{1},\quad \theta(t_{2}) = \theta_{2}\quad and\quad \theta(t_{3}) = \theta_{3}
\end{equation}
Similarly,
\begin{equation}
\phi(t_{1}) = \phi_{1},\quad \phi(t_{2}) = \phi_{2}\quad and\quad \phi(t_{3}) = \phi_{3}
\end{equation}
We note that
\begin{equation}
\theta_{1} = h^{x}_{D}t_{1},\quad \theta_{2} = h^{x}_{D}t_{2}\quad and\quad \theta_{3} = h^{x}_{D}t_{3}
\end{equation}
and 
\begin{equation}
\phi_{1} = h^{x}_{D}\left(T - t_{1}\right),\quad \phi_{2} = h^{x}_{D}\left(T - t_{2}\right)\quad and\quad \phi_{3} = h^{x}_{D}\left(T - t_{3}\right)
\end{equation}
We can write
\begin{equation}
U^{I}_{3}(T,0) = \left(-i\right)^{3}\left(I_{A} + I_{B} + I_{C} + I_{D}\right)
\end{equation}
where we need to calculate the following three integrals
\begin{equation}
I_{A} = \int_{0}^{T/2}\int_{0}^{t_{1}}\int_{0}^{t_{2}} dt_{1}dt_{2}dt_{3} U(t_{1})U(t_{2})U(t_{3})
\end{equation}
\begin{equation}
I_{B} = \int_{T/2}^{T}\int_{0}^{T/2}\int_{0}^{t_{2}} dt_{1}dt_{2}dt_{3} W(t_{1})U(t_{2})U(t_{3})
\end{equation}
\begin{equation}
I_{C} = \int_{T/2}^{T}\int_{T/2}^{t_{1}}\int_{0}^{T/2} dt_{1}dt_{2}dt_{3} W(t_{1})W(t_{2})U(t_{3})
\end{equation}
\begin{equation}
I_{D} = \int_{T/2}^{T}\int_{T/2}^{t_{1}}\int_{T/2}^{t_{2}} dt_{1}dt_{2}dt_{3} W(t_{1})W(t_{2})W(t_{3})
\end{equation}
where
\begin{equation}
U(t) = H^{x}_{0} - h^{z}cos\left(2\theta(t)\right)S^{z} + h^{z}sin\left(2\theta(t)\right)S^{y}
\end{equation}
\begin{equation}
W(t) = H^{x}_{0} - h^{z}cos\left(2\phi(t)\right)S^{z} + h^{z}sin\left(2\phi(t)\right)S^{y}
\end{equation}
While evaluating the integrals, we will write down two expressions for each integral. One of them is the general result of the integral, the other is the form which this result takes after imposing the freezing condition. \\

Now, evaluating the integrals, we get
\begin{multline}
    \left(H^{x}_{0}\right)^{3}\left[ \int_{0}^{T/2}\int_{0}^{t_{1}}\int_{0}^{t_{2}} dt_{1}dt_{2}dt_{3} +  \int_{T/2}^{T}\int_{0}^{T/2}\int_{0}^{t_{2}} dt_{1}dt_{2}dt_{3} + \int_{T/2}^{T}\int_{T/2}^{t_{1}}\int_{0}^{T/2} dt_{1}dt_{2}dt_{3}\right. \\ + \left. \int_{T/2}^{T}\int_{T/2}^{t_{1}}\int_{T/2}^{t_{2}} dt_{1}dt_{2}dt_{3} \right] = \frac{\left(H^{x}_{0}T\right)^{3}}{6}
\label{EqSup:UI_Sys_3rdOrd_Term1}
\end{multline}
\begin{align}
    & -h^{z}\left(H^{x}_{0}\right)^{2}S^{z}\left[ \int_{0}^{T/2}\int_{0}^{t_{1}}\int_{0}^{t_{2}} dt_{1}dt_{2}dt_{3}cos\left(2\theta_{3}\right) +  \int_{T/2}^{T}\int_{0}^{T/2}\int_{0}^{t_{2}} dt_{1}dt_{2}dt_{3}cos\left(2\theta_{3}\right) \right. \nonumber \\ & \left. + \int_{T/2}^{T}\int_{T/2}^{t_{1}}\int_{0}^{T/2} dt_{1}dt_{2}dt_{3}cos\left(2\theta_{3}\right)  +  \int_{T/2}^{T}\int_{T/2}^{t_{1}}\int_{T/2}^{t_{2}} dt_{1}dt_{2}dt_{3}cos\left(2\phi_{3}\right) \right] \nonumber \\ & = -\frac{h^{z}}{8\left(h^{x}_{D}\right)^{3}}\left[2h^{x}_{D}T + \left(\left(h^{x}_{D}T\right)^{2} - 2\right)sin\left(h^{x}_{D}T\right)\right]\left(H^{x}_{0}\right)^{2}S^{z} \nonumber \\ & \xrightarrow{h^{x}_{D} = k\omega} \left[-\frac{h^{z}T}{4\left(h^{x}_{D}\right)^{2}}\right]\left(H^{x}_{0}\right)^{2}S^{z}
\label{EqSup:UI_Sys_3rdOrd_Term2}
\end{align}
\begin{align}
    & h^{z}\left(H^{x}_{0}\right)^{2}S^{y}\left[ \int_{0}^{T/2}\int_{0}^{t_{1}}\int_{0}^{t_{2}} dt_{1}dt_{2}dt_{3}sin\left(2\theta_{3}\right) +  \int_{T/2}^{T}\int_{0}^{T/2}\int_{0}^{t_{2}} dt_{1}dt_{2}dt_{3}sin\left(2\theta_{3}\right) \right. \nonumber \\ & \left. + \int_{T/2}^{T}\int_{T/2}^{t_{1}}\int_{0}^{T/2} dt_{1}dt_{2}dt_{3}sin\left(2\theta_{3}\right)  +  \int_{T/2}^{T}\int_{T/2}^{t_{1}}\int_{T/2}^{t_{2}} dt_{1}dt_{2}dt_{3}sin\left(2\phi_{3}\right) \right] \nonumber \\ & = \frac{h^{z}}{8\left(h^{x}_{D}\right)^{3}}\left[ 2\left(h^{x}_{D}T\right)^{2} - 2 + \left(2 - \left(h^{x}_{D}T\right)^{2}\right)cos\left(h^{x}_{D}T\right)\right]\left(H^{x}_{0}\right)^{2}S^{y} \nonumber \\ & \xrightarrow{h^{x}_{D} = k\omega} \left[\frac{h^{z}T^{2}}{8h^{x}_{D}}\right]\left(H^{x}_{0}\right)^{2}S^{y}
\end{align}
\begin{align}
    & -h^{z}\left(H^{x}_{0}S^{z}H^{x}_{0}\right)\left[ \int_{0}^{T/2}\int_{0}^{t_{1}}\int_{0}^{t_{2}} dt_{1}dt_{2}dt_{3}cos\left(2\theta_{2}\right) +  \int_{T/2}^{T}\int_{0}^{T/2}\int_{0}^{t_{2}} dt_{1}dt_{2}dt_{3}cos\left(2\theta_{2}\right) \right. \nonumber \\ & \left. + \int_{T/2}^{T}\int_{T/2}^{t_{1}}\int_{0}^{T/2} dt_{1}dt_{2}dt_{3}cos\left(2\phi_{2}\right)  +  \int_{T/2}^{T}\int_{T/2}^{t_{1}}\int_{T/2}^{t_{2}} dt_{1}dt_{2}dt_{3}cos\left(2\phi_{2}\right) \right] \nonumber \\ & = -\frac{h^{z}}{4\left(h^{x}_{D}\right)^{3}}\left[ \left(2 + \left(h^{x}_{D}T\right)^{2}\right)sin\left(h^{x}_{D}T\right) - 2h^{x}_{D}T \right] H^{x}_{0}S^{z}H^{x}_{0} \nonumber \\ & \xrightarrow{h^{x}_{D} = k\omega} \left[\frac{h^{z}T}{2\left(h^{x}_{D}\right)^{2}}\right]H^{x}_{0}S^{z}H^{x}_{0}
\end{align}
\begin{align}
    & \left(h^{z}\right)^{2}H^{x}_{0}\left(S^{z}\right)^{2}\left[ \int_{0}^{T/2}\int_{0}^{t_{1}}\int_{0}^{t_{2}} dt_{1}dt_{2}dt_{3}cos\left(2\theta_{2}\right)cos\left(2\theta_{3}\right) +  \int_{T/2}^{T}\int_{0}^{T/2}\int_{0}^{t_{2}} dt_{1}dt_{2}dt_{3}cos\left(2\theta_{2}\right)cos\left(2\theta_{3}\right) \right. \nonumber \\ & \left. + \int_{T/2}^{T}\int_{T/2}^{t_{1}}\int_{0}^{T/2} dt_{1}dt_{2}dt_{3}cos\left(2\phi_{2}\right)cos\left(2\theta_{3}\right)  +  \int_{T/2}^{T}\int_{T/2}^{t_{1}}\int_{T/2}^{t_{2}} dt_{1}dt_{2}dt_{3}cos\left(2\phi_{2}\right)cos\left(2\phi_{3}\right) \right] \nonumber \\ & = \frac{\left(h^{z}\right)^{2}}{32\left(h^{x}_{D}\right)^{3}}\left[ 6h^{x}_{D}T - 4h^{x}_{D}T cos\left(2h^{x}_{D}T\right) - 8sin\left(h^{x}_{D}T\right) + 3sin\left(2h^{x}_{D}T\right) \right] H^{x}_{0}\left(S^{z}\right)^{2} \nonumber \\ & \xrightarrow{h^{x}_{D} = k\omega} \left[\frac{\left(h^{z}\right)^{2}T}{16\left(h^{x}_{D}\right)^{2}}\right]H^{x}_{0}\left(S^{z}\right)^{2}
\end{align}
\begin{align}
    & -\left(h^{z}\right)^{2}H^{x}_{0}S^{z}S^{y}\left[ \int_{0}^{T/2}\int_{0}^{t_{1}}\int_{0}^{t_{2}} dt_{1}dt_{2}dt_{3}cos\left(2\theta_{2}\right)sin\left(2\theta_{3}\right) +  \int_{T/2}^{T}\int_{0}^{T/2}\int_{0}^{t_{2}} dt_{1}dt_{2}dt_{3}cos\left(2\theta_{2}\right)sin\left(2\theta_{3}\right) \right. \nonumber \\ & \left. + \int_{T/2}^{T}\int_{T/2}^{t_{1}}\int_{0}^{T/2} dt_{1}dt_{2}dt_{3}cos\left(2\phi_{2}\right)sin\left(2\theta_{3}\right)  +  \int_{T/2}^{T}\int_{T/2}^{t_{1}}\int_{T/2}^{t_{2}} dt_{1}dt_{2}dt_{3}cos\left(2\phi_{2}\right)sin\left(2\phi_{3}\right) \right] \nonumber \\ & = \frac{\left(h^{z}\right)^{2}}{32\left(h^{x}_{D}\right)^{3}}\left[ 5 + 2\left(h^{x}_{D}T\right)^{2} - 8cos\left(h^{x}_{D}T\right) + 3cos\left(2h^{x}_{D}T\right) + 4h^{x}_{D}T \left( sin\left(2h^{x}_{D}T\right) - 2sin\left(h^{x}_{D}T\right)\right) \right] H^{x}_{0}S^{z}S^{y} \nonumber \\ & \xrightarrow{h^{x}_{D} = k\omega} \left[\frac{\left(h^{z}\right)^{2}T^{2}}{16h^{x}_{D}}\right]H^{x}_{0}S^{z}S^{y}
\end{align}
\begin{align}
    & h^{z}\left(H^{x}_{0}S^{y}H^{x}_{0}\right)\left[ \int_{0}^{T/2}\int_{0}^{t_{1}}\int_{0}^{t_{2}} dt_{1}dt_{2}dt_{3}sin\left(2\theta_{2}\right) +  \int_{T/2}^{T}\int_{0}^{T/2}\int_{0}^{t_{2}} dt_{1}dt_{2}dt_{3}sin\left(2\theta_{2}\right) \right. \nonumber \\ & \left. + \int_{T/2}^{T}\int_{T/2}^{t_{1}}\int_{0}^{T/2} dt_{1}dt_{2}dt_{3}sin\left(2\phi_{2}\right)  +  \int_{T/2}^{T}\int_{T/2}^{t_{1}}\int_{T/2}^{t_{2}} dt_{1}dt_{2}dt_{3}sin\left(2\phi_{2}\right) \right] \nonumber \\ & = \frac{h^{z}}{4\left(h^{x}_{D}\right)^{3}}\left[ 2 - \left(2 + \left(h^{x}_{D}T\right)^{2}\right)cos\left(h^{x}_{D}T\right) \right] H^{x}_{0}S^{y}H^{x}_{0} \nonumber \\ & \xrightarrow{h^{x}_{D} = k\omega} \left[ -\frac{h^{z}T^{2}}{4h^{x}_{D}}\right]H^{x}_{0}S^{y}H^{x}_{0}
\end{align}
\begin{align}
    & -\left(h^{z}\right)^{2}H^{x}_{0}S^{y}S^{z}\left[ \int_{0}^{T/2}\int_{0}^{t_{1}}\int_{0}^{t_{2}} dt_{1}dt_{2}dt_{3}sin\left(2\theta_{2}\right)cos\left(2\theta_{3}\right) +  \int_{T/2}^{T}\int_{0}^{T/2}\int_{0}^{t_{2}} dt_{1}dt_{2}dt_{3}sin\left(2\theta_{2}\right)cos\left(2\theta_{3}\right) \right. \nonumber \\ & \left. + \int_{T/2}^{T}\int_{T/2}^{t_{1}}\int_{0}^{T/2} dt_{1}dt_{2}dt_{3}sin\left(2\phi_{2}\right)cos\left(2\theta_{3}\right)  +  \int_{T/2}^{T}\int_{T/2}^{t_{1}}\int_{T/2}^{t_{2}} dt_{1}dt_{2}dt_{3}sin\left(2\phi_{2}\right)cos\left(2\phi_{3}\right) \right] \nonumber \\ & = -\frac{\left(h^{z}\right)^{2}}{32\left(h^{x}_{D}\right)^{3}}\left[ 3 + 2\left(h^{x}_{D}T\right)^{2} - 3cos\left(2h^{x}_{D}T\right) - 4h^{x}_{D}Tsin\left(2h^{x}_{D}T\right) \right] H^{x}_{0}S^{y}S^{z} \nonumber \\ & \xrightarrow{h^{x}_{D} = k\omega} \left[-\frac{\left(h^{z}\right)^{2}T^{2}}{16h^{x}_{D}}\right]H^{x}_{0}S^{y}S^{z}
\end{align}
\begin{align}
    & \left(h^{z}\right)^{2}H^{x}_{0}\left(S^{y}\right)^{2}\left[ \int_{0}^{T/2}\int_{0}^{t_{1}}\int_{0}^{t_{2}} dt_{1}dt_{2}dt_{3}sin\left(2\theta_{2}\right)sin\left(2\theta_{3}\right) +  \int_{T/2}^{T}\int_{0}^{T/2}\int_{0}^{t_{2}} dt_{1}dt_{2}dt_{3}sin\left(2\theta_{2}\right)sin\left(2\theta_{3}\right) \right. \nonumber \\ & \left. + \int_{T/2}^{T}\int_{T/2}^{t_{1}}\int_{0}^{T/2} dt_{1}dt_{2}dt_{3}sin\left(2\phi_{2}\right)sin\left(2\theta_{3}\right)  +  \int_{T/2}^{T}\int_{T/2}^{t_{1}}\int_{T/2}^{t_{2}} dt_{1}dt_{2}dt_{3}sin\left(2\phi_{2}\right)sin\left(2\phi_{3}\right) \right] \nonumber \\ & = \frac{\left(h^{z}\right)^{2}}{32\left(h^{x}_{D}\right)^{3}}\left[ 2h^{x}_{D}T\left( 5 - 4cos\left(h^{x}_{D}T\right) + 2cos\left(2h^{x}_{D}T\right)\right) - 3sin\left(2h^{x}_{D}T\right) \right] H^{x}_{0}\left(S^{y}\right)^{2} \nonumber \\ & \xrightarrow{h^{x}_{D} = k\omega} \left[\frac{3\left(h^{z}\right)^{2}T}{16\left(h^{x}_{D}\right)^{2}}\right]H^{x}_{0}\left(S^{y}\right)^{2}
\end{align}
\begin{align}
    & -h^{z}S^{z}\left(H^{x}_{0}\right)^{2}\left[ \int_{0}^{T/2}\int_{0}^{t_{1}}\int_{0}^{t_{2}} dt_{1}dt_{2}dt_{3}cos\left(2\theta_{1}\right) +  \int_{T/2}^{T}\int_{0}^{T/2}\int_{0}^{t_{2}} dt_{1}dt_{2}dt_{3}cos\left(2\phi_{1}\right) \right. \nonumber \\ & \left. + \int_{T/2}^{T}\int_{T/2}^{t_{1}}\int_{0}^{T/2} dt_{1}dt_{2}dt_{3}cos\left(2\phi_{1}\right)  +  \int_{T/2}^{T}\int_{T/2}^{t_{1}}\int_{T/2}^{t_{2}} dt_{1}dt_{2}dt_{3}cos\left(2\phi_{1}\right) \right] \nonumber \\ & = -\frac{h^{z}}{8\left(h^{x}_{D}\right)^{3}}\left[2h^{x}_{D}T + \left(\left(h^{x}_{D}T\right)^{2} - 2\right)sin\left(h^{x}_{D}T\right)\right] S^{z}\left(H^{x}_{0}\right)^{2} \nonumber \\ & \xrightarrow{h^{x}_{D} = k\omega} \left[-\frac{h^{z}T}{4\left(h^{x}_{D}\right)^{2}}\right] S^{z}\left(H^{x}_{0}\right)^{2}
\end{align}
\begin{align}
    & \left(h^{z}\right)^{2}S^{z}H^{x}_{0}S^{z}\left[ \int_{0}^{T/2}\int_{0}^{t_{1}}\int_{0}^{t_{2}} dt_{1}dt_{2}dt_{3}cos\left(2\theta_{1}\right)cos\left(2\theta_{3}\right) +  \int_{T/2}^{T}\int_{0}^{T/2}\int_{0}^{t_{2}} dt_{1}dt_{2}dt_{3}cos\left(2\phi_{1}\right)cos\left(2\theta_{3}\right) \right. \nonumber \\ & \left. + \int_{T/2}^{T}\int_{T/2}^{t_{1}}\int_{0}^{T/2} dt_{1}dt_{2}dt_{3}cos\left(2\phi_{1}\right)cos\left(2\theta_{3}\right)  +  \int_{T/2}^{T}\int_{T/2}^{t_{1}}\int_{T/2}^{t_{2}} dt_{1}dt_{2}dt_{3}cos\left(2\phi_{1}\right)cos\left(2\phi_{3}\right) \right] \nonumber \\ & = -\frac{\left(h^{z}\right)^{2}}{16\left(h^{x}_{D}\right)^{3}}\left[ 2h^{x}_{D}T - 8sin\left(h^{x}_{D}T\right) + 3sin\left(2h^{x}_{D}T\right) \right] S^{z}H^{x}_{0}S^{z} \nonumber \\ & \xrightarrow{h^{x}_{D} = k\omega} \left[-\frac{\left(h^{z}\right)^{2}T}{8\left(h^{x}_{D}\right)^{2}}\right] S^{z}H^{x}_{0}S^{z}
\end{align}
\begin{align}
    & -\left(h^{z}\right)^{2}S^{z}H^{x}_{0}S^{y}\left[ \int_{0}^{T/2}\int_{0}^{t_{1}}\int_{0}^{t_{2}} dt_{1}dt_{2}dt_{3}cos\left(2\theta_{1}\right)sin\left(2\theta_{3}\right) +  \int_{T/2}^{T}\int_{0}^{T/2}\int_{0}^{t_{2}} dt_{1}dt_{2}dt_{3}cos\left(2\phi_{1}\right)sin\left(2\theta_{3}\right) \right. \nonumber \\ & \left. + \int_{T/2}^{T}\int_{T/2}^{t_{1}}\int_{0}^{T/2} dt_{1}dt_{2}dt_{3}cos\left(2\phi_{1}\right)sin\left(2\theta_{3}\right)  +  \int_{T/2}^{T}\int_{T/2}^{t_{1}}\int_{T/2}^{t_{2}} dt_{1}dt_{2}dt_{3}cos\left(2\phi_{1}\right)sin\left(2\phi_{3}\right) \right] \nonumber \\ & = -\frac{\left(h^{z}\right)^{2}}{16\left(h^{x}_{D}\right)^{3}}\left[ 1 - 4cos\left(h^{x}_{D}T\right) + 3cos\left(2h^{x}_{D}T\right) + 4h^{x}_{D}T sin\left(h^{x}_{D}T\right) \right] S^{z}H^{x}_{0}S^{y} \nonumber \\ & \xrightarrow{h^{x}_{D} = k\omega} \left[ 0 \right] S^{z}H^{x}_{0}S^{y}
\end{align}
\begin{align}
    & \left(h^{z}\right)^{2}\left(S^{z}\right)^{2}H^{x}_{0}\left[ \int_{0}^{T/2}\int_{0}^{t_{1}}\int_{0}^{t_{2}} dt_{1}dt_{2}dt_{3}cos\left(2\theta_{1}\right)cos\left(2\theta_{2}\right) +  \int_{T/2}^{T}\int_{0}^{T/2}\int_{0}^{t_{2}} dt_{1}dt_{2}dt_{3}cos\left(2\phi_{1}\right)cos\left(2\theta_{2}\right) \right. \nonumber \\ & \left. + \int_{T/2}^{T}\int_{T/2}^{t_{1}}\int_{0}^{T/2} dt_{1}dt_{2}dt_{3}cos\left(2\phi_{1}\right)cos\left(2\phi_{2}\right)  +  \int_{T/2}^{T}\int_{T/2}^{t_{1}}\int_{T/2}^{t_{2}} dt_{1}dt_{2}dt_{3}cos\left(2\phi_{1}\right)cos\left(2\phi_{2}\right) \right] \nonumber \\ & = \frac{\left(h^{z}\right)^{2}}{32\left(h^{x}_{D}\right)^{3}}\left[ 6h^{x}_{D}T - 4h^{x}_{D}T cos\left(2h^{x}_{D}T\right) - 8sin\left(h^{x}_{D}T\right) + 3sin\left(2h^{x}_{D}T\right) \right] \left(S^{z}\right)^{2}H^{x}_{0} \nonumber \\ & \xrightarrow{h^{x}_{D} = k\omega} \left[\frac{\left(h^{z}\right)^{2}T}{16\left(h^{x}_{D}\right)^{2}}\right] \left(S^{z}\right)^{2}H^{x}_{0}
\end{align}
\begin{align}
    & -\left(h^{z}\right)^{3}\left(S^{z}\right)^{3}\left[ \int_{0}^{T/2}\int_{0}^{t_{1}}\int_{0}^{t_{2}} dt_{1}dt_{2}dt_{3}cos\left(2\theta_{1}\right)cos\left(2\theta_{2}\right)cos\left(2\theta_{3}\right) \right. \nonumber \\ & \left. +  \int_{T/2}^{T}\int_{0}^{T/2}\int_{0}^{t_{2}} dt_{1}dt_{2}dt_{3}cos\left(2\phi_{1}\right)cos\left(2\theta_{2}\right)cos\left(2\theta_{3}\right) \right. \nonumber \\ & \left. + \int_{T/2}^{T}\int_{T/2}^{t_{1}}\int_{0}^{T/2} dt_{1}dt_{2}dt_{3}cos\left(2\phi_{1}\right)cos\left(2\phi_{2}\right)cos\left(2\theta_{3}\right) \right. \nonumber \\ & \left. +  \int_{T/2}^{T}\int_{T/2}^{t_{1}}\int_{T/2}^{t_{2}} dt_{1}dt_{2}dt_{3}cos\left(2\phi_{1}\right)cos\left(2\phi_{2}\right)cos\left(2\phi_{3}\right) \right] \nonumber \\ & = -\frac{\left(h^{z}\right)^{3}}{6\left(h^{x}_{D}\right)^{3}}\left[ sin^{3}\left(h^{x}_{D}T\right) \right] \left(S^{z}\right)^{3} \nonumber \\ & \xrightarrow{h^{x}_{D} = k\omega} \left[ 0 \right] \left(S^{z}\right)^{3}
\end{align}
\begin{align}
    & \left(h^{z}\right)^{3}\left(S^{z}\right)^{2}S^{y}\left[ \int_{0}^{T/2}\int_{0}^{t_{1}}\int_{0}^{t_{2}} dt_{1}dt_{2}dt_{3}cos\left(2\theta_{1}\right)cos\left(2\theta_{2}\right)sin\left(2\theta_{3}\right) \right. \nonumber \\ & \left. +  \int_{T/2}^{T}\int_{0}^{T/2}\int_{0}^{t_{2}} dt_{1}dt_{2}dt_{3}cos\left(2\phi_{1}\right)cos\left(2\theta_{2}\right)sin\left(2\theta_{3}\right) \right. \nonumber \\ & \left. + \int_{T/2}^{T}\int_{T/2}^{t_{1}}\int_{0}^{T/2} dt_{1}dt_{2}dt_{3}cos\left(2\phi_{1}\right)cos\left(2\phi_{2}\right)sin\left(2\theta_{3}\right) \right. \nonumber \\ & \left. +  \int_{T/2}^{T}\int_{T/2}^{t_{1}}\int_{T/2}^{t_{2}} dt_{1}dt_{2}dt_{3}cos\left(2\phi_{1}\right)cos\left(2\phi_{2}\right)sin\left(2\phi_{3}\right) \right] \nonumber \\ & = \frac{\left(h^{z}\right)^{3}}{24\left(h^{x}_{D}\right)^{3}}\left[ 5 - 3cos\left(h^{x}_{D}T\right) - 3cos\left(2h^{x}_{D}T\right) + cos\left(3h^{x}_{D}T\right) - 3h^{x}_{D}T sin\left(h^{x}_{D}T\right) \right] \left(S^{z}\right)^{2}S^{y} \nonumber \\ & \xrightarrow{h^{x}_{D} = k\omega} \left[ 0 \right] \left(S^{z}\right)^{2}S^{y}
\end{align}
\begin{align}
    & -\left(h^{z}\right)^{2}S^{z}S^{y}H^{x}_{0}\left[ \int_{0}^{T/2}\int_{0}^{t_{1}}\int_{0}^{t_{2}} dt_{1}dt_{2}dt_{3}cos\left(2\theta_{1}\right)sin\left(2\theta_{2}\right) +  \int_{T/2}^{T}\int_{0}^{T/2}\int_{0}^{t_{2}} dt_{1}dt_{2}dt_{3}cos\left(2\phi_{1}\right)sin\left(2\theta_{2}\right) \right. \nonumber \\ & \left. + \int_{T/2}^{T}\int_{T/2}^{t_{1}}\int_{0}^{T/2} dt_{1}dt_{2}dt_{3}cos\left(2\phi_{1}\right)sin\left(2\phi_{2}\right)  +  \int_{T/2}^{T}\int_{T/2}^{t_{1}}\int_{T/2}^{t_{2}} dt_{1}dt_{2}dt_{3}cos\left(2\phi_{1}\right)sin\left(2\phi_{2}\right) \right] \nonumber \\ & = -\frac{\left(h^{z}\right)^{2}}{32\left(h^{x}_{D}\right)^{3}}\left[ 3 + 2\left(h^{x}_{D}T\right)^{2} - 3cos\left(2h^{x}_{D}T\right) - 4h^{x}_{D}T sin\left(2h^{x}_{D}T\right) \right] S^{z}S^{y}H^{x}_{0} \nonumber \\ & \xrightarrow{h^{x}_{D} = k\omega} \left[-\frac{\left(h^{z}\right)^{2}T^{2}}{16h^{x}_{D}}\right]S^{z}S^{y}H^{x}_{0}
\end{align}
\begin{align}
    & \left(h^{z}\right)^{3}S^{z}S^{y}S^{z}\left[ \int_{0}^{T/2}\int_{0}^{t_{1}}\int_{0}^{t_{2}} dt_{1}dt_{2}dt_{3}cos\left(2\theta_{1}\right)sin\left(2\theta_{2}\right)cos\left(2\theta_{3}\right) \right. \nonumber \\ & \left. +  \int_{T/2}^{T}\int_{0}^{T/2}\int_{0}^{t_{2}} dt_{1}dt_{2}dt_{3}cos\left(2\phi_{1}\right)sin\left(2\theta_{2}\right)cos\left(2\theta_{3}\right) \right. \nonumber \\ & \left. + \int_{T/2}^{T}\int_{T/2}^{t_{1}}\int_{0}^{T/2} dt_{1}dt_{2}dt_{3}cos\left(2\phi_{1}\right)sin\left(2\phi_{2}\right)cos\left(2\theta_{3}\right) \right. \nonumber \\ & \left. +  \int_{T/2}^{T}\int_{T/2}^{t_{1}}\int_{T/2}^{t_{2}} dt_{1}dt_{2}dt_{3}cos\left(2\phi_{1}\right)sin\left(2\phi_{2}\right)cos\left(2\phi_{3}\right) \right] \nonumber \\ & = \frac{\left(h^{z}\right)^{3}}{12\left(h^{x}_{D}\right)^{3}}\left[ 2cos^{3}\left(h^{x}_{D}T\right) - 2 + 3h^{x}_{D}T sin\left(h^{x}_{D}T\right) \right] S^{z}S^{y}S^{z} \nonumber \\ & \xrightarrow{h^{x}_{D} = k\omega} \left[ 0 \right] S^{z}S^{y}S^{z}
\end{align}
\begin{align}
    & -\left(h^{z}\right)^{3}S^{z}\left(S^{y}\right)^{2}\left[ \int_{0}^{T/2}\int_{0}^{t_{1}}\int_{0}^{t_{2}} dt_{1}dt_{2}dt_{3}cos\left(2\theta_{1}\right)sin\left(2\theta_{2}\right)sin\left(2\theta_{3}\right) \right. \nonumber \\ & \left. +  \int_{T/2}^{T}\int_{0}^{T/2}\int_{0}^{t_{2}} dt_{1}dt_{2}dt_{3}cos\left(2\phi_{1}\right)sin\left(2\theta_{2}\right)sin\left(2\theta_{3}\right) \right. \nonumber \\ & \left. + \int_{T/2}^{T}\int_{T/2}^{t_{1}}\int_{0}^{T/2} dt_{1}dt_{2}dt_{3}cos\left(2\phi_{1}\right)sin\left(2\phi_{2}\right)sin\left(2\theta_{3}\right) \right. \nonumber \\ & \left. +  \int_{T/2}^{T}\int_{T/2}^{t_{1}}\int_{T/2}^{t_{2}} dt_{1}dt_{2}dt_{3}cos\left(2\phi_{1}\right)sin\left(2\phi_{2}\right)sin\left(2\phi_{3}\right) \right] \nonumber \\ & = -\frac{\left(h^{z}\right)^{3}}{24\left(h^{x}_{D}\right)^{3}}\left[ 6sin\left(h^{x}_{D}T\right) - 3h^{x}_{D}T cos\left(h^{x}_{D}T\right) - 3sin\left(2h^{x}_{D}T\right) + sin\left(3h^{x}_{D}T\right) \right] S^{z}\left(S^{y}\right)^{2} \nonumber \\ & \xrightarrow{h^{x}_{D} = k\omega} \left[ \frac{\left(h^{z}\right)^{3}T}{8\left(h^{x}_{D}\right)^{2}} \right] S^{z}\left(S^{y}\right)^{2}
\end{align}
\begin{align}
    & h^{z}S^{y}\left(H^{x}_{0}\right)^{2}\left[ \int_{0}^{T/2}\int_{0}^{t_{1}}\int_{0}^{t_{2}} dt_{1}dt_{2}dt_{3}sin\left(2\theta_{1}\right) +  \int_{T/2}^{T}\int_{0}^{T/2}\int_{0}^{t_{2}} dt_{1}dt_{2}dt_{3}sin\left(2\phi_{1}\right) \right. \nonumber \\ & \left. + \int_{T/2}^{T}\int_{T/2}^{t_{1}}\int_{0}^{T/2} dt_{1}dt_{2}dt_{3}sin\left(2\phi_{1}\right)  +  \int_{T/2}^{T}\int_{T/2}^{t_{1}}\int_{T/2}^{t_{2}} dt_{1}dt_{2}dt_{3}sin\left(2\phi_{1}\right) \right] \nonumber \\ & = \frac{h^{z}}{8\left(h^{x}_{D}\right)^{3}}\left[ 2\left(h^{x}_{D}T\right)^{2} - 2 + \left(2 - \left(h^{x}_{D}T\right)^{2}\right)cos\left(h^{x}_{D}T\right)\right]S^{y}\left(H^{x}_{0}\right)^{2} \nonumber \\ & \xrightarrow{h^{x}_{D} = k\omega} \left[\frac{h^{z}T^{2}}{8h^{x}_{D}}\right]S^{y}\left(H^{x}_{0}\right)^{2}
\end{align}
\begin{align}
    & -\left(h^{z}\right)^{2}S^{y}H^{x}_{0}S^{z}\left[ \int_{0}^{T/2}\int_{0}^{t_{1}}\int_{0}^{t_{2}} dt_{1}dt_{2}dt_{3}cos\left(2\theta_{1}\right)sin\left(2\theta_{3}\right) +  \int_{T/2}^{T}\int_{0}^{T/2}\int_{0}^{t_{2}} dt_{1}dt_{2}dt_{3}cos\left(2\phi_{1}\right)sin\left(2\theta_{3}\right) \right. \nonumber \\ & \left. + \int_{T/2}^{T}\int_{T/2}^{t_{1}}\int_{0}^{T/2} dt_{1}dt_{2}dt_{3}cos\left(2\phi_{1}\right)sin\left(2\theta_{3}\right)  +  \int_{T/2}^{T}\int_{T/2}^{t_{1}}\int_{T/2}^{t_{2}} dt_{1}dt_{2}dt_{3}cos\left(2\phi_{1}\right)sin\left(2\phi_{3}\right) \right] \nonumber \\ & = -\frac{\left(h^{z}\right)^{2}}{16\left(h^{x}_{D}\right)^{3}}\left[ 1 - 4cos\left(h^{x}_{D}T\right) + 3cos\left(2h^{x}_{D}T\right) + 4h^{x}_{D}T sin\left(h^{x}_{D}T\right) \right] S^{y}H^{x}_{0}S^{z} \nonumber \\ & \xrightarrow{h^{x}_{D} = k\omega} \left[ 0 \right] S^{y}H^{x}_{0}S^{z}
\end{align}
\begin{align}
    & \left(h^{z}\right)^{2}S^{y}H^{x}_{0}S^{y}\left[ \int_{0}^{T/2}\int_{0}^{t_{1}}\int_{0}^{t_{2}} dt_{1}dt_{2}dt_{3}sin\left(2\theta_{1}\right)sin\left(2\theta_{3}\right) +  \int_{T/2}^{T}\int_{0}^{T/2}\int_{0}^{t_{2}} dt_{1}dt_{2}dt_{3}sin\left(2\phi_{1}\right)sin\left(2\theta_{3}\right) \right. \nonumber \\ & \left. + \int_{T/2}^{T}\int_{T/2}^{t_{1}}\int_{0}^{T/2} dt_{1}dt_{2}dt_{3}sin\left(2\phi_{1}\right)sin\left(2\theta_{3}\right)  +  \int_{T/2}^{T}\int_{T/2}^{t_{1}}\int_{T/2}^{t_{2}} dt_{1}dt_{2}dt_{3}sin\left(2\phi_{1}\right)sin\left(2\phi_{3}\right) \right] \nonumber \\ & = \frac{\left(h^{z}\right)^{2}}{16\left(h^{x}_{D}\right)^{3}}\left[ 2h^{x}_{D}T - 8h^{x}_{D}T cos\left(h^{x}_{D}T\right) + 3sin\left(2h^{x}_{D}T\right) \right] S^{y}H^{x}_{0}S^{y} \nonumber \\ & \xrightarrow{h^{x}_{D} = k\omega} \left[-\frac{3\left(h^{z}\right)^{2}T}{8\left(h^{x}_{D}\right)^{2}}\right] S^{y}H^{x}_{0}S^{y}
\end{align}
\begin{align}
    & -\left(h^{z}\right)^{2}S^{y}S^{z}H^{x}_{0}\left[ \int_{0}^{T/2}\int_{0}^{t_{1}}\int_{0}^{t_{2}} dt_{1}dt_{2}dt_{3}sin\left(2\theta_{1}\right)cos\left(2\theta_{2}\right) +  \int_{T/2}^{T}\int_{0}^{T/2}\int_{0}^{t_{2}} dt_{1}dt_{2}dt_{3}sin\left(2\phi_{1}\right)cos\left(2\theta_{2}\right) \right. \nonumber \\ & \left. + \int_{T/2}^{T}\int_{T/2}^{t_{1}}\int_{0}^{T/2} dt_{1}dt_{2}dt_{3}sin\left(2\phi_{2}\right)cos\left(2\phi_{2}\right)  +  \int_{T/2}^{T}\int_{T/2}^{t_{1}}\int_{T/2}^{t_{2}} dt_{1}dt_{2}dt_{3}sin\left(2\phi_{1}\right)cos\left(2\phi_{2}\right) \right] \nonumber \\ & = \frac{\left(h^{z}\right)^{2}}{32\left(h^{x}_{D}\right)^{3}}\left[ 5 + 2\left(h^{x}_{D}T\right)^{2} - 8cos\left(2h^{x}_{D}T\right) + 3cos\left(2h^{x}_{D}T\right) + 4h^{x}_{D}T \left(sin\left(2h^{x}_{D}T\right) - 2sin\left(h^{x}_{D}T\right)\right) \right] S^{y}S^{z}H^{x}_{0} \nonumber \\ & \xrightarrow{h^{x}_{D} = k\omega} \left[\frac{\left(h^{z}\right)^{2}T^{2}}{16h^{x}_{D}}\right]S^{y}S^{z}H^{x}_{0}
\end{align}
\begin{align}
    & \left(h^{z}\right)^{3}S^{y}\left(S^{z}\right)^{2}\left[ \int_{0}^{T/2}\int_{0}^{t_{1}}\int_{0}^{t_{2}} dt_{1}dt_{2}dt_{3}sin\left(2\theta_{1}\right)cos\left(2\theta_{2}\right)cos\left(2\theta_{3}\right) \right. \nonumber \\ & \left. +  \int_{T/2}^{T}\int_{0}^{T/2}\int_{0}^{t_{2}} dt_{1}dt_{2}dt_{3}sin\left(2\phi_{1}\right)cos\left(2\theta_{2}\right)cos\left(2\theta_{3}\right) \right. \nonumber \\ & \left. + \int_{T/2}^{T}\int_{T/2}^{t_{1}}\int_{0}^{T/2} dt_{1}dt_{2}dt_{3}sin\left(2\phi_{1}\right)cos\left(2\phi_{2}\right)cos\left(2\theta_{3}\right) \right. \nonumber \\ & \left. +  \int_{T/2}^{T}\int_{T/2}^{t_{1}}\int_{T/2}^{t_{2}} dt_{1}dt_{2}dt_{3}sin\left(2\phi_{1}\right)cos\left(2\phi_{2}\right)cos\left(2\phi_{3}\right) \right] \nonumber \\ & = \frac{\left(h^{z}\right)^{3}}{24\left(h^{x}_{D}\right)^{3}}\left[ 5 - 3cos\left(h^{x}_{D}T\right) - 3cos\left(2h^{x}_{D}T\right) + cos\left(3h^{x}_{D}T\right) - 3h^{x}_{D}T sin\left(h^{x}_{D}T\right) \right] S^{y}\left(S^{z}\right)^{2} \nonumber \\ & \xrightarrow{h^{x}_{D} = k\omega} \left[ 0 \right] S^{y}\left(S^{z}\right)^{2}
\end{align}
\begin{align}
    & -\left(h^{z}\right)^{3}S^{y}S^{z}S^{y}\left[ \int_{0}^{T/2}\int_{0}^{t_{1}}\int_{0}^{t_{2}} dt_{1}dt_{2}dt_{3}sin\left(2\theta_{1}\right)cos\left(2\theta_{2}\right)sin\left(2\theta_{3}\right) \right. \nonumber \\ & \left. +  \int_{T/2}^{T}\int_{0}^{T/2}\int_{0}^{t_{2}} dt_{1}dt_{2}dt_{3}sin\left(2\phi_{1}\right)cos\left(2\theta_{2}\right)sin\left(2\theta_{3}\right) \right. \nonumber \\ & \left. + \int_{T/2}^{T}\int_{T/2}^{t_{1}}\int_{0}^{T/2} dt_{1}dt_{2}dt_{3}sin\left(2\phi_{1}\right)cos\left(2\phi_{2}\right)sin\left(2\theta_{3}\right) \right. \nonumber \\ & \left. +  \int_{T/2}^{T}\int_{T/2}^{t_{1}}\int_{T/2}^{t_{2}} dt_{1}dt_{2}dt_{3}sin\left(2\phi_{1}\right)cos\left(2\phi_{2}\right)sin\left(2\phi_{3}\right) \right] \nonumber \\ & = -\frac{\left(h^{z}\right)^{3}}{24\left(h^{x}_{D}\right)^{3}}\left[ 6h^{x}_{D}Tcos\left(h^{x}_{D}T\right) + 3sin\left(h^{x}_{D}T\right) - 6sin\left(2h^{x}_{D}T\right) + sin\left(3h^{x}_{D}T\right) \right] S^{y}S^{z}S^{y} \nonumber \\ & \xrightarrow{h^{x}_{D} = k\omega} \left[ -\frac{\left(h^{z}\right)^{3}T}{4\left(h^{x}_{D}\right)^{2}} \right] S^{y}S^{z}S^{y}
\end{align}
\begin{align}
    & \left(h^{z}\right)^{2}\left(S^{y}\right)^{2}H^{x}_{0}\left[ \int_{0}^{T/2}\int_{0}^{t_{1}}\int_{0}^{t_{2}} dt_{1}dt_{2}dt_{3}sin\left(2\theta_{1}\right)sin\left(2\theta_{2}\right) +  \int_{T/2}^{T}\int_{0}^{T/2}\int_{0}^{t_{2}} dt_{1}dt_{2}dt_{3}sin\left(2\phi_{1}\right)sin\left(2\theta_{2}\right) \right. \nonumber \\ & \left. + \int_{T/2}^{T}\int_{T/2}^{t_{1}}\int_{0}^{T/2} dt_{1}dt_{2}dt_{3}sin\left(2\phi_{1}\right)sin\left(2\phi_{2}\right)  +  \int_{T/2}^{T}\int_{T/2}^{t_{1}}\int_{T/2}^{t_{2}} dt_{1}dt_{2}dt_{3}sin\left(2\phi_{1}\right)sin\left(2\phi_{2}\right) \right] \nonumber \\ & = \frac{\left(h^{z}\right)^{2}}{32\left(h^{x}_{D}\right)^{3}}\left[ 2h^{x}_{D}T\left( 5 - 4cos\left(h^{x}_{D}T\right) + 2cos\left(2h^{x}_{D}T\right)\right) - 3sin\left(2h^{x}_{D}T\right) \right] \left(S^{y}\right)^{2}H^{x}_{0} \nonumber \\ & \xrightarrow{h^{x}_{D} = k\omega} \left[\frac{3\left(h^{z}\right)^{2}T}{16\left(h^{x}_{D}\right)^{2}}\right]\left(S^{y}\right)^{2}H^{x}_{0}
\end{align}
\begin{align}
    & -\left(h^{z}\right)^{3}\left(S^{y}\right)^{2}S^{z}\left[ \int_{0}^{T/2}\int_{0}^{t_{1}}\int_{0}^{t_{2}} dt_{1}dt_{2}dt_{3}sin\left(2\theta_{1}\right)sin\left(2\theta_{2}\right)cos\left(2\theta_{3}\right) \right. \nonumber \\ & \left. +  \int_{T/2}^{T}\int_{0}^{T/2}\int_{0}^{t_{2}} dt_{1}dt_{2}dt_{3}sin\left(2\phi_{1}\right)sin\left(2\theta_{2}\right)cos\left(2\theta_{3}\right) \right. \nonumber \\ & \left. + \int_{T/2}^{T}\int_{T/2}^{t_{1}}\int_{0}^{T/2} dt_{1}dt_{2}dt_{3}sin\left(2\phi_{1}\right)sin\left(2\phi_{2}\right)cos\left(2\theta_{3}\right) \right. \nonumber \\ & \left. +  \int_{T/2}^{T}\int_{T/2}^{t_{1}}\int_{T/2}^{t_{2}} dt_{1}dt_{2}dt_{3}sin\left(2\phi_{1}\right)sin\left(2\phi_{2}\right)cos\left(2\phi_{3}\right) \right] \nonumber \\ & = -\frac{\left(h^{z}\right)^{3}}{24\left(h^{x}_{D}\right)^{3}}\left[ 6sin\left(h^{x}_{D}T\right) - 3h^{x}_{D}T cos\left(h^{x}_{D}T\right) - 3sin\left(2h^{x}_{D}T\right) + sin\left(3h^{x}_{D}T\right) \right] \left(S^{y}\right)^{2}S^{z} \nonumber \\ & \xrightarrow{h^{x}_{D} = k\omega} \left[ \frac{\left(h^{z}\right)^{3}T}{8\left(h^{x}_{D}\right)^{2}} \right] \left(S^{y}\right)^{2}S^{z}
\end{align}
\begin{align}
    & \left(h^{z}\right)^{3}\left(S^{y}\right)^{3}\left[ \int_{0}^{T/2}\int_{0}^{t_{1}}\int_{0}^{t_{2}} dt_{1}dt_{2}dt_{3}cos\left(2\theta_{1}\right)cos\left(2\theta_{2}\right)cos\left(2\theta_{3}\right) \right. \nonumber \\ & \left. +  \int_{T/2}^{T}\int_{0}^{T/2}\int_{0}^{t_{2}} dt_{1}dt_{2}dt_{3}cos\left(2\phi_{1}\right)cos\left(2\theta_{2}\right)cos\left(2\theta_{3}\right) \right. \nonumber \\ & \left. + \int_{T/2}^{T}\int_{T/2}^{t_{1}}\int_{0}^{T/2} dt_{1}dt_{2}dt_{3}cos\left(2\phi_{1}\right)cos\left(2\phi_{2}\right)cos\left(2\theta_{3}\right) \right. \nonumber \\ & \left. +  \int_{T/2}^{T}\int_{T/2}^{t_{1}}\int_{T/2}^{t_{2}} dt_{1}dt_{2}dt_{3}cos\left(2\phi_{1}\right)cos\left(2\phi_{2}\right)cos\left(2\phi_{3}\right) \right] \nonumber \\ & = \frac{\left(h^{z}\right)^{3}}{3\left(h^{x}_{D}\right)^{3}}\left[ 4sin^{6}\left(\frac{h^{x}_{D}T}{2}\right) \right] \left(S^{y}\right)^{3} \nonumber \\ & \xrightarrow{h^{x}_{D} = k\omega} \left[ 0 \right] \left(S^{y}\right)^{3}
\label{EqSup:UI_Sys_3rdOrd_Term27}
\end{align}
\paragraph{The Third Order Floquet Unitary and Effective Hamiltonian}
Collecting all the above terms and grouping them appropriately, one can write down the third-order contribution to the Floquet unitary. Here we do not write it down explicitly. Rather we concentrate on the 3rd-order contribution to the effective Hamiltonian which is given as
\begin{equation}
H_{F}^{\left(3\right)} = \frac{i}{T}\left[ U^{I}_{3}(T,0) - U^{I}_{1}(T,0)U^{I}_{2}(T,0) + \frac{1}{3}\left(U^{I}_{1}(T,0)\right)^{3} \right]
\end{equation} 
But we already know that in our case, we have
\begin{equation}
    U^{I}_{2}(T,0) = \frac{1}{2}\left(U^{I}_{1}(T,0)\right)^{2}
\end{equation}
Using this, the formula for 3rd order contribution to the effective Hamiltonian simplifies to
\begin{equation}
    H_{F}^{\left(3\right)} = \frac{i}{T}\left[ U^{I}_{3}(T,0) - \frac{1}{6}\left(U^{I}_{1}(T,0)\right)^{3} \right]
\label{EqSup:UI_Sys_3rdOrd_Def}
\end{equation}

\begin{figure*}[htb]
\begin{center}
\includegraphics[width=0.7\linewidth]{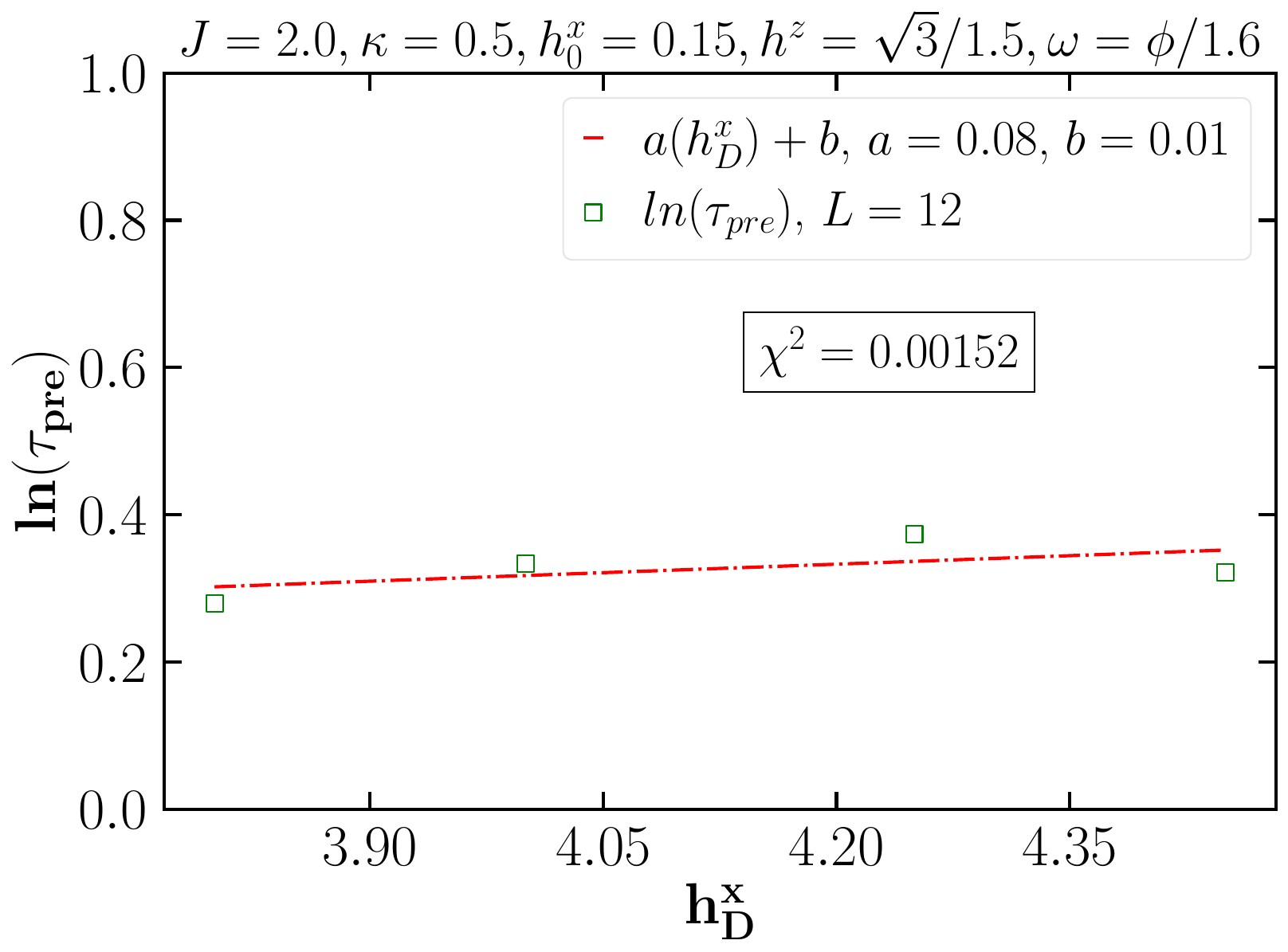}
\end{center}
\caption{Fitting the prethermal time scale $\tau_{pre}$ of exponential decay in the thermalizing regime. 
}
\label{FigSup_Method:2:tau_pre_Fitting}
\end{figure*}    
%
%
Even after this simplification, the form of the 3rd order contribution to the effective Hamiltonian remains very complicated. So, instead of writing down the 3rd-order contribution in general, we calculate it after imposing the freezing condition $h^{x}_{D}$ = $k\omega$, where $k$ is an integer (this is the condition under which the ECOs are most accurate).
Now, after imposing the freezing condition, the effective Hamiltonian up to 2nd order consists of only $H^x_0$. It is quite clear from Eq. (\ref{EqSup:UI_Sys_3rdOrd_Def}) that this term exactly cancels the 3rd order term in Eq. (\ref{EqSup:UI_Sys_3rdOrd_Term1}). So, apart from the term $H^x_0$ which comes from the first order, the third order contribution to the effective Hamiltonian consists of all the terms from Eq. (\ref{EqSup:UI_Sys_3rdOrd_Term2}) to Eq. (\ref{EqSup:UI_Sys_3rdOrd_Term27}). From Eq. (\ref{EqSup:UI_3rdOrd_Def}) and Eq. (\ref{EqSup:Heff_3rdOrd_Def}) it is clear that while contributing to the effective Hamiltonian, all the 3rd-order terms from Eq. (\ref{EqSup:UI_Sys_3rdOrd_Term2}) to Eq. (\ref{EqSup:UI_Sys_3rdOrd_Term27}) have to be multiplied by (-1/T). Keeping all these in mind, in the next section, we write down the effective Hamiltonian up to 3rd order.

\subsection{The Effective Hamiltonian}
Here we give the expression for the effective Hamiltonian ($H_{eff}$) up to 3rd order in Floquet Perturbation Theory. Note that we are not providing the most general expression of $H_{eff}$. Rather, we are writing down only those terms that survive after the freezing condition ($h^{x}_{D}$ = k$\omega$, where k is an integer) is imposed. \\

We split the entire effective Hamiltonian into 12 parts (as shown in Eq. (\ref{EqSup:Heff_Sys_3rdOrd_FC_Def})) and then write down the 12 parts separately. The expression for $H_{eff}$ is as follows
\begin{equation}
H_{eff} = H_A + H_B + H_C + H_D + H_E + H_F + H_G + H_H + H_I + H_J + H_K + H_L
\label{EqSup:Heff_Sys_3rdOrd_FC_Def}
\end{equation}
where
\begin{equation}
H_A = H^{x}_{0}
\label{EqSup:Heff_Sys_3rdOrd_FC_TermA}
\end{equation}
\begin{equation}
H_B = \frac{h^{z}}{4\left(h^{x}_{D}\right)^{2}}\left[\left(H^{x}_{0}\right)^{2}S^{z} + S^{z}\left(H^{x}_{0}\right)^{2}\right]
\label{EqSup:Heff_Sys_3rdOrd_FC_TermB}
\end{equation}
\begin{equation}
H_C = -\frac{h^{z}T}{8h^{x}_{D}}\left[\left(H^{x}_{0}\right)^{2}S^{y} + S^{y}\left(H^{x}_{0}\right)^{2}\right]
\end{equation}
\begin{equation}
H_D = -\frac{h^{z}}{2\left(h^{x}_{D}\right)^{2}}\left[H^{x}_{0}S^{z}H^{x}_{0}\right]
\end{equation}
\begin{equation}
H_E = \frac{h^{z}T}{4h^{x}_{D}}\left[H^{x}_{0}S^{y}H^{x}_{0}\right]
\end{equation}
\begin{equation}
H_F = -\frac{1}{16}\left(\frac{h^{z}}{h^{x}_{D}}\right)^{2}\left[H^{x}_{0}\left(S^{z}\right)^{2} + \left(S^{z}\right)^{2}H^{x}_{0}\right]
\end{equation}
\begin{equation}
H_G = -\frac{3}{16}\left(\frac{h^{z}}{h^{x}_{D}}\right)^{2}\left[H^{x}_{0}\left(S^{y}\right)^{2} + \left(S^{y}\right)^{2}H^{x}_{0}\right]
\end{equation}
\begin{equation}
H_H = \frac{\left(h^{z}\right)^{2}T}{16h^{x}_{D}}\left[H^{x}_{0}\left(S^{y}S^{z} - S^{z}S^{y}\right) + \left(S^{z}S^{y} - S^{y}S^{z}\right)H^{x}_{0} \right]
\end{equation}
\begin{equation}
H_I = \frac{1}{8}\left(\frac{h^{z}}{h^{x}_{D}}\right)^{2}\left[S^{z}H^{x}_{0}S^{z}\right]
\end{equation}
\begin{equation}
H_J = \frac{3}{8}\left(\frac{h^{z}}{h^{x}_{D}}\right)^{2}\left[S^{y}H^{x}_{0}S^{y}\right]
\end{equation}
\begin{equation}
H_K = -\frac{\left(h^{z}\right)^{3}}{8\left( h^{x}_{D} \right)^{2}}\left[\left(S^{y}\right)^{2}S^{z} + S^{z}\left(S^{y}\right)^{2}\right]
\end{equation}
\begin{equation}
H_L = \frac{\left(h^{z}\right)^{3}}{4\left( h^{x}_{D} \right)^{2}}\left[S^{y}S^{z}S^{y} \right]
\label{EqSup:Heff_Sys_3rdOrd_FC_TermL}
\end{equation}
The term $H_A$ = $H^{x}_{0}$ in $H_{eff}$ is actually a first order contribution and it commutes with $m^x$, $C^{1}_{x}$ and $C^{2}_{x}$. The second order contribution to $H_{eff}$ is zero. The 3rd order contributes all the remaining (Eq. (\ref{EqSup:Heff_Sys_3rdOrd_FC_TermB}) to Eq. (\ref{EqSup:Heff_Sys_3rdOrd_FC_TermL})) terms which do not commute with $m^x$, $C^{1}_{x}$ and $C^{2}_{x}$.
%
\begin{figure*}[t!]
\begin{center}
\includegraphics[width=0.3253\linewidth]{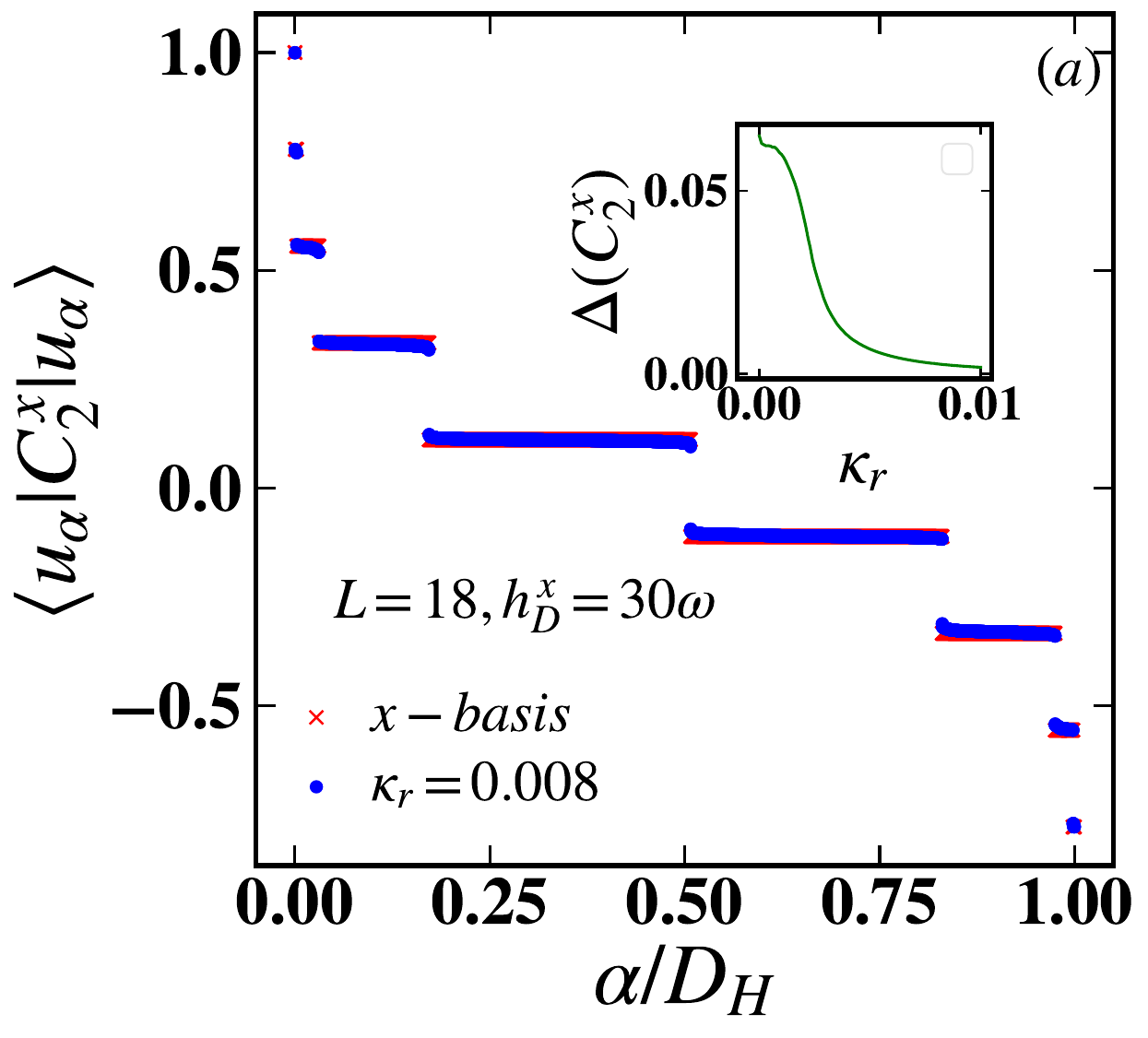}
\includegraphics[width=0.3253\linewidth]{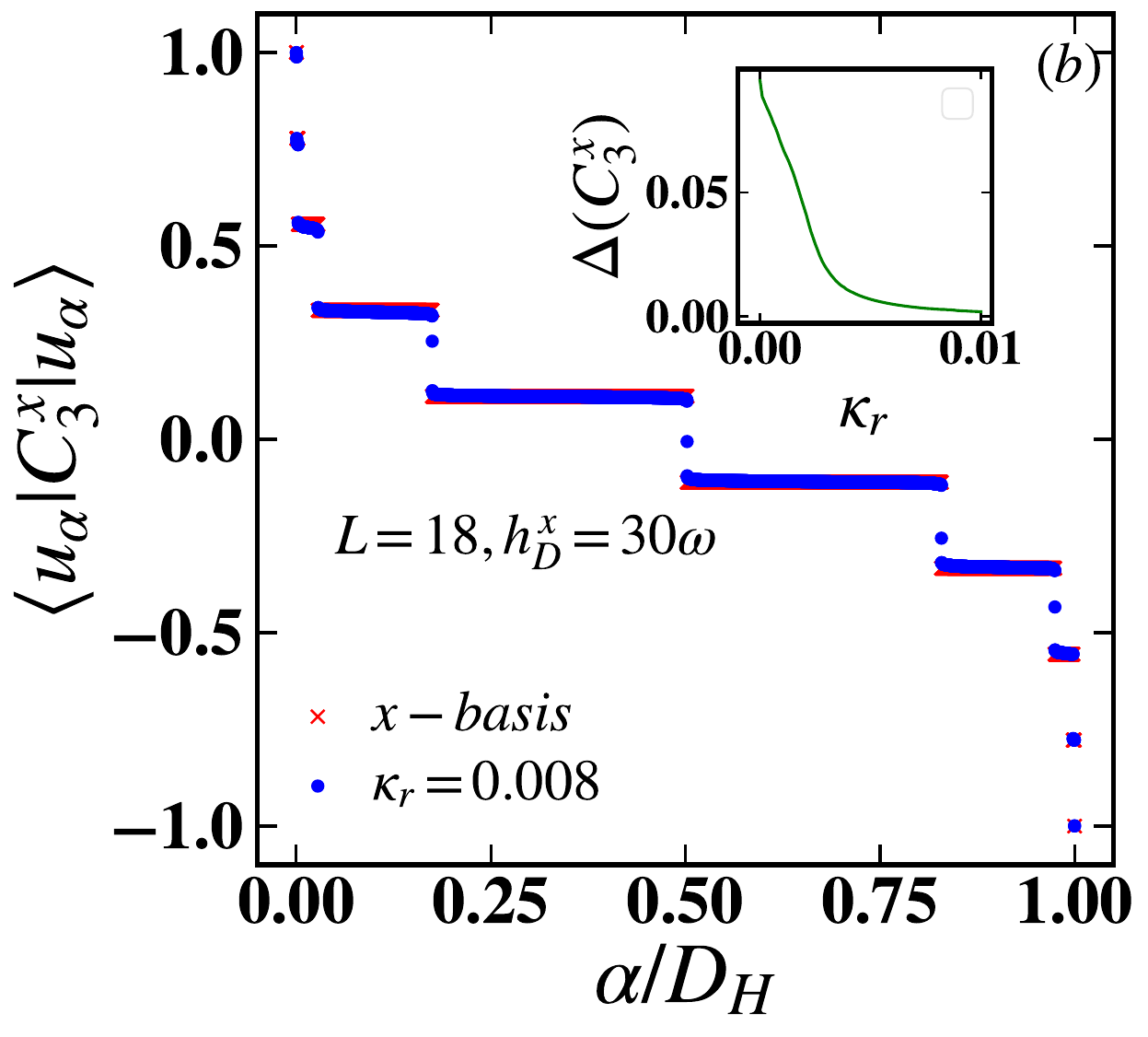}
\includegraphics[width=0.3253\linewidth]{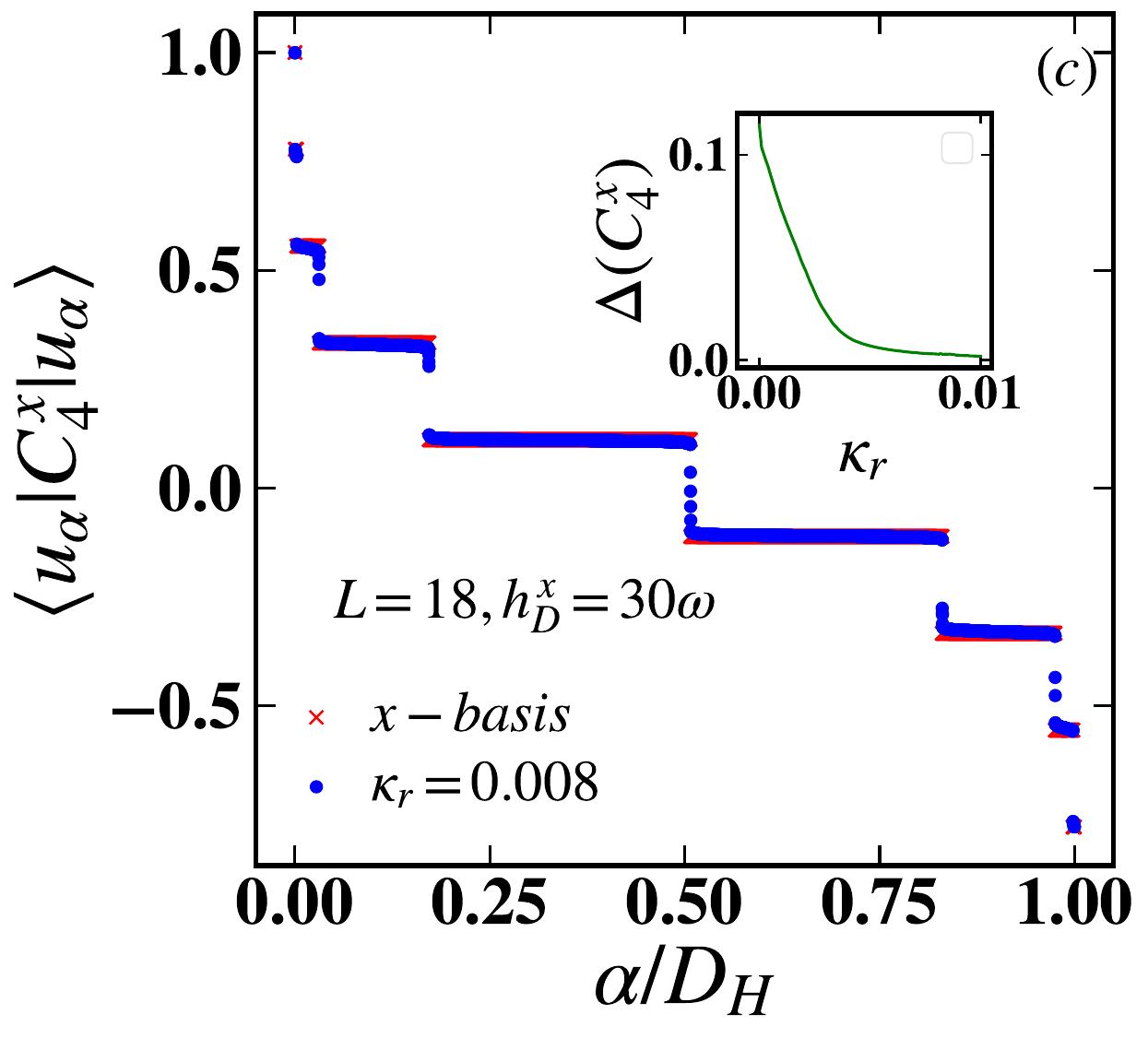}
\includegraphics[width=0.3253\linewidth]{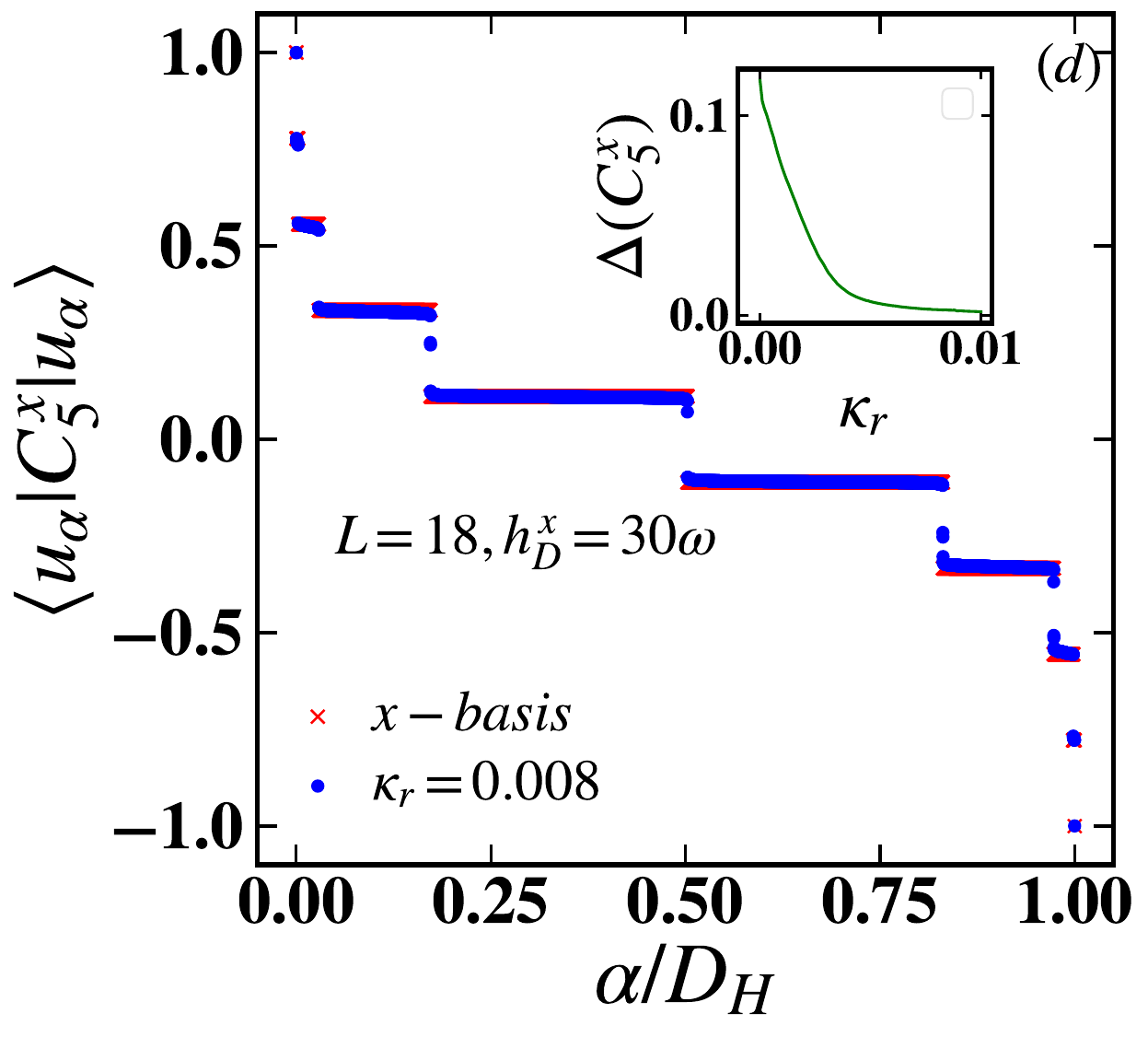}
\includegraphics[width=0.3253\linewidth]{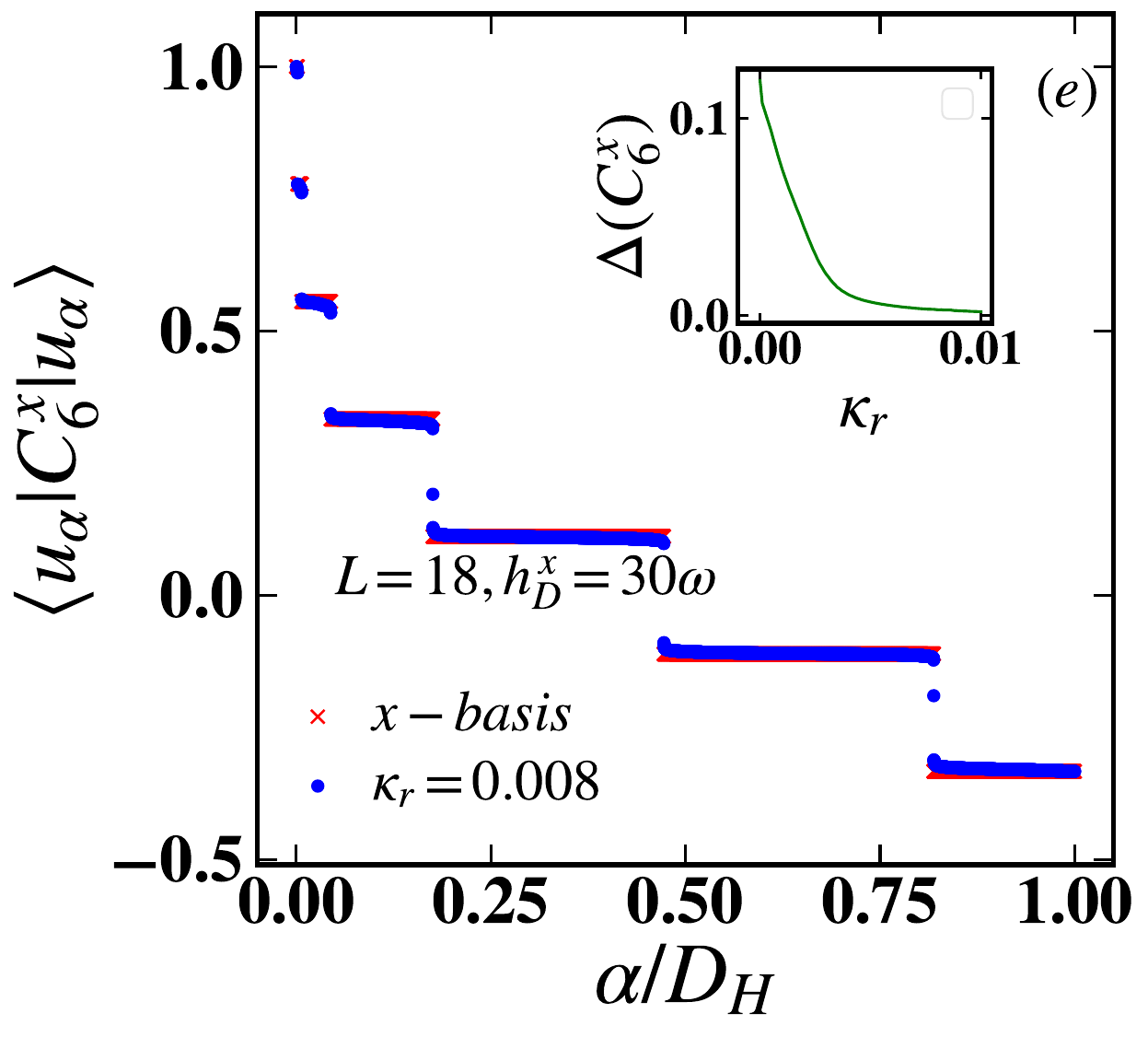}
\includegraphics[width=0.3253\linewidth]{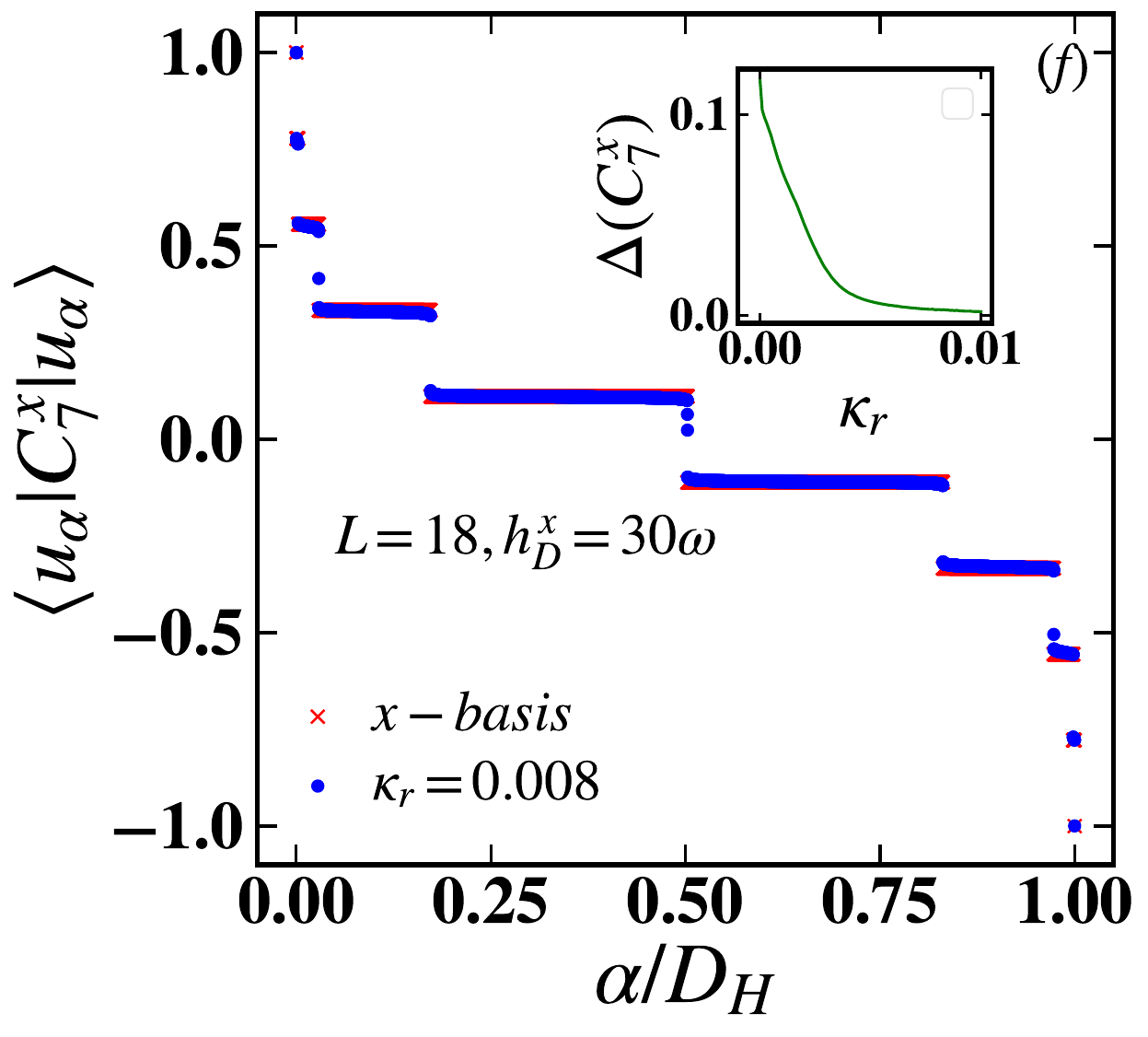}
\end{center}
\caption{{\bf Designing short and long-ranged ECOs $C_{r}^{x}$:} 
Replacing the static $C_{2}^{x}$ term in $H(t)$ by $C_{r}^{x}$
with a small coupling elevates $C_{r}^{x}$ to the status of an ECO. In each frame,
the main plot shows the step-like structure of the 
Floquet expectation-values of $C_{r}^{x},$ 
compared with the eigenvalues of $C_{r}^{x}.$ Insets show the rapid decline of 
$\Delta(C_{r}^{x})$ as a the function of strength $\kappa_{r}$ of the coupling of $C_{r}^{x}$
in the Hamiltonian.
Parameter values: $J=2.0, h^x_0 = 0.15, h^{z}=\sqrt{3}/1.5, \omega = \phi/1.6,$ where $\phi =$ Golden-mean, $L=18$. In this Fig. $\vert u_\alpha \rangle$s denote the Floquet eigenstates.
}
\label{FigSup:5:Cr_ECO_all}
\end{figure*}    
\section{Extracting the Prethermalization Time $\tau_{pre}$: The $h_{D}^{x}$ vs $\tau_{pre}$ Plot}

Here we estimate the prethermal time following the approach described in~\cite{Ho_Mori_Abanin_Pretherm_Rev}.
The prethermal $\tau_{pre}$ is given as follows.
\begin{equation}
    \tau_{pre} = \left(\frac{A}{\Lambda}\right)e^{C(\Omega / \Lambda)},
\label{EqSup:TakDef_TauPre}     
\end{equation}
\noindent
where $\Omega$ is the driving frequency, $\Lambda$ is the local
bandwidth estimated from the norm of the driven Hamiltonian (Eq.(1) in the Main text),
and $C$ \& $A$ are parameters that do not depend on $\Omega,$ but can depend on other parameters of the Hamiltonian, and those are extracted from 
the fitting shown in Fig.~\ref{FigSup_Method:2:tau_pre_Fitting}.
In terms of the fitting parameters. We fit
\begin{equation}
    \ln{(\tau_{pre})} = a~h_{D}^{x} + b.
    \label{EqSup:tau_pre_Fitting_Eqn}
\end{equation}
\noindent
Comparing Eqs.~(\ref{EqSup:TakDef_TauPre}) and~(\ref{EqSup:tau_pre_Fitting_Eqn}), we have
\begin{equation}
    C = a\Lambda/2; ~ {\rm and} ~ 
    A = \Lambda e^{b}.
\end{equation}
We use those values of $A$ and $C$ to along with
the value of $\Lambda$ evaluated from $H(t)$ (see Methods Sec. in the Main text) to estimate
$\tau_{pre}.$
\section{Designing Emergent Conserved Operators (ECOs)}

Here, in Fig.~\ref{FigSup:5:Cr_ECO_all},
we show the results for emergent conservation of the operators 
\begin{equation}
C_{r}^{x} = \frac{1}{L}\sum_{i}\sigma_{i}^{x}
\sigma_{i+r}^{x},
    \label{Supp_Eq:ECO_Cr}
\end{equation}
\noindent as the function of their coupling
strength $\kappa_{r}$ 
in the static part of the driven Hamiltonian $H_{r}(t)$ given below.
\bea H_{r}(t) &=& H_{0}(t) ~+~ V, ~~ {\rm where} \non \\
H_{0} (t) &=& H_{0}^{x} ~+~ \Sgn (\sin (\om t)) ~H_{D}, ~~ {\rm with} 
\non \\
H_{0}^{x} = &-& ~\sum_{n=1}^L ~J \si_n^x \si_{n+1}^x - \sum_{n=1}^L ~\ka_{r} 
\si_n^x \si_{n+r}^x - h_{0}^{x}~\sum_{n=1}^L\si_n^x, \non \\
H_{D} = &-& ~ h_D^x ~\sum_{n=1}^L \si_n^x, ~~ {\rm and} \non \\
V = &-& ~ h^z \sum_{n=1}^L \si_n^z, 
\label{Sup_Eq:Hr} 
\eea
\noindent where, $\si_{n}^{x/y/z}$ are the Pauli matrices, and $\Sgn{( ~ )}$ denotes the sign of its
argument.

\end{document}